%% file: main.tex
\documentclass[manuscript,screen,natbib=true]{acmart}

\setcopyright{acmlicensed}
\acmJournal{TORS}\acmYear{2025} \acmVolume{1} \acmNumber{1} \acmArticle{1} \acmMonth{1}\acmDOI{10.1145/3721298}

\input{config}
\begin{document}
\thanks{This is the author's version of the work. It is posted here for your personal use. 
Not for redistribution. The definitive version was published in \textit{Transactions on Recommender Systems}, 
DOI: \href{https://doi.org/10.1145/3721298}{10.1145/3721298}.}

\input{macros}
\title{Modeling and Analyzing the Influence of Non-Item Pages on Sequential Next-Item Prediction}

\author{Elisabeth Fischer} \email{elisabeth.fischer@informatik.uni-wuerzburg.de}
\orcid{https://orcid.org/0009-0000-7371-052X}
\author{Albin Zehe}
\orcid{https://orcid.org/0000-0002-9472-0783}
\email{zehe@informatik.uni-wuerzburg.de}
\author{Andreas Hotho}
\orcid{https://orcid.org/0000-0002-0483-5772}
\email{hotho@informatik.uni-wuerzburg.de}
\author{Daniel Schlör}
\email{schloer@informatik.uni-wuerzburg.de}
\orcid{https://orcid.org/0009-0001-6983-3719}
\affiliation{
  \institution{Data Science Chair, Center for Artificial Intelligence and Data Science (CAIDAS), Julius-Maximilians-Universität (JMU)}
  \city{Würzburg}
  \country{Germany}
}
\renewcommand{\shortauthors}{Fischer et al.}

\input{sections/00_abstract}




\maketitle

\input{sections/10_introduction}

\input{sections/20_related_work}
\input{sections/30_methodology}
\input{sections/40_dataset}
\input{sections/50_experiments}

\input{sections/60_discussion}
\input{sections/70_conclusion}

\begin{acks}
This work is partially supported by both the German Research Foundation (DFG) under grant number 438232455 (HydrAS) and the German Federal Ministry of Education and Research (BMBF) under grant number 01IS22051B (KILiMod). The authors are responsible for the content of this publication.
\end{acks}

\bibliographystyle{ACM-Reference-Format}
\bibliography{references-2}
\FloatBarrier

\appendix
\input{sections/80_appendix}
\end{document}

%% file: config.tex
\usepackage{xspace}
\usepackage{amsfonts}
\usepackage{amsmath}
\usepackage{hyperref}
\usepackage{xcolor}
\usepackage{placeins}
\usepackage{siunitx}
\usepackage{microtype}

\usepackage{caption}
\usepackage{subcaption}
\usepackage[inline]{enumitem}
\usepackage{cleveref}
\usepackage{multirow}
\usepackage{numprint}
\npdecimalsign{.}

%% file: macros.tex
\newcommand{\chairx}{Data Science Chair, Julius-Maximilians University Würzburg}

\newcommand{\revised}[1]{{#1}}

\newcommand{\todo}[2][TODO]{\marginpar{\fbox{\parbox{.9\marginparwidth}{\raggedright\textbf{#1:} #2}}}}
\newcommand{\komm}[2]{\textbf{/* #2 (#1) */}}
\newcommand{\aho}[1]{\komm{aho}{#1}}\newcommand{\maho}[1]{\todo[aho]{#1}}
\newcommand{\dzo}[1]{\komm{dzo}{#1}}\newcommand{\mdzo}[1]{\todo[dzo]{#1}}
\newcommand{\tni}[1]{\komm{tni}{#1}}\newcommand{\mtni}[1]{\todo[tni]{#1}}
\newcommand{\lsc}[1]{\komm{lsc}{#1}}\newcommand{\mlsc}[1]{\todo[lsc]{#1}}
\newcommand{\mbe}[1]{\komm{mbe}{#1}}\newcommand{\mmbe}[1]{\todo[mbe]{#1}}
\newcommand{\ada}[1]{\komm{ada}{#1}}\newcommand{\mada}[1]{\todo[ada]{#1}}
\newcommand{\mri}[1]{\komm{mri}{#1}}\newcommand{\mmri}[1]{\todo[mri]{#1}}
\newcommand{\dsc}[1]{\komm{dsc}{#1}}\newcommand{\mdsc}[1]{\todo[dsc]{#1}}
\newcommand{\aze}[1]{\komm{aze}{#1}}\newcommand{\maze}[1]{\todo[aze]{#1}}
\newcommand{\lfi}[1]{\komm{lfi}{#1}}\newcommand{\mlfi}[1]{\todo[lfi]{#1}}

\newcommand{\final}[1]{\textbf{/* for camera ready/long version: #1  */}}
\newcommand{\journal}[1]{\textbf{/* TODO for journal version: #1  */}}
\newcommand{\done}[2]{\emph{#1} (see {#2})}


\newcommand{\eg}{e.g.,\xspace}
\newcommand{\ie}{i.e.,\xspace}
\newcommand{\Eg}{E.g.,\xspace}
\newcommand{\Ie}{I.e.,\xspace}
\newcommand{\wrt}{w.r.t.\xspace}
\newcommand{\cf}{cf.\xspace}

\newcommand{\bibs}{BibSonomy\xspace}

\newcommand{\bertforrec}{BERT4Rec\xspace}
\newcommand{\userbert}{$\text{U-BERT4Rec}$\xspace}
\newcommand{\usersasrec}{$\text{U-SASRec}^{cross}$\xspace}
\newcommand{\newsasrec}{$\text{SASRec}^{cross}$\xspace}
\newcommand{\originalsas}{$\text{SASRec}^{neg}$\xspace}
\newcommand{\sasrec}{SASRec\xspace}
\newcommand{\bert}{BERT\xspace}
\newcommand{\keybertforrec}{Ke\bertforrec}
\newcommand{\kesasrec}{Ke\newsasrec}
\newcommand{\variantone}{$\text{KE}_{m}$\xspace}
\newcommand{\varianttwo}{$\text{KE}_{l}$\xspace}
\newcommand{\pagesasrec}{$\text{P-SASRec}^{c}$\xspace}
\newcommand{\pagebert}{$\text{P-Bert4Rec}$\xspace}
\newcommand{\nextitnet}{NextItNet\xspace}
\newcommand{\caser}{CASER\xspace}
\newcommand{\narm}{NARM\xspace}
\newcommand{\core}{CORE\xspace}
\newcommand{\gruforrec}{GRU4Rec\xspace}
\newcommand{\lightsans}{LightSANs\xspace}

\newcommand{\hr}[1]{HR@#1}

\newcommand{\movielenssystem}{Movielens\xspace}
\newcommand{\movielens}{Movielens-1m\xspace}
\newcommand{\coveo}{Coveo\xspace}
\newcommand{\coveopage}{Coveo-Pageview\xspace}
\newcommand{\coveosearch}{Coveo-Search\xspace}
\newcommand{\movielenslarge}{ML-20m\xspace}
\newcommand{\syndata}{SynDS\xspace}
\newcommand{\onlineshop}{Fashion\xspace}

\newcommand{\furl}[1]{\footnote{\url{#1}}}

%% file: sections/00_abstract.tex
\begin{abstract}

Analyzing sequences of interactions between users and items, sequential recommendation models can learn user intent and make predictions about the next item. Next to item interactions, most systems also have interactions with what we call non-item pages: these pages are not related to specific items but still can provide insights into the user's interests, as, for example, navigation pages. 
We therefore propose a general way to include these non-item pages in sequential recommendation models to enhance next-item prediction.

First, we demonstrate the influence of non-item pages on following interactions using the hypotheses testing framework HypTrails and propose methods for representing non-item pages in sequential recommendation models.
Subsequently, we adapt popular sequential recommender models to integrate non-item pages and investigate their performance with different item representation strategies as well as their ability to handle noisy data. 
To show the general capabilities of the models to integrate non-item pages, we create a synthetic dataset for a controlled setting and then evaluate the improvements from including non-item pages on two real-world datasets.

Our results show that non-item pages are a valuable source of information, and incorporating them in sequential recommendation models increases the performance of next-item prediction across all analyzed model architectures.

\end{abstract}

%% file: sections/10_introduction.tex
\section{Introduction}
\label{sec:introduction}

\begin{figure*}
\centering
\includegraphics[width=1\textwidth]{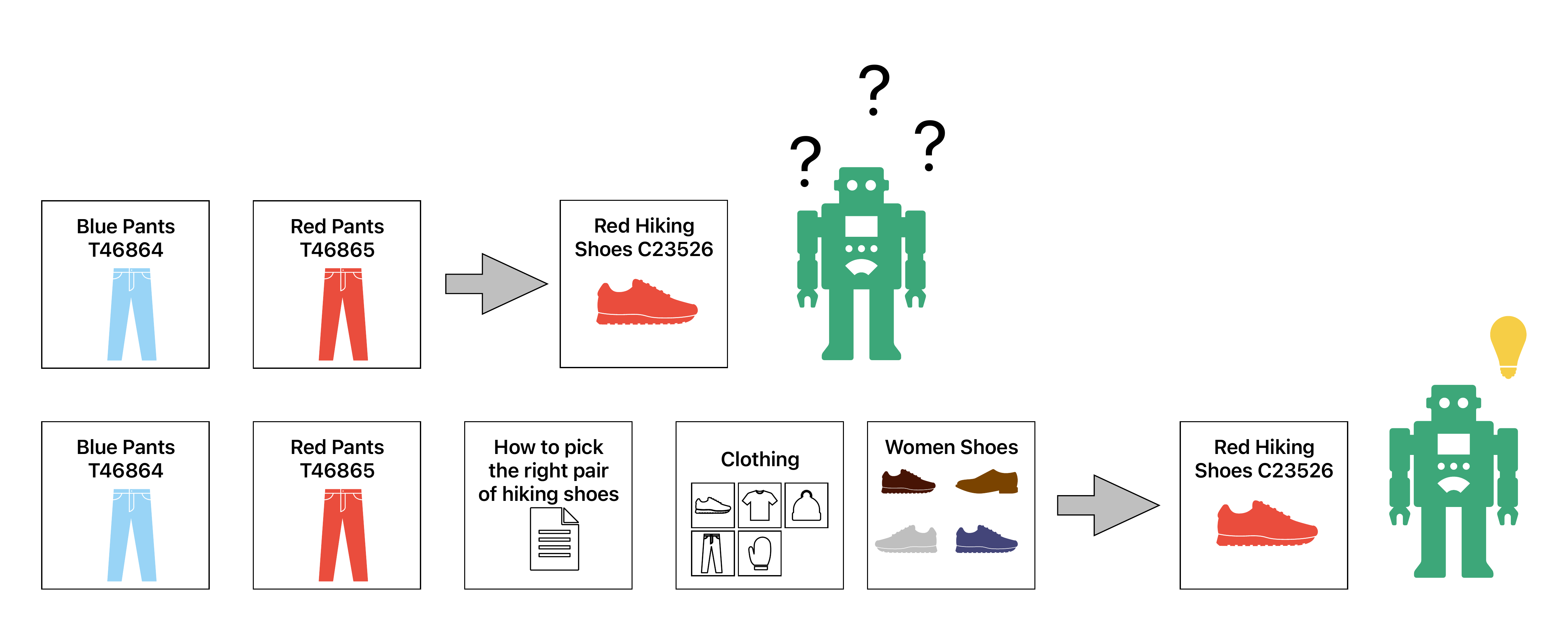}
\caption{Exemplary click sequence in an online store. 
The inclusion of non-item pages provides an explanation for the shift in user interest, enabling a more accurate prediction of their preferences.}
\label{fig:fig1}
\Description{Exemplary click sequence in an online store. }
\end{figure*}

Sequential recommender models are useful building blocks for providing personalized recommendations to users. 
By analyzing the sequence of historical interactions of the user with items,
sequential recommendation models can learn and infer user preferences and make predictions about the next item of interest. Recently, especially deep learning approaches based on recurrent neural networks (RNNs)~\citep{hidasi_session-based_2016}, convolutional neural networks (CNNs)~\citep{tang_personalized_2018}, and Transformer models~\citep{kang_self-attentive_2018,sun_bert4rec_2019} have received attention, typically modeling each interaction with an item as one step in the sequence.
Significant amounts of research have been conducted that incorporates additional item information, for example,  textual and visual item features~\cite{hidasi_parallel_2016,tuan_3d_2017}, item attribute embeddings~\cite{fischer_integrating_2020,liu_non-invasive_2021}, and categorical and numerical item features~\cite{de_souza_pereira_moreira_transformers4rec_2021}.
Apart from click sequences on item pages, most systems utilizing recommenders also incorporate additional forms of information. This includes data about users and pages that do not directly represent an item but still involve a user interaction.
Although there are many papers exploring strategies to incorporate user-specific information \cite{tang_personalized_2018,chen_behavior_2019,wu_sse-pt_2020,fischer_integrating_2020}, the aspect of leveraging information from non-items in sequences to further improve recommendation has been less explored. 
However, in many applications, interactions with pages that are not items themselves or cannot even be associated with a particular item (hereafter referred to as non-item pages, cf. \Cref{sec:nonitempages}) still provide useful information in the context of sequential recommendation.
Examples of these pages include search results or blog posts that contain descriptions of entire item categories, which can be very helpful in detecting the user's intent when browsing through a website.
\Cref{fig:fig1} shows a case where including these non-item pages can be useful:
When only looking at the item history, we see that the user switches from pants to hiking shoes without any explanation.
Including non-item pages, we see that the user is interested in hiking equipment, which explains their next step of looking at hiking shoes and enables the recommender to specifically suggest other hiking shoes as opposed to more pants.

We argue that ignoring this information, which is available in most real-world settings (cf. \Cref{sec:nonitempages}), limits the potential of recommender systems to follow the user's intent.
To verify our intuition that non-item pages provide valuable information, we analyze our datasets with the HypTrails framework~\cite{singer_hyptrails_2015}. We formalize and test our hypothesis that non-item pages influence the following interaction, finding that this is indeed the case.
To leverage this information, we propose incorporating non-item pages and present various modeling approaches for non-item pages for sequential next-item prediction for a variety of commonly used recommender models. 
As a proof of concept for our approach in a controlled setting, we create a synthetic dataset with artificial non-item pages, which are highly related to the following item.
We additionally study the influence of noisy non-item pages with varying degree of relatedness in this setting.
Finally, we evaluate the effectiveness of our modeling on two real-world datasets from the e-commerce domain.

We compare different types of non-item page representations (based on their content or a unique id) and evaluate eight popular sequential recommender models on click sequences enriched by these representations.
Specifically, we consider three types of non-item pages: (1) a list of items, (2) a single non-item, e.g. a blog post, and (3) a list of non-items.

To cover a wide variety of popular or recent architectures, we adapt several models in the RecBole \cite{zhao_recbole_2021} framework: \gruforrec, a popular RNN baseline model, \caser and \nextitnet as models based on CNNs and several models based on the attention mechanism: \narm combining RNNs and attention, \sasrec as the first Transformer-based model, and \bertforrec adapting bidirectional Transformers. We also include \core, which couples the encoding and decoding representation spaces and \lightsans, which introduces a low-rank decomposed self-attention.
We extend our evaluation by shedding light on the specific representations of items and non-items across different models and representation types, including an analysis of item embedding similarities across various configurations.

In summary, our proposed modeling approaches and evaluation address the gap in current research regarding the modeling of non-item pages in sequential recommendation systems and show the utility of including such information. 

To sum up our contributions, we:
\begin{enumerate}[label=\arabic*)]
\item validate our hypothesis that non-item pages influence subsequent interactions and
\item propose various modeling approaches for non-item pages.
\item We adapt a wide set of models to include categorical and vector representations of pages in a generic way and
\item  analyze the capabilities of different architectures to include these on one synthetic and two real-world datasets.
\end{enumerate}
Our results show that non-items can reveal valuable information about the user's intent, but also the importance of finding good representations for non-item pages. All adapted models are capable of taking advantage of the additional non-item pages, depending on the representations, showing the general applicability of our approach.

This paper is a direct extension of our work in \cite{fischer_enhancing_2023}, which included only \sasrec and \bertforrec. We expand our work by generalizing the approach and applying it to a wider set of sequential recommender architectures in a popular recommender framework. We expand our experiments to examine the behavior of models on noisy data and provide deeper insights on the effects of different models and embedding variants on the resulting item representations. Our results confirm previous findings and underline the general utility of including non-item pages. 

The remainder of our paper is structured as follows: \Cref{sec:related_work} provides an overview of related work. We introduce the concept and proposed modeling approaches for non-item pages in \Cref{sec:methodology}, as well as our proposed adaptions to sequential recommender models to include them. In \Cref{sec:data} we present the datasets, while \Cref{sec:experiments} covers our experiments. We start by testing our hypotheses about non-item pages, before analyzing their inclusion in sequential recommenders in different setups. We discuss some of the general findings and compare them with the results of our previous paper in \Cref{sec:discussion}. Our conclusion and closing remarks can be found in \Cref{sec:conclusion}.


%% file: sections/20_related_work.tex
\section{Related Work}
\label{sec:related_work}

In this section, we first present an overview on deep sequential recommendation and then focus on architectures handling information apart from the sequence of item interactions themselves.
With the increasing popularity of neural networks, neural architectures have been widely employed for modeling and learning sequences of user interactions~\cite{hidasi_session-based_2016,tang_personalized_2018,xu_recurrent_2019,li_neural_2017,kang_self-attentive_2018,sun_bert4rec_2019}. 
While these works focused on the dependencies on a per-item level, some studies aimed at incorporating additional information along with the items, which can be sorted into two specific groups, either adding additional information for items or adding additional information to model the user.

The inclusion of additional item information has been the goal of several modifications of the models mentioned above. 
Hidasi et al.~\cite{hidasi_parallel_2016}, for example, model textual and visual item features in separate RNNs and combine their output afterward, while Tuan et al.~\cite{tuan_3d_2017} include character-wise item descriptions using CNNs with 3D convolutions.
NovaBert~\cite{liu_non-invasive_2021} uses bidirectional transformers and merges item attributes directly in the attention mechanism as query and value, as do the authors of ~\cite{xie_decoupled_2022}. Fischer et al.~\cite{fischer_integrating_2020} allow the integration of categorical item attributes in \bertforrec by adding attribute embeddings to the item embeddings, while AttriBERT ~\cite{jagatap_attribert_2023} uses attribute dictionaries to represent products. Transformers4Rec~\cite{de_souza_pereira_moreira_transformers4rec_2021} uses models like XLNet~\cite{yang_xlnet_2019} or Electra \cite{clark_electra_2020} from the Natural Language Processing domain and integrates categorical and numerical item features. Liu et al.~\cite{liu_context-aware_2016} show how the inclusion of context (being time and location) can be utilized in a RNN. 

Regarding user information,  several approaches have been proposed, for example, CASER~\cite{tang_personalized_2018}, SSE-PT~\cite{wu_sse-pt_2020}, the work of Chen et al.~\cite{chen_behavior_2019} and TARN \cite{zhang_neural_2022}, which combines user, item, and temporal context. They treat the user separately from the sequence, while the approach by Fischer et al.~\cite{fischer_integrating_2020} integrates the user representation as the first item in a sequence. This is more similar to our usecase, but in contrast to user representations, interactions with non-item pages can appear several times and have specific positions in the sequence. 

Some works that participated in the SIGIR Coveo Challenge 2021~\cite{tagliabue_sigir_2021} are notable exceptions, as they include additional information that is not related to items or users.
The Coveo dataset contains page views that are not related to items, as well as search results. Three participants in the challenge,~\cite{moreira_transformers_2021,ishihara_adversarial_2021,fischer_comparison_2021}, embed page views as an item in the sequence for transformer models, while~\cite{sakatani_session-based_2021} uses the count of page views and the first page view as features for an ensemble model of LSTMs and Matrix Factorization. Moreira et al.~\cite{moreira_transformers_2021} also used search queries as context for the whole sequence. The effect of including page events in the sequence is only reported by~\cite{fischer_comparison_2021}, 
which can be considered to be most similar to our work.
However, since the focus of their work is particularly on the challenge, the benefits of non-item information and different modeling approaches are not systematically explored and cannot be investigated completely from outside the challenge's evaluation framework.
\revised{
Following this, \cite{fischer_enhancing_2023} introduces the notion of non-items to be included in sequential recommender models, which is the work we directly build upon here.
This modeling has since then also been adapted to detect mentions of entities in chat conversations~\cite{zehe_adapting_2024}.
}

In contrast to previous works, we verify our modeling approaches on several datasets and recommender architectures and systematically model various types of non-item pages in the sequence to consider the dependencies to items in chronological sequence order.


%% file: sections/30_methodology.tex
\section{Methodology}
\label{sec:methodology}

In this section we discuss the problem setting, including our modeling approach for various types of non-item pages. We introduce the setting for HypTrails as well as the foundations of sequential recommender models and our adaptations.

\subsection{Problem Setting}
\label{sec:setting}
\newcommand{\userset}{\mathcal{U}}
\newcommand{\itemset}{\mathcal{V}}
\newcommand{\interactionset}{\mathcal{\itemset \cup \pageset}}
\newcommand{\pageset}{\mathcal{LP}}
\newcommand{\sessionset}{\mathcal{S}}
\newcommand{\transitionset}{\mathcal{M}}
\newcommand{\hypothesismatrix}{H}
\newcommand{\hypstates}{\mathcal{K}}
\newcommand{\categoryset}{\mathcal{\hat{A}}}
\newcommand{\permset}{\mathcal{\hat{C}}}
\newcommand{\categorysetcat}{\mathcal{A}_{cat}}
\newcommand{\categorysetfilter}{\mathcal{A}_{fil}}
\newcommand{\sessioninteraction}{i_t}
\newcommand{\pageinteraction}{p_t^{u}}
\newcommand{\attribute}{\hat{a}}
\newcommand{\itemattributes}[1]{\hat{c}_{#1}}

To formalize the problem setting, we closely follow \cite{fischer_integrating_2020} and introduce our notation as follows:
Our work aims to solve the task of recommending items to a user given the sequence of previous items and non-items.
We define the item set as $\itemset=\{v_1, v_2, \ldots, v_{|\itemset|}\}$ and the set of non-items as $\pageset=\{p_1, p_2, \ldots, p_{|\pageset|}\}$.
An interaction $i$ is either an item $v \in \itemset$ or a non-item page $p \in \pageset$.
The set of users is defined as $\userset= \{u_1, u_2, \ldots, u_{|\userset|}\}$.
For each user $u$, we can denote the sequence of interactions as 
$s_u = [i_1, i_2, \ldots,  i_{n}]\subseteq \{\interactionset \}^n $. We use the term `\textit{session}' synonymously for a sequence of interactions. The set of all sequences is defined as $\sessionset = \{s_{u_1}, s_{u_2}, \ldots, s_{u_{|\userset|}}\}$. 
Formally written, our goal is to solve the task of predicting the next item $i_{n + 1} \in \itemset $ for every interaction $i_n$ in each sequence $s_u \in \sessionset$.

Additionally, we have an optional set $\categoryset$ of categorical attributes $\attribute$.
These attributes can be, for example, tags like ``category:shoes'' or ``genre:horror''.
The set of all possible combinations of attributes that occur in the dataset is $\permset \subset 2^\categoryset$.
An interaction $i$ can be associated with one or multiple attributes $\itemattributes{i} \in \permset$.
\input{sections/31_background}

\subsection{Modeling Non-Item Pages}
\label{sec:modnonitems}

For each of the three types of non-item pages (lists of items, lists of non-item pages and other non-item pages) we explore three options for representing the page: using or generating a unique id, generating a content-based id for the page, or generating an embedding for the page's content.

\paragraph{Unique Page ID (UPID)}
The simplest solution is to assign a unique id to each non-item page $p_n$.
This is possible for any kind of page, for example by employing the URL used to access the page, the title of the page or - as a last resort - by assigning a random id to each unique page.
However, this comes with several drawbacks: First, there are instances where the generated unique id is not informative. This is for example the case if a URL is generic and content is generated on the fly, so the relevant information is lost.
On the other hand, if the URL is \revised{too} specific (e.g. because it contains a search query), the associated id may be \revised{too} rare for a model to extract useful information.
The vocabulary size will increase and become unmanageable, and the models will learn to predict non-items, even though the goal is item recommendation. While predictions can be filtered afterwards, valuable capacity of the model will be used for a task irrelevant to next-item prediction. Nevertheless, we include UPID as a baseline.

\paragraph{Content-based Page ID (CPID)}
To tackle these challenges, we utilize the content of non-item pages.
We can use categorical attributes to create meaningful and shared ids for similar non-item pages.
In the easiest case, a non-item page $p_n \in \pageset$ already has some categorical attributes $\itemattributes{p_n} = \{\attribute_1, \attribute_2, \ldots, \attribute_{|\itemattributes{p_n}|} \}$, $\attribute \in \categoryset$, assigned, for example because it is a category list page (all items from category ``shoes'') or because it has been manually tagged.
For list pages we can also construct page attributes from the attributes of the items contained in the list.
Therefore the set of non-items becomes 
$
\pageset_\text{CPID} = \{ \itemattributes{p} \mid p \in \pageset \}$.
For non-list pages, this strategy is not applicable if there are no attributes assigned directly to the page.

\paragraph{Page Embedding (PE)}
By building ids from attributes, as proposed in the previous paragraph, our vocabulary still grows by size $|\pageset_\text{CPID}|$.
Therefore, we propose to use one single placeholder id to represent all non item pages and encode the content of a page with an embedding, mitigating the issue of increasing the vocabulary size.
This also allows the use of any vector representation for the page.
This representation $\hat{r}_{p_n} \in \mathbb{R}^{m}$ ($m$: dimension of the embedding vector) can be extracted from the categorical attributes of a page $p_n$, from textual content, images, or a search query or be given directly. In general, this embedding representation can be applied to any kind of non-item page.

\subsection{Verifying the Influence of Non-Item Pages using HypTrails}
\label{sec:meth:hyptrails}

\revised{
Following our intuition, it is reasonable to assume that non-item have an influence over the users' navigation behavior, as argued in \Cref{sec:introduction}.
Verifying this influence in a statistical way would provide a clear signal that incorporating non-items into a sequential recommender should enable the model to better understand the users' intent. 
To achieve this, we can formulate different hypotheses about the users' navigation behavior and compare them to see which one fits our data best.
In our case, we are mostly interested in two hypotheses, where one assumes that non-items pages are influential and the other that they are not.
Assuming that non-item pages are influential, we construct a ‘structural’ hypothesis, which expresses the belief that users only move from a non-item page to items of the page’s category.
The opposite hypothesis (`uniform') is that there is no influence, and that users move uniformly between pages of the shop.
\\
A suitable method to compare these hypotheses is the HypTrails ~\cite{singer_hyptrails_2015} framework. 
It was developed to compare hypotheses about navigation trails of users on the web.
A hypothesis in this framework describes a belief about the transition behavior between states, in our case between the pages of a web shop.
HypTrails enables us, given several such hypotheses, to rank them according to how well they explain the observed data.
If the `structural' hypothesis is ranked higher than the uniform hypothesis, we can conclude that non-item pages indeed contain valuable information about the following user interactions.
\\
\subsubsection{Background}
The HypTrails approach utilizes Bayesian inference to compute the likelihood of the observed data $D$, in our case the interaction sequences $s \in \sessionset$, under the assumption of a hypothesis $H$. 
For data $D$, items and non-items are modeled as \emph{states} and the navigation behavior is represented by a first-order Markov chain in form of a \emph{transition} matrix, containing the likelihood of moving from one state to another.
For a hypothesis $H$, we manually construct a pseudo transition matrix, representing our belief about the likelihood of transitions. For example, according to the `uniform' hypothesis, all transitions are equally likely (see \Cref{eq:hyp:uni}); or for the `structural' hypothesis, the likelihood of going from a non-item page to all pages of the same genre is equal, while the likelihood to go to pages of different genres is zero (see \Cref{eq:hyp:struct}).
\\
HypTrails can then be used to automatically construct a Bayesian Dirichlet prior from these hypotheses.
The Dirichlet parameters $\alpha$ are directly influenced by a concentration factor $k$, which represents the strength of our belief in a hypothesis. 
\\
We can now compute the marginal likelihood $P(D \mid H)$ as evidence for a hypothesis given different concentration factors $k$.
Comparing the evidence between multiple hypotheses yields a ranking: The hypothesis with the highest evidence provides the best explanation for the data.
This allows us to compare hypotheses for a range of beliefs, from ``allowing some errors'' to ``fitting accurately''. For a higher concentration factor $k$ the difference in evidence between two hypotheses becomes more pronounced if one hypothesis fits the data better.
\\
To rank the hypotheses, we plot their evidence for different values of $k$ in a line plot. If the lines do not cross, it is evident that one hypothesis fits better than the other.
\\
Formally, the evidence for a hypothesis $H$ is calculated as follows:
\begin{equation}
    P(D \mid H) = \prod_i \frac{\Gamma(\sum_j \alpha_{ij}) \prod_j \Gamma(\alpha_{ij} + t_{ij})}{\prod_j \Gamma(\alpha_{ij}) \ \Gamma(\sum_j (\alpha_{ij} + t_{ij}))}
    \label{eq:hyptrails}
\end{equation}
Here, $t$ represents the actual transition counts in our data $D$ and $\alpha$ are Dirichlet parameters that incorporate the prior, corresponding to a specific hypothesis $H$ (cf. \Cref{sec:hyp:approach}).
The resulting evidence scores can now be used to compare multiple hypotheses with each other and rank them with respect to the concentration factor $k$. 
Note that the evidence values of a hypothesis can not be interpreted in isolation, but can only be used for comparison with another hypothesis.
For further details, we refer the reader to Singer et. al \cite{singer_hyptrails_2015}.}

\revised{\subsubsection{Approach}
\label{sec:hyp:approach}
To use HypTrails, we need to define states, transitions, and hypotheses.}
In our case, items and non-items will be mapped to states and transitions are obtained from the sessions in our dataset. 

\paragraph{States}
As states in HypTrails are discrete and we need some knowledge about the states to formulate meaningful hypotheses, we employ the previously introduced Content-Based Page ID (CPID) as the only way to map non-item pages to informative states. 
To represent items as states, we follow the idea of network abstraction from
~\cite{koopmann_comptrails_2024} and aggregate all items of the same attribute combination into one single state. This allows us to better represent the more abstract hypothesis that a non-item page is followed by any item of the same category and also reduces computation time significantly. We write this set of aggregated items with content-based ids as
$\itemset_\text{CID} = \{ \itemattributes{v} \mid v \in \itemset\}$. Now, we can define the states $\mathcal{K}$ as the disjoint union of both the aggregated items and non-items represented by their combination of categories $\itemattributes{} $:
\[\hypstates = \itemset_\text{CID} \bigsqcup \pageset_\text{CPID}\]

\paragraph{Transitions}
First, we gather the multi-sets $\transitionset_{s_u}$ of all transitions in a session
$s_u = [i_1, i_2, \ldots,  i_{n}]$ mapped to their respective Content-based ID $\hat{\imath} \in \hypstates$ as $\transitionset_{s_u} = \{(\hat{\imath}_t,\hat{\imath}_{t+1}) | t \in [0,n)\}$.
We can now construct the multi-set of all transitions in the dataset: $$\transitionset = \bigcup\limits_{s_u \in S} \transitionset_{s_u}.$$

\paragraph{Hypotheses}
For each of our hypotheses, we define a matrix $\hypothesismatrix \in [0, 1] \subset \mathbb{R}^{{(|\hypstates|})^2}$ which contains the transition probabilities between our states, which are both items from $\itemset_\text{CID}$ and non-item pages from $\pageset_\text{CPID}$.


We describe and formalize four hypotheses as follows: 
\begin{itemize}
    \item \textbf{uniform:} Non-Item pages have no special influence on the following items, therefore all transition probabilities are distributed uniformly. Therefore we set:
    \begin{equation}
            \hypothesismatrix^\text{uni}_{x,y}= \frac{1}{|\hypstates|} 
             \label{eq:hyp:uni}
    \end{equation}

    \item \textbf{structural:} Non-Item pages directly influence the following interaction - we assign equal transition probabilities from each non-item page to the corresponding items of the same category $\hat{a}$ and zero probability to items of other categories and other non-item pages as follows:

        \begin{equation}
        \hypothesismatrix^\text{struct}_{x,y}= \begin{cases}
        \frac{1}{|\hypstates|}  &, \text{if } x \in \itemset_\text{CID}  \\
        1   &, \text{if } x \in \pageset_\text{CPID}, y \in \itemset_\text{CID} \text{ and } \hat{a}_x = \hat{a}_y\\
         0                  &, \text{otherwise} \\
        \end{cases}
        \label{eq:hyp:struct}
        \end{equation}
    
    \item \textbf{uniform+structural:} A mixture of both previous hypotheses: non-item pages have a higher transition probability to items of the same category than to other pages and we set the transition probabilities to the normalized sum of the two previous hypotheses:
        \begin{equation}
            \hypothesismatrix^\text{uni+struct}_{x,y}= \begin{cases}
        \frac{1}{|\hypstates|}  &, \text{if } x \in \itemset_\text{CID}  \\
        \frac{2}{1+|\hypstates|}    &, \text{if } x \in \pageset_\text{CPID}, y \in \itemset_\text{CID}  \text{ and } \hat{a}_x = \hat{a}_y\\
        \frac{1}{1+|\hypstates|}    &, \text{otherwise} \\
        \end{cases}
        \end{equation}
    
    \item \textbf{data:} The transition probabilities actually found in the data, serving as an upper bound. We count transitions from state $x$ to state $y$, normalizing by the count of any transition from $x$ and assign the following probabilities:


    \begin{equation}
    \hypothesismatrix^\text{data}_{x,y} = \frac{
        \lvert \left\{ (x,y) \right\} \cap \transitionset \rvert
    }
    {
        \sum_{y' \in \hypstates} \lvert \left\{ (x,y') \right\} \cap \transitionset \rvert
    }
    \end{equation}
\end{itemize}

\subsection{Adapting Sequential Recommender Models for Non-Item Representations}

\begin{figure*}
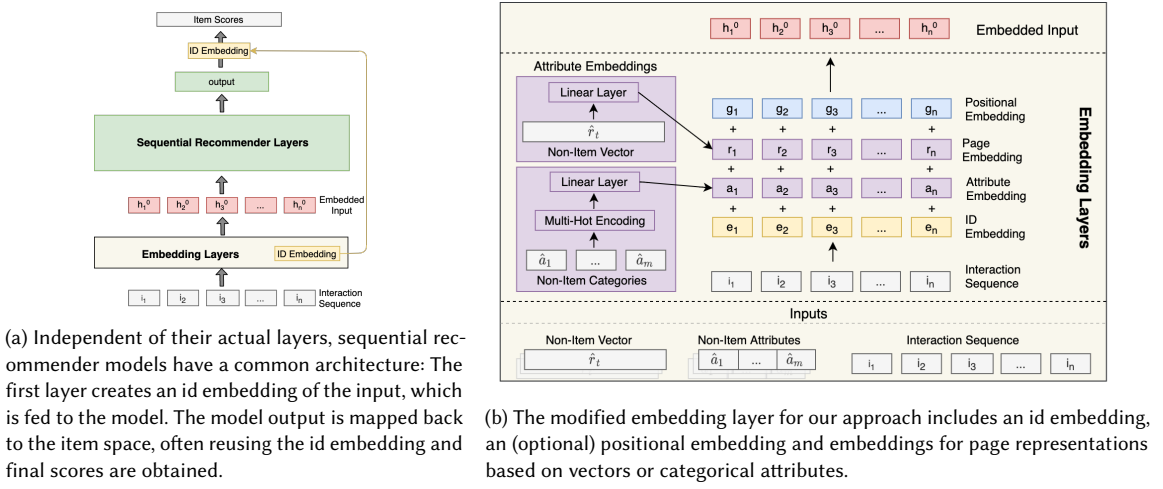

 \begin{subfigure}{0.40\textwidth}
         \centering
         \includegraphics[height=0.20\textheight]{images/PLP_journal2a.png}
         \caption{Independent of their actual  layers, sequential recommender models have a common architecture: The first layer creates an id embedding of the input, which is fed to the model. The model output is mapped back to the item space, often reusing the id embedding and final scores are obtained.}
         \label{fig:models_general}
     \end{subfigure}
     \hfill
  \begin{subfigure}{0.58\textwidth}
     \centering
     \includegraphics[height=0.26\textheight]{images/PLP_journal2b.png}
     \caption{The modified embedding layer for our approach includes an id embedding, an (optional) positional embedding and embeddings for page representations based on vectors or categorical attributes.}
     \label{fig:models_layer}
 \end{subfigure}
\centering
\caption{The common setup for sequential recommendation models and our approach for including non-item representations. }
\Description[Common setup for sequential recommendation models and our approach for including non-item representations.]{The common setup for sequential recommendation models and our approach for including non-item representations.}
\label{fig:models}
\end{figure*}

In general, most sequential recommender models follow a similar setup to the one shown in Fig. \ref{fig:models_general}. 
The input of the model is a sequence of item ids, which are fed to an \textit{embedding layer} for learning the representation of the sequence of interactions. 
This layer includes at least an embedding of the item ids and an optional positional embedding. The embedded input is then forwarded to the actual architecture of the model, which can be seen as a black box of \textit{sequential recommender layers} here. The output of these layers is mapped back to the item space through a \textit{projection layer}. Oftentimes, the original item embedding from the first layer is reused here.

As we want to study the general impact of including non-item pages, we include a variety of sequential recommender models fitting this setup in our experiments. We utilize RecBole~\cite{zhao_recbole_2021}, a popular recommendation framework, which already includes a number of sequential recommendation models used for next-item prediction. We further limit our selection to models without user or item features. This leaves us with the following models, many of them popular baselines:
\begin{itemize}
    \item{GRU4Rec\cite{tan_improved_2016}} is a popular baseline, which utilizes multiple layers of Gated Recurrent Units for sequential recommendation.
    \item{CASER \cite{tang_personalized_2018}} is a convolution-based network, which employs both horizontal and vertical convolution layers to model sequences as high-order Markov Chains. 
    \item{NARM \cite{li_neural_2017}} enhances recommendations by combining recurrent neural networks with an attention mechanism.
    \item{\revised{NextITNet \cite{yuan_simple_2019} analyzes and addresses some limitations of CASER by stacking dilated 1D-convolutional layers to improve representations of both short- and long-range item dependencies.}}
    \item{\sasrec \cite{kang_self-attentive_2018}} were the first to introduce a model based on self-attention for sequential recommendation.  
    \item{\bertforrec \cite{sun_bert4rec_2019}} adopts the famous BERT~\cite{devlin_bert_2019} setup for recommendation by using bi-directional transformer layers and the cloze task~ \cite{taylor_cloze_1953}, which masks items randomly for training.
    \item{CORE \cite{hou_core_2022}} introduces a representation-consistent encoder, which couples the representation space for both the encoding and decoding by using a linear combination of the item embeddings as session embedding. The items themselves are also encoded with self-attention blocks.
    \item{LightSANs \cite{fan_lighter_2021}} also uses self attention, but adapts a low-rank decomposed self-attention to capture latent factors in the interactions and a decoupled positional encoding in addition.
\end{itemize}

All these models can be used to include non-items with unique ids directly. To adapt them to different types of non-item representations, we follow the approach presented in~\cite{fischer_enhancing_2023} for \bertforrec and \sasrec and their variants for including non-item pages called \pagesasrec and \pagebert. The embedding layer in the basic form for all models contains

\newcounter{enum_layer}
\begin{enumerate}[label=(\roman*)]
    \item an embedding $E_{\itemset} \in \mathbb{R}^{|\itemset| \times d}$ of the identifier and
    \item (optionally) an embedding $E_G \in \mathbb{R}^{N \times d}$ to encode the position of the items in the sequence, with $N$ as the maximum input sequence length.
    \setcounter{enum_layer}{\value{enumi}}
\end{enumerate}

When we create unique ids for each page $p$, the embedding of the id will be $E_{\itemset} \in \mathbb{R}^{|\interactionset| \times d}$ instead or 
$E_{\itemset} \in \mathbb{R}^{|\itemset \cup \pageset_\text{CPID}| \times d}$ if the ids are based on the categorical attributes.

For including the page information as categorical attributes, the embedding is defined as $E_{\itemset} \in \mathbb{R}^{(|\itemset|+1) \times d}$ instead, as we add only one general token for all pages $p$. To include the categorical attributes of the items directly, we therefore add 

\begin{enumerate}[label=(\roman*)]
    \setcounter{enumi}{\value{enum_layer}}
    \item a multi-hot encoding of the attributes $E_{\categoryset} \in \{0,1\}^{|\categoryset|}$ and
    a linear layer for resizing $E_\categoryset$ to the hidden size $d$: 
    $l_\categoryset(x) = Wx+b $ with the weight matrix $W \in \mathbb{R}^{|\categoryset| \times d}$ and bias $b \in \mathbb{R}^{d} $
    \setcounter{enum_layer}{\value{enumi}}
\end{enumerate}

to our embedding layer. Further, we adapt the models to utilize arbitrary non-item representations as follows: Assuming that there is a latent representation of size $R$ for $i$ called $\hat{r}_i \in \mathbb{R}^{|R|}$, we add 

\begin{enumerate}[label=(\roman*)]
    \setcounter{enumi}{\value{enum_layer}}
    \item one linear layer $l_R(x) = Wx+b $ with the weight matrix $W \in \mathbb{R}^{R \times d}$ and bias $b \in \mathbb{R}^{d} $ to scale the latent representation to the hidden size $d$.
\end{enumerate}

To create the final output of the embedding layer, all layers are summed to include the additional outputs of $l_\categoryset$ and $l_R$. For each timestep $t$, we use the id embedding 
$e_t = E_{\itemset}(i_t)$ of interaction $i_t$, the positional embedding $g_t = E_G(t)$, the keyword embedding $a_t = l_\categoryset( E_\categoryset(\hat{a}_{p_t}))$ and the scaled latent representation $r_t = l_R(\hat{r}_{p_t})$ to compute the sum $e_t + a_t + r_t + g_t = h^0_t$ as summarized in \Cref{fig:models_layer}, which is then used as the input to the model. The other layers of the models remain unchanged. For the masked training of \bertforrec, we make sure to also mask attributes and representations whenever an interaction is masked. \revised{Finally, training and calculation of the loss remain the same as in the original models, without differentiation between non-items and items, i.e., non-items are also masked in \bertforrec. However, we do ensure in our data preprocessing that all sequences end with an actual item.}

%% file: sections/31_background.tex

\subsection{Non-Item Pages}
\label{sec:nonitempages}
In this section we describe the types of non-item pages that we aim to include in our models in more detail.

In almost any setting where sequential recommendation is usually applied, the users' navigation does not consist exclusively of items or item pages.
For example, web-shops often have additional pages, some of which are not relevant for inferring the users' intent (e.g., payment information or legal notes), while others are strongly related to items and therefore to the users' intent.
Examples of these pages include search results (where a user has either typed out a search query or selected a set of filters like color or size), category pages (e.g., ``women shoes'' or ``men shirts''), or blog pages containing additional content provided by the shop's owner (e.g., pages comparing different types of hiking shoes regarding their suitability for travelling in mountains or other areas).

These pages often provide even clearer hints towards the user's intent (``I want to go hiking in the mountains'') than could be inferred only from their item clicks.
Similar types of non-item pages exist also in other settings: for example, streaming services often provide info pages for actors (listing all of the movies or shows that the actor stars in), category pages (listing all content for a specific genre like horror, drama, ...) or tags (e.g., ``high suspense'').
While these kinds of pages are often not included in public datasets, they usually exist within the respective companies and can be used to improve recommendations.

In this paper, we consider three types of non-item pages: 
\begin{enumerate}
    \item a list of items: e.g., a search result page
    \item a single non-item page: e.g., a blog post with information about items
    \item a list of non-item pages: e.g., a category page for all kinds of clothing, which lists category pages for pants, shoes etc.
\end{enumerate}

All of these pages can be represented by different strategies, for example based on their content (the text in the case of blog posts and the contained items in the case of list pages) or separate unique ids.


%% file: sections/40_dataset.tex
\section{Datasets}
\label{sec:data}
This section introduces the datasets used in our experiments. The statistics for all datasets can be found in \Cref{tab:dataset_stats}. We generally follow the data processing steps from \cite{fischer_enhancing_2023}. We filter sessions to end with an actual item, but in contrast to \cite{fischer_enhancing_2023} we include this step in our preprocessing instead of the dataloader leading to slightly different dataset statistics.

\subsection{Synthetic Dataset \syndata}
To assess the effectiveness of our proposed approaches in enabling sequential recommender models to learn from non-item pages, it is crucial to use a dataset that ensures that the non-item pages contain valuable information. Therefore, we create a synthetic dataset \syndata, based on the \movielenslarge dataset.
The \movielenslarge \furl{https://grouplens.org/datasets/movielens/20m/} dataset consists of movie ratings from users and is a popular benchmark for sequential recommendation. For each movie there is metadata in the form of genres available. \revised{We use the ratings as interactions and build interaction sequences, as for example in ~\cite{sun_bert4rec_2019}: For each user we construct a sequence of movie interactions by using each rated movie as an item and sorting them by the timestamp of the rating. We also remove users and items with less than four ratings.
\\
We split the data randomly by users, with \num{80}{\%} of users for training, \num{10}{\%} for validation and \num{10}{\%} for testing, utilizing the exact same split as in \cite{fischer_enhancing_2023} for comparability (cf. \Cref{sec:discussion:comparison}).}

As the dataset does not contain any non-item interaction, we create three dataset variants, containing different levels of information based on the movie genres.
Note that the results obtained on these datasets are not comparable to the original \movielenslarge, as we leak information about the following movie.
The datasets are designed to demonstrate how models can leverage information from additional non-item pages in a controlled setting, where we ensure that these pages contain valuable information, or in the case of Random-\syndata confirm that the models are not negatively impacted by the introduction of noise.

For each dataset, we represent non-item pages either by building a Content-based Page ID (CPID) with the genre combination as the id or by a page embedding with the genres as embedded categorical keywords. 

\subsubsection{Prev-\syndata}
\begin{figure}
    \centering
     \begin{subfigure}[b]{0.48\textwidth}
         \centering
         \includegraphics[width=\textwidth]{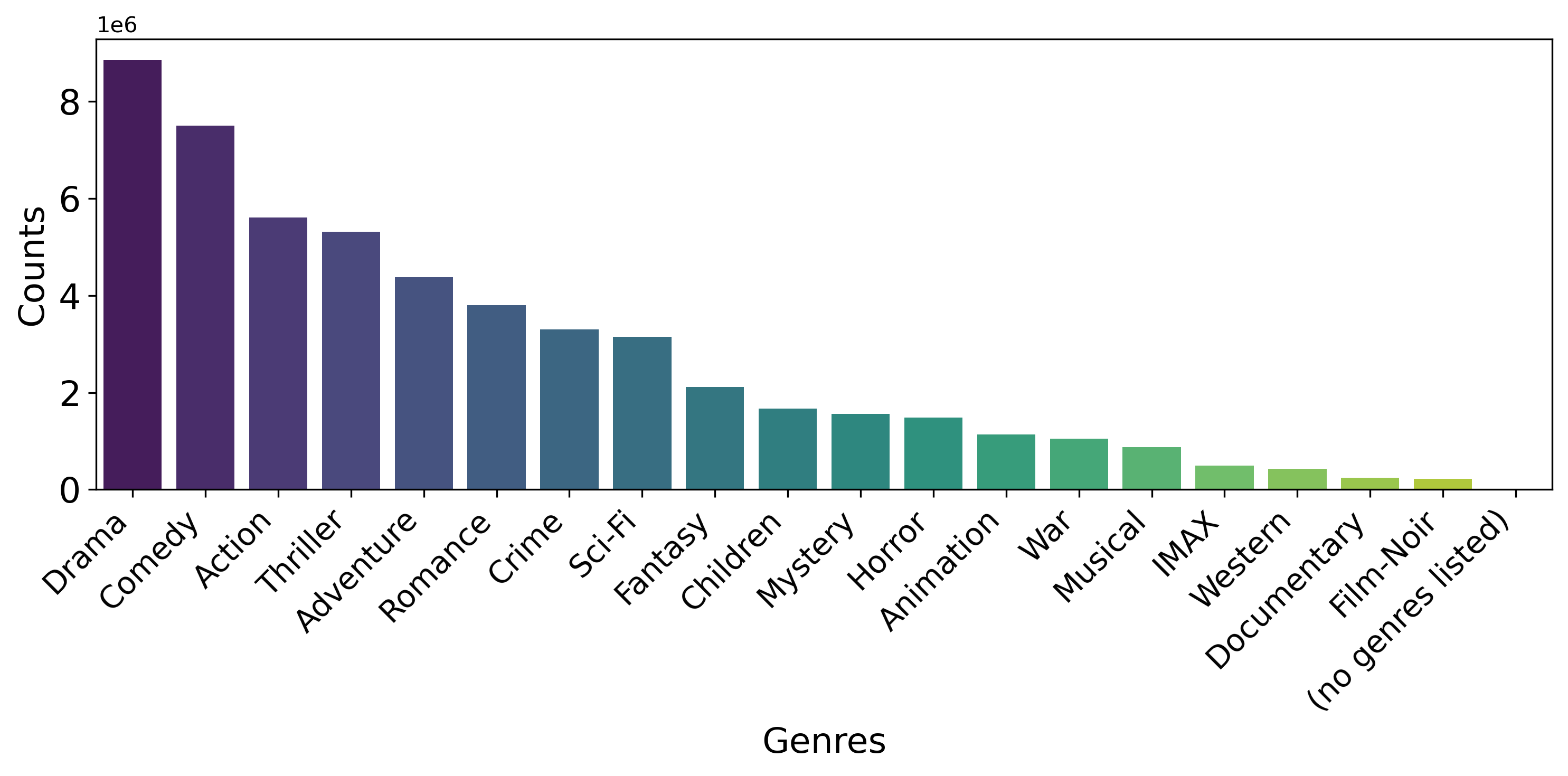}
         \caption{Distribution of genres over all interactions within the Prev-\syndata dataset.}
         \label{fig:data:movie_att}
     \end{subfigure}
     \hfill
     \begin{subfigure}[b]{0.48\textwidth}
         \centering
         \includegraphics[width=\textwidth]{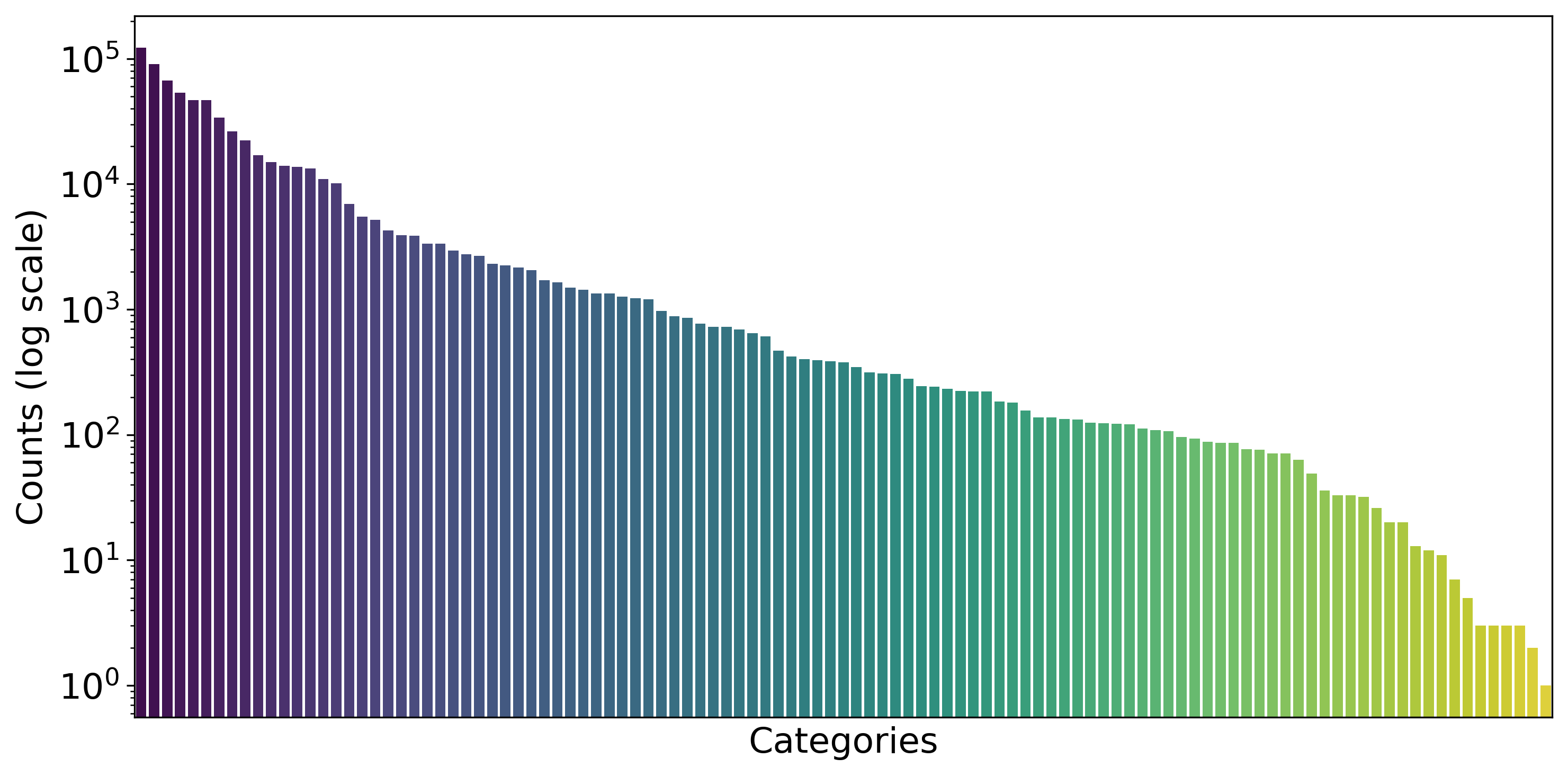}
         \caption{Distribution of categories (using the first results categories) over all search pages on the \coveosearch dataset.}
         \label{fig:data:coveo_att}
     \end{subfigure}
     \caption{Distribution of attributes within \syndata and \coveosearch dataset.}
     \Description[Attribute distribution]{Distribution of attributes within \syndata and \coveosearch dataset.}
    \label{fig:data:data_atts}
\end{figure}

In our first experiment, we want to explore the benefit of highly informative non-item pages. Therefore, for each movie interaction in a sequence, we add a non-item page tagged with the movie's genres immediately prior to that movie.
This simulates a setting in which the user first navigates to a category page listing all movies of a particular genre and selects a specific movie afterwards.
Each added non-item page contains highly relevant information for the recommendation task, as the genre for the next movie is now already known. 
For example, instead of the movie sequence ``Alien $\rightarrow$ Jaws $\rightarrow$ Shrek'' the enriched sequence would be ``Horror$|$Sci-Fi $\rightarrow$ Alien $\rightarrow$ Action$|$Horror $\rightarrow$ Jaws $\rightarrow$ Adventure$|$Children $\rightarrow$ Shrek''. We show the distribution of the genres within the complete dataset in \Cref{fig:data:movie_att}. The distribution is highly skewed, with Drama and Comedy covering more than \SI{35}{\%} of the interactions. While the popular genres might not be very informative on their own, a non-item page has $\approx 2.66$ genres on average and therefore the genre combinations are far more distinctive.

\subsubsection{Group-\syndata}
As Prev-\syndata contains a very strong prior, we also explore a more realistic setting by only adding a non-item-page when the genre of the following movie is from different genres than the movies before.  For example, instead of ``Horror$|$Sci-Fi $\rightarrow$ The Thing $\rightarrow$ Horror$|$Sci-Fi $\rightarrow$ Alien $\rightarrow$ Action$|$Horror $\rightarrow$ Jaws'' as in Prev-\syndata, we generate ``Horror$|$Sci-Fi $\rightarrow$ The Thing $\rightarrow$ Alien $\rightarrow$ Action$|$Horror $\rightarrow$ Jaws''. This reduces the number of non-item interactions to about 19 million (cf. \Cref{tab:dataset_stats}).

\subsubsection{Random-(X-)\syndata}
To cover the case of non-item pages containing very noisy signals or no useful information at all, we generate a set of datasets with randomized non-item pages to explore the susceptibility of the models to noise potentially introduced by adding these pages. We therefore add a non-item page before each movie interaction as in Prev-\syndata and shuffle a given percentage of the pages including their genres. We denote the ratio in the name, e.g. Random-0.1-\syndata contains 10\% non-item pages with randomized genres. Please note that some of them will still be assigned correct genres by chance, so the actual number of `wrong' genres might be lower.
For the example from Prev-\syndata, the sequence could look like ``Musical$|$Children $\rightarrow$ Alien $\rightarrow$ Comedy$|$Adventure $\rightarrow$ Jaws $\rightarrow$ Documentary $\rightarrow$ Shrek'' in Random-1.0-\syndata.

\subsection{Coveo}
Published for the SIGIR 21 challenge, the Coveo\furl{https://www.coveo.com/en/ailabs/sigir-ecom-data-challenge} dataset \cite{tagliabue_sigir_2021} is an extensive e-commerce dataset that captures browsing and search interactions.
The captured browsing behavior consists of interactions with a product page or the view of a web page. A view can be related to a product, but is often only represented by its hashed URL. The dataset also provides search requests, including the embedded search query as a vector of length $45$ and the list of retrieved items.

Moreover, the dataset includes metadata such as hierarchical categories, price, and preprocessed text and image embeddings for products. As our focus in this paper is on incorporating non-item pages, we are not directly utilizing them. However, we use the categories of items to build representations for the lists of retrieved items from search requests.
\revised{For preprocessing, we first drop all duplicate interactions (which are erroneous artifacts of the dataset), keeping only the first one. 
Additionally, a product view often creates a page view event linked to the product with the same timestamp.
To condense these interactions, we drop the page view event, which is less informative, and only keep the corresponding product view.
We ensure that each session ends with an actual item by removing page views from the tail of each session until the last interaction is a product view. We then drop sessions which contain less than two interactions.}
Furthermore, items and page views that occur less than five times in the dataset are dropped from the sequences. 

As we want to examine both types of non-item pages in the dataset separately, we split the data into two sets: For \textit{\coveopage}, we exclusively use item interactions and page views as non-item pages. As there is no further information available, we can only represent a page view with separate ids, namely their hashed URL.
For \textit{\coveosearch}, we reduce the dataset to sessions containing at least one search event. For each search event, we use the list of retrieved items as a non-item page. We create several representations for search pages:
First we compute the three most frequent categories of the products in the retrieved list.
As a second categorical representation, we use the categories of the first product in the search result list.
We show the distribution of the categories (based on the first search result) in \Cref{fig:data:coveo_att}.
Both can now be represented by a category-based id or by embedding the categories.
We also use the product id of the first search result directly as a representation, therefore using an item to represent the non-item page and not increasing the vocabulary at all.
As all these representations depend on the quality of the retrieved item list, we also use the search query embedding vector, which is a direct expression of a user's interest.
In comparison to the genres in \syndata, the distribution seems to be even more skewed.
This could be explained by the categories being hierarchical - the limited number of `root' categories will be present in every search page, categories in the `mid level' far less, and the finest granularity of categories will be even rarer.

\revised{We split both \coveopage and \coveosearch by time.
We sort all item and non-item interactions by time and divide them at two specific points in time, such that about 70\% of interactions are in the first split. These are grouped by user to build sequences for training.
The second split covers about 15\% of interactions and is used for validation, while the remaining interactions are for used testing, both again grouped by user to build sequences.
}

\subsection{Fashion}
The \onlineshop dataset is composed of interactions with product and list pages from a major online fashion store, collected over 20 days. 
A product page shows a single item, while a list page displays multiple items of particular (sub-)categories with additional filters applied by the user. Categories and filters are available as attributes for each page, and while the information is partially overlapping, filters usually contain more detailed information. 

For example, a list page with the category ``shoes'' can have more specific filters, such as ``type:shoes'', ``color:black'' and ``material:leather''. 
Multiple interactions with the same product or page in a row are condensed into a single interaction. Furthermore, we ensure a minimum sequence length of $3$, a maximum of $200$ and ensure an item occurs at least five times in the data as in \cite{sun_bert4rec_2019}.

\revised{
We sort all item and non-item interactions by time and divide them at two specific points in time, such that about 70\% of interactions are in the first split. These are grouped by user to build sequences for training.
The second split covers about 15\% of interactions and is used for validation, while the remaining interactions are for used testing, both again grouped by user to build sequences.
}

\begin{table*}
    \centering
    \caption{Statistics of the preprocessed datasets \syndata, \coveosearch, \coveopage and \onlineshop, showing the number of users, items and unique pages, page attributes and their permutations as CPIDs as well as the number of interactions with items and pages and the average session lengths. \coveosearch attributes are based on the first or the most frequent categories, \onlineshop attributes on categories and filters.}
    \input{data/data_recbole/dataset_stats}
    \label{tab:dataset_stats}
\end{table*}

%


%% file: data/data_recbole/dataset_stats.tex
\resizebox{\textwidth}{!}{\begin{tabular}{lr|rrrrrrrrr}
\toprule
\multicolumn{2}{c|}{Dataset:}  & \multicolumn{1}{c}{Users:} & \multicolumn{1}{c}{Items:} & \multicolumn{1}{c}{UPID:} & \multicolumn{1}{c}{Attributes:} & \multicolumn{1}{c}{CPID:} & \multicolumn{2}{c}{Interactions:}

& \multicolumn{2}{c}{Avg. session length:} \\
Name & Variant & \multicolumn{1}{c}{$|U|$} 
& \multicolumn{1}{c}{$|\itemset|$}
& \multicolumn{1}{c}{$|\pageset|$}
& \multicolumn{1}{c}{$|\categoryset|$}
& \multicolumn{1}{c}{$|\pageset_\text{CPID}|$}
& \multicolumn{1}{c}{|$i_\itemset$|} 
& \multicolumn{1}{c}{|$i_\pageset$|} 
& $i_\itemset$ 
& with $i_\pageset$ \\
\midrule
\syndata 
& Previous/Random
& 138 493
& 26 744
& -
& 20 
& 1 329
& 20 m 
& 20 m
& 144.4
& 288.8  \\

& Group
& 138 493
& 26 744
& -
& 20 
& 1 329
& 20 m 
& 19 m
& 144.4
& 281.9  \\
Coveo & PageViews 
& 1 318 922
& 29 268
& 73 217
& -
& -
& 8.6 m
& 6.6 m
& 6.6
& 10.3 \\
 & Search
& 134 881
& 27 013
& -
& 112/112
& 166/166
& 895 760
& 185 259
& 6.6
& 8.0\\
\onlineshop   &
& 199 474 
& 8 412 
& -
& 204/5 647
& 7 207/27 863 
& 4.3 m 
& 1.3 m
& 21.7
& 28.2\\
\bottomrule
\end{tabular}}

%% file: sections/50_experiments.tex
\section{Experiments}
\label{sec:experiments}

In this section, we introduce the research questions we aim to answer and an explanation of the experimental setup, present the conducted experiments, and subsequently discuss our results.

In our experiments, we aim to address the following five research questions: 
\begin{enumerate}[label=RQ\arabic*]
    \item Do non-item pages have an influence on the following item interaction?
    \item Can sequential recommender models utilize non-item pages for predicting the next item?
    \item Are there gains from including non-item pages even without access to their content?
    \item What representations are useful to integrate search requests as non-item interactions?
    \item Can we find good representations to integrate list pages as non-item pages?
\end{enumerate}

We aim to answer these questions in our following experiments and provide the common experimental setups next:

\subsection{Experimental Setup}
\revised{
We train all our models with the Adam optimizer and cross-entropy loss.
In line with previous research~\cite{fischer_enhancing_2023}, we use the same hyperparameter settings for all models and keep the embedding and inner size, as well as the number of layers the same for all models.
We use the following hyperparameter settings, following \cite{fischer_enhancing_2023} where applicable:
The hidden or embedding size for all models is $64$ and the inner size $256$.
For all transformer-based models and GRU4Rec, we set the number of layers (and heads) to $2$ and the dropout to $0.2$.
For \bertforrec the mask ratio is $0.2$ and the finetuning probability $0.1$.
\\
For \caser and \nextitnet, we base the additional hyperparameters on the RecBole defaults:
we employ a dropout rate of $0.4$, a Markov chain length of $5$ and $4$ vertical and $8$ horizontal filters for \caser.
For \nextitnet this means a kernel size of $3$, a width of the convolutional filter of $5$ and dilations of $[1, 4]$.
\\
As \syndata is based on the MovieLens data, we also use this dataset's established sequence length of $200$ (cf. \cite{sun_bert4rec_2019}).
For the remaining datasets, we aim to include at least $95$\% of sessions in their entirety while keeping the sequence length reasonable for efficiency and choose the sequence lengths accordingly, rounding up.
The number of epochs was determined based on preliminary experiments, where the loss curve showed convergence with no further meaningful improvements.
\\
Overall, we do not aim to find the overall optimal model, but to systematically investigate the utility of non-item pages and the effect of their inclusion in several different models.
Therefore, we do not conduct hyperparameter studies, but rely on the parameter settings described above, which are anchored in previous research.
\\
We train all models on the seeds [212, 6, 10, 404, 42] and report the average and standard deviation over the five runs, except our experiments on Random-X-\syndata and \onlineshop due to limited computational resources, for which we report only the results on seed $212$.
We keep our batch size consistent for all experiments and choose $64$ to utilize our GPUs best.
}

\paragraph{Evaluation Settings}
\revised{
For each sequence in the test (and validation) data, we use the last item as the prediction target and the rest of the test (or validation) sequence as input, following previous studies~\cite{hidasi_recurrent_2018, yuan_simple_2019}.
\\
Note that we have already ensured in our data preprocessing that the target is an item and not a non-item.
To evaluate the quality of recommendation for actual items, we set the probability of all non-items in the model output to $0$ before computing the metrics.
We report the full \textit{Hit Rate (HR)} and the \textit{Normalized Discounted Cumulative Gain (NDCG)} at $k \in \{1,5,10\}$ without any sampling.
Additionally, we use the dependent Student's t-test for paired samples to calculate $p$-values on user-level for significance testing and report significant increases over the baselines for $p<0.01$ and $p<0.05$ for our main experiments.
}

We provide the code and configurations for our experiments in our repository \footnote{ \url{https://github.com/LSX-UniWue/non-items-recbole}}.  

\newcommand{\sig}{*}
\newcommand{\lesssig}{\textsuperscript{+}}

\input{sections/51_exp_0_hyptrails}

\input{sections/51_exp_1_syndata}

\input{sections/52_exp_2_pageview}

\input{sections/53_exp_3_search}
\input{sections/54_exp_4_fashion}

%% file: sections/51_exp_0_hyptrails.tex
\subsection{RQ1: Do Non-Item Pages have an influence on the following item interaction?}

\begin{figure}
 \begin{subfigure}[b]{0.40\textwidth}
         \centering
         \includegraphics[width=\textwidth]{images/hyptrails/hyptrails-ml-20m-train-agg-data.png}
         \caption{Evidence for hypotheses on Prev-\syndata}
         \label{fig:hyptrails_1}
     \end{subfigure}
     \hfill
  \begin{subfigure}[b]{0.40\textwidth}
     \centering
     \includegraphics[width=\textwidth]{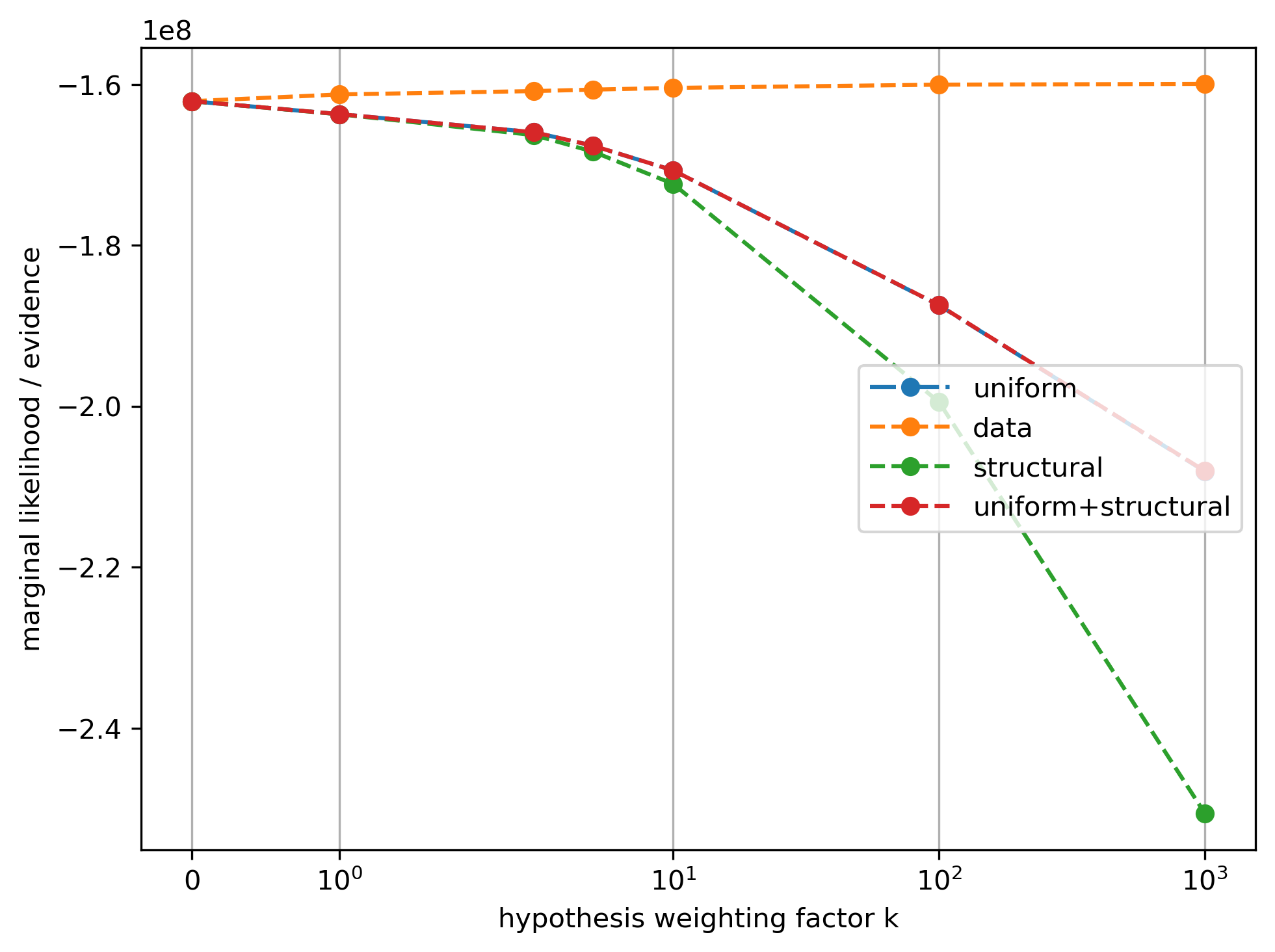}
     \caption{Evidence for hypotheses on Random-1.0-\syndata}
 \end{subfigure}
 \medskip
 \begin{subfigure}[b]{0.40\textwidth}
         \centering
         \includegraphics[width=\textwidth]{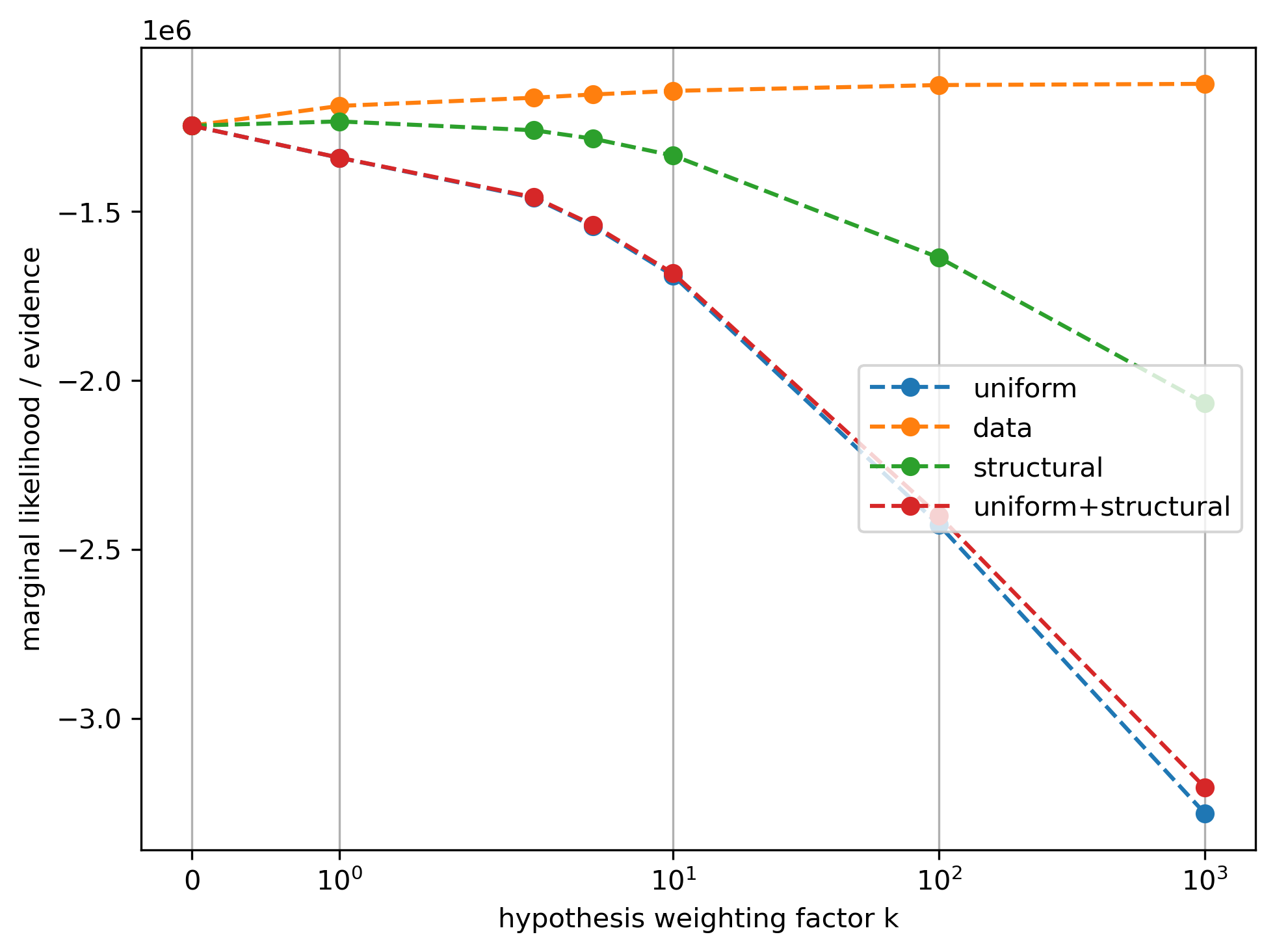}
         \caption{Evidence for hypotheses on \coveosearch.}
     \end{subfigure}
     \hfill
  \begin{subfigure}[b]{0.40\textwidth}
     \centering
     \includegraphics[width=\textwidth]{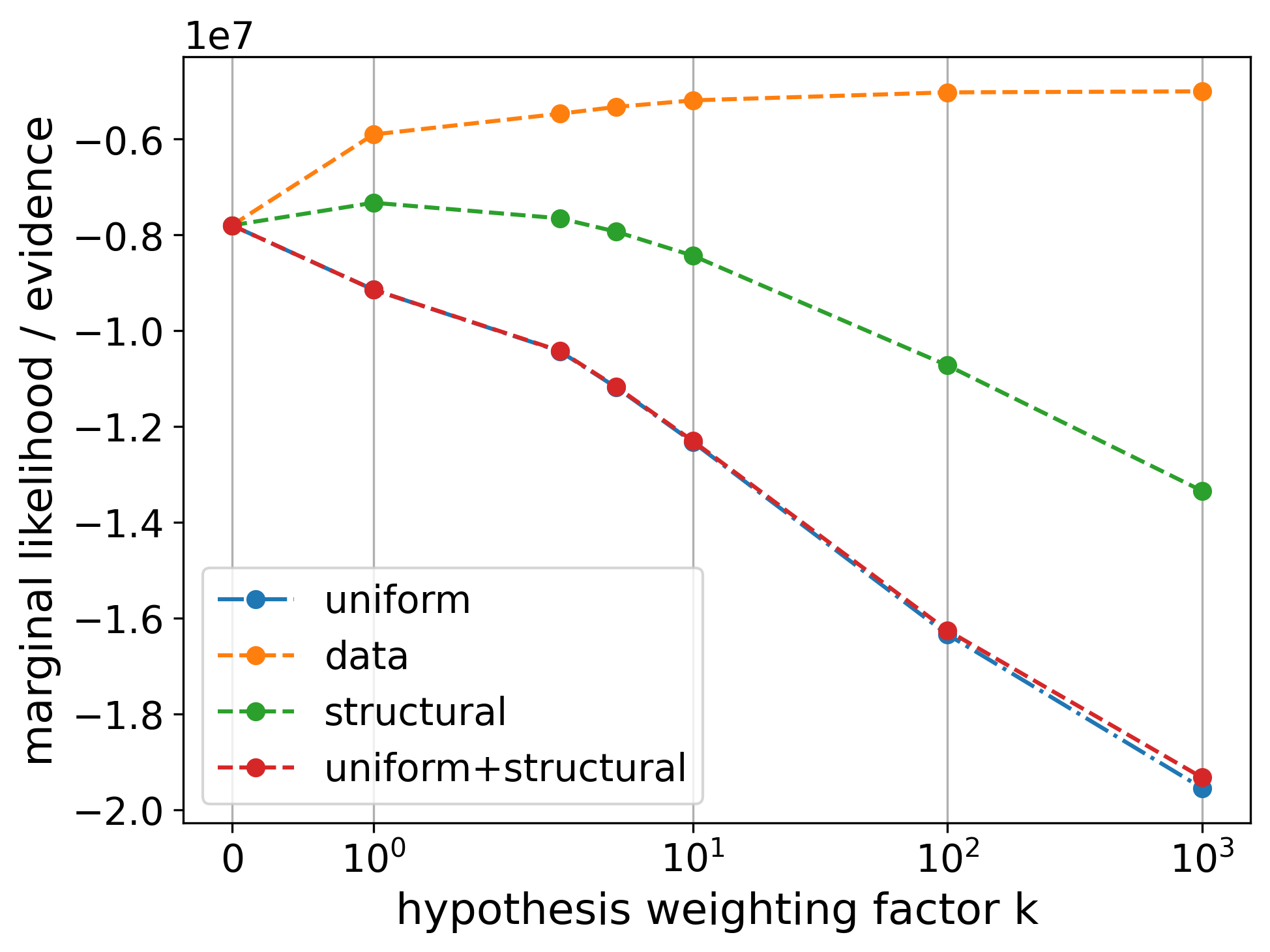}
     \caption{Evidence for hypotheses on \onlineshop}
 \end{subfigure}
\centering
\caption{The evidence or marginal likelihood for our hypotheses on \syndata, \coveosearch and \onlineshop for different hypothesis weighting factor $k$. A higher evidence means a hypothesis is a better explanation for the data, while $k$ can be interpreted as how strongly you believe in the hypothesis.}
\label{fig:hyptrails}
\Description[Evidence for hypotheses]{The evidence or marginal likelihood for our hypotheses on \syndata, \coveosearch and \onlineshop for different hypothesis weighting factor $k$. A higher evidence means a hypothesis is a better explanation to the data, while $k$ can be interpreted as how strongly you believe in the hypothesis.}
\end{figure}

Our first goal is to verify our hypothesis about the importance of non-item pages for following items within click sequences. We utilize the HypTrails framework (cf. \Cref{sec:meth:hyptrails}) on four of our datasets to test whether a non-item page with a specific category is generally followed by items of the same category. We use \syndata and Random-X-\syndata with genre-based CPIDs and \coveosearch and \onlineshop with category-based CPIDs.

\revised{In \Cref{fig:hyptrails} we plot the evidence or marginal likelihood against the concentration factor $k$ for our datasets and compare four hypotheses against each other: The `structural' hypothesis (non-items influence following interactions), the `uniform' hypothesis (non-items have no special influence on following interactions), the `uniform-structural' hypothesis (non-items have a small influence on following interactions) and the `data' hypothesis, which serves as an upper bound as it fits the data perfectly.
\\
A higher evidence shows that a hypothesis provides a better explanation for the observed data. Recall from \Cref{sec:meth:hyptrails} that the $k$-factor represents the amount of belief in a hypothesis or how strictly deviations from the actual data are penalized. 
For $k=0$ we do not believe in a hypothesis at all and allow any amount of deviations from the hypothesis in the data.
Therefore all hypotheses yield the same evidence in our plots. Once we increase $k$ and become more strict in our belief, we see greater differences in evidence between the hypotheses, allowing us to rank them.
\\
On the Prev-\syndata (see \Cref{fig:hyptrails_1}) the structural hypothesis ranks second-best, as it can explain most of the observed transitions. 
At higher concentration factor $k$ the evidence drops, as it assumes uniformly distributed probabilities from items to other states.
\\
Only the data hypothesis performs better, as it exactly fits the transition probabilities.
Overall, the uniform hypothesis has a lower evidence compared to the structural hypothesis. The combined hypothesis ranks slightly higher, although only by a small difference. 
In comparison, the structural hypothesis gives the best explanation for the transitions in our dataset (except for the upper bound data hypothesis), which verifies that our dataset indeed contains influential non-item pages.
On Random-1.0-\syndata, we see that the signal of the non-item pages has been diluted as intended. For a lower concentration factor, the differences between the three hypotheses are minimal, while for higher values, the structural hypothesis shows a harsher drop.
For both \coveosearch and \onlineshop we can also verify the meaningfulness of non-item pages, similar to Prev-\syndata. The evidence for the structural hypothesis is constantly above the uniform hypothesis. The combined hypothesis is also slightly better than the uniform hypothesis for both datasets, as the structural part of the hypothesis includes at least some explanation about the transitions.}
\\
One limitation of HypTrails in our setting is the need for discrete states, which means that we can only represent non-item pages by ids and not their content. Additionally, we need some knowledge about the non-item pages to build meaningful hypotheses. Therefore, we can neither apply HypTrails for \coveopage, as we have no information about the non-item pages there, nor to non-item representations based on, for example, queries. 
A second restriction is that only first-order dependencies are modeled, but we would expect non-item pages to have influence on more than the single next interaction. Nevertheless, our results already show a positive influence for first-order dependencies, so we estimate the actual influence to be even higher.
Overall, our results give us the necessary evidence to consider non-item pages as a valuable information source for our following experiments.

%% file: sections/51_exp_1_syndata.tex
\subsection{RQ2: Can sequential recommender models utilize non-item pages for predicting the next item?}
\label{sec:experiment1}

\begin{table}
    \centering
    \caption{\revised{Average Hitrate and NDCG with standard deviation over five random seeds for non-item models on the \syndata dataset. This includes the genre-specific popularity baseline (Genre-POP) and the baseline model on item interactions only (Items-BL). We report results for both Prev- and Group-\syndata datasets for including Non-Item Pages with both page embeddings (PE) and id embeddings (ID). Results marked with * show a significantly higher performance compared to the items-only baseline for $p<0.01$.}}
    \input{data/data_recbole/ml-experiments}
    \label{tab:ml_table-rec}
\end{table}

Our next objective is to investigate whether sequential recommender models can effectively utilize non-item pages. 
To accomplish this, we use the \syndata datasets based on \movielenslarge with different variants as introduced in \Cref{sec:data}: Prev-\syndata with non-item pages contains the exact genre information for the next movie and Group-\syndata includes fewer non-item pages with the genre information for the following movies condensed. In addition, we evaluate Random-\syndata  non-item-pages with pure noise and Random-(X)-\syndata with varying percentage of noisy and meaningful non-item pages.
For all of the artificial pages, we either generate content-based ids by concatenating categories, or assign one placeholder id and encode the categories as attributes. We train all models for both cases. As baseline, we train a model utilizing the sequence with items only. We also include a most popular baseline given the correct genres of the target movie.
We train all models for $100$ epochs with early stopping, a batch size of $64$, and a maximum sequence length of $200$.

\subsubsection{\syndata Results}

\Cref{tab:ml_table-rec} illustrates the results of our experiments on both Prev-\syndata and Group-\syndata. The genre-specific most popular baseline already performs nearly half as well as most of the models trained on items only, showing that genres are indeed a good indicator for the next movie. 
The benefit of including meaningful non-item pages for each movie interaction in Prev-\syndata compared to the models trained on pure item sequences is therefore not surprising, as the genre of a movie is given through the preceding non-item page. 
Over all metrics we can see significant improvements between the Items-Only baseline and Prev-\syndata with non-item pages represented by
ids (CPID). Looking at the \hr{10}, \bertforrec has the smallest increase with  $0.22$, followed by \nextitnet with an increase of $\approx 0.52$, while the other models even improve by about $0.70$ in \hr{10}. NDCG is also higher over all models compared to the Items-Only baseline, but the improvements are slightly smaller than those observed for the respective HR. For example, NDCG@10 for \bertforrec has increased by $0.16$, and \sasrec, which has the highest improvement of the other models, by $0.47$. But in general, all models are able to leverage the additional information in the non-item pages and increase their performance over all metrics.

Using page embeddings (PE) of the attributes instead gives very similar results for \lightsans, \narm and \sasrec, with less than $0.04$ difference in \hr{10} compared to the CPID embedding. For \gruforrec we see a slight improvement with PE. For \caser and \core the metrics are lower when using PE, but overall they still show a substantial increase in performance compared to the baseline. Notable exceptions are \bertforrec and \nextitnet. Both models can learn from page embeddings but exhibit a high variance in their performance as the model collapses for some runs. Out of the five seeds, three models are collapsing with a performance worse than the baseline. In the case of \nextitnet, only one model is able to reach the performance of the baseline, while the others perform similar to the CPID embedding ($0.564$ on average). The introduction of page embeddings might make these models more unstable, as we do not see a similar behavior using CPIDs.

Group-\syndata contains the same information as Prev-\syndata, but in a more condensed form. When using CPID, the performance is slightly lower, but still very similar to Prev-\syndata for most models, even though the information is less explicitly given. The big exception here again is \bertforrec, as the \hr{10} reaches $0.624$, which is far higher than on Prev-\syndata and an increase more similar to the other models. On Group-\syndata, we can observe that \bertforrec is also able to leverage page embeddings, with even higher performance than with CPIDs and similar performance over all seeds. For \nextitnet, the results are slightly lower than on Prev-\syndata and there is again one collapsing model (on the same seed as before). For \narm and \caser the setting in Group-\syndata and page embeddings even yields the best overall performance.

We conclude that all models are capable of utilizing information from non-item pages, as they show significant improvements. The integration of non-item pages as separate ids yields slightly better results than the integration via attributes (PE) for half of the models, but the differences remain relatively small overall. A reason for the better performance could be the relatively small number of non-item pages ($\approx1.300$), as it might be easier for the models to learn a fixed number of ids instead of a more complex embedding space. Some models exhibit instabilities with page embeddings, especially \bertforrec. In contrast to all other models, this is the only model trained with the cloze task \cite{taylor_cloze_1953}, which could be one reason for the different behavior. \nextitnet with page embeddings also shows a high variance. While the model overall achieves good results, one specific seed gives results similar to the baseline. 
In general using page embeddings has a comparable performance for most models and is a more flexible method, therefore both representations for non-item pages seem to be valid approaches.

We can further observe that some of the baselines perform surprisingly well. While we did not optimize parameters for any models specifically, baseline models like \gruforrec, \caser as a simpler convolution neural network or \sasrec as a transformer-based architecture still perform comparably. \sasrec gives the best overall results, which is in line with previous findings of Fischer et al.~\cite{fischer_integrating_2020} for training the model with a cross-entropy loss.

\subsubsection{Noise in Non-Item Pages}
\label{sec:exp1:noise}

\begin{figure}
\centering
\includegraphics[width=0.74\textwidth]{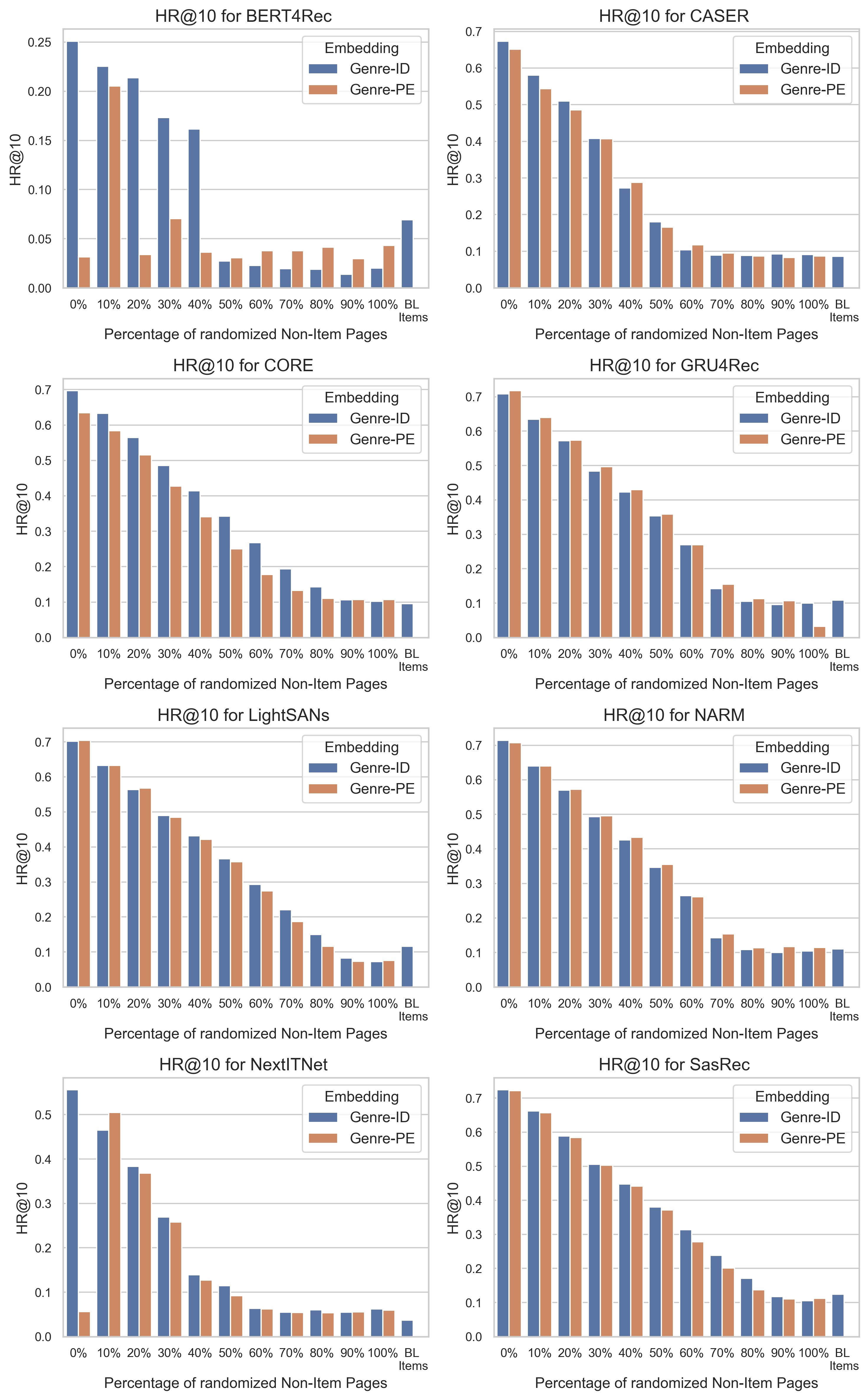}
\centering
\caption{\revised{HR@10 for models on Rand-\syndata with seed 212 on datasets with a growing percentage of shuffled non-item pages. Genre-ID shows the performance of models including non-item pages based on genres with an ID embedding, while Genre-PE shows those with an Page Embedding. The baseline utilizing items only is marked by  ``BL Items''.}}
\label{fig:ml-random-exp}
\Description[HR@10 for Rand-\syndata]{HR@10 for all models on Rand-\syndata with seed 212 on datasets a growing percentage of shuffled Non-Item-Pages and the baseline utilizing items only. We use the genre to create either CPIDs or Page Embeddings.}
\end{figure}

While our previous experiment indicates that models can profit from additional information in the form of non-item pages, in a realistic setting more data will introduce additional noise. We therefore experiment with the Random-(X)-\syndata datasets, which contain different levels of noise. In \Cref{fig:ml-random-exp} we show the performance of all models on the datasets ranging from 0\% to 100\% shuffled non-item pages and the baseline model using items only for one random seed ($212$). 

While performance naturally drops with increased noise in all models, there are clear differences in the amount of noise different models can handle and the steepness and curve of the drop.

An obvious outlier is \bertforrec, as we can see again the model's instability. For page embeddings, only models with $10$\% and $30$\% shuffled pages are improving in performance. For the id based approach, the first four models of \bertforrec perform relatively well.
In contrast to this, \nextitnet appears more stable with only one exception. The models trained with highly informative non-item pages can leverage this information with both page embeddings and id representation. The performance drops in a steep curve until $50$\% shuffled pages are reached, but even with more noise the models surpass the baseline. 
The drop with \caser is also of similar steepness, but overall the model achieves a slightly higher performance.
\narm and \gruforrec show a more linear decline with a sharper drop in performance at about $70$\%.
\lightsans and \sasrec even show improved performance up to $80$\% of shuffled non-item pages. \core is in between, as CPID and PE embeddings perform quite differently.

We can also observe differences between models with page embeddings and CPIDs. For \caser and \core the integration with page ids performs better. We can explain this behavior for \core, as the model explicitly links encoding and decoding for the items, which our page embedding will most likely interfere with. For \lightsans and \sasrec CPIDs perform better with more noise. The other results are mixed, but for \gruforrec page embeddings are performing slightly better.

Overall, we can see that all models are able to handle a reasonable amount of noisy non-item pages. For most models the added noise even has no negative impact, as the performance is still similar to the items-only baseline. In a more realistic setting, non-item pages would be  selected avoiding such high levels of noise. Therefore we would not expect a decrease in performance from non-item pages. The complete results are reported in \Cref{tab:ml_table-rand-id} for CPID and \Cref{tab:ml_table-rand-pe} for page embeddings.

\subsubsection{Visualizing Items and Non-Items in the Embedding Space}
\label{sec:exp1_vis}

\begin{figure}
    \begin{subfigure}[t]{0.49\textwidth}
    \centering
    \includegraphics[width=\linewidth]{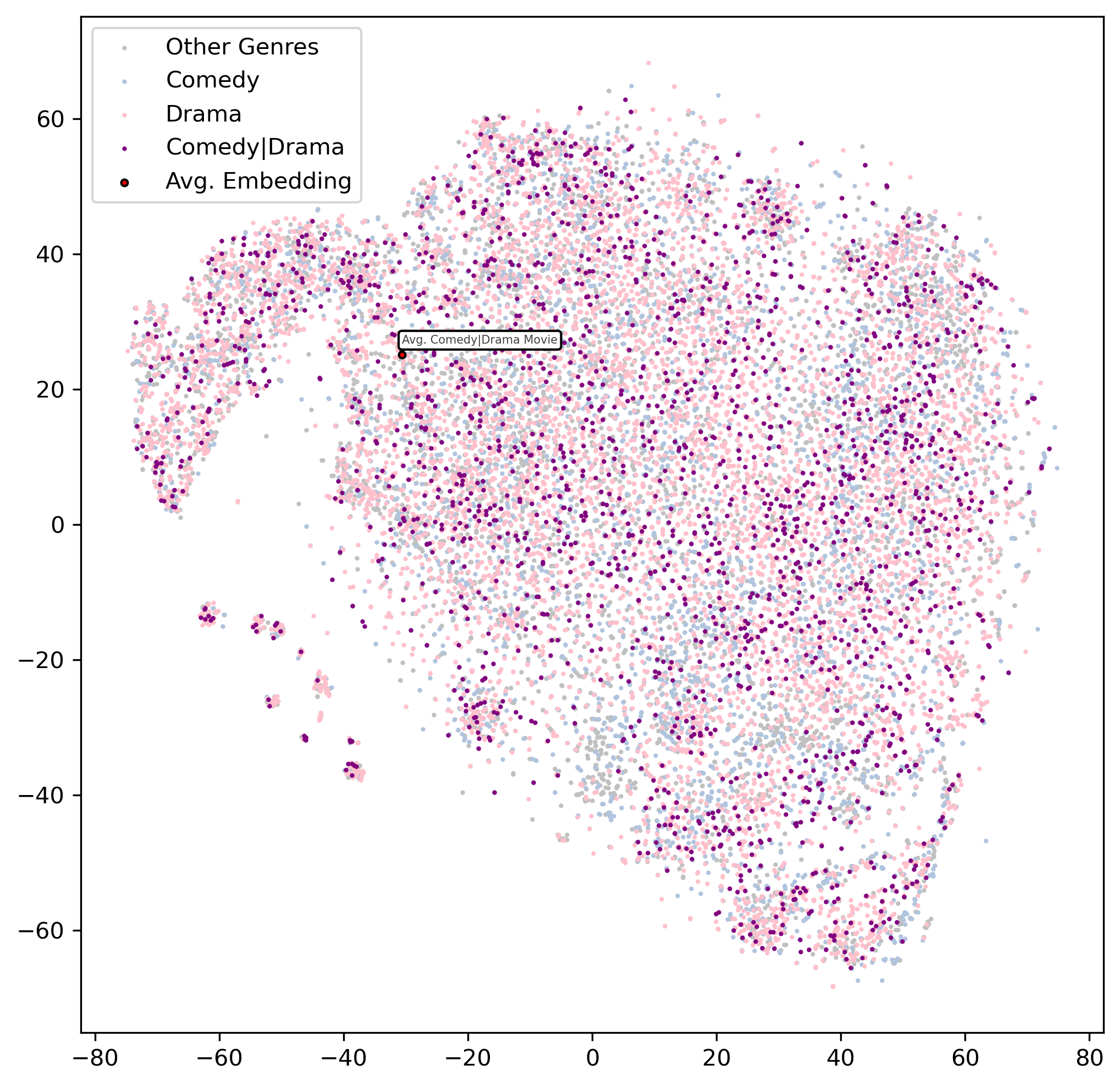}
    \caption{SASRec for the Items-Only baseline.}
    \label{fig:exp1-tsne-big-a}
    \end{subfigure}
    \hfill
    \begin{subfigure}[t]{0.49\textwidth}
    \centering
    \includegraphics[width=\linewidth]{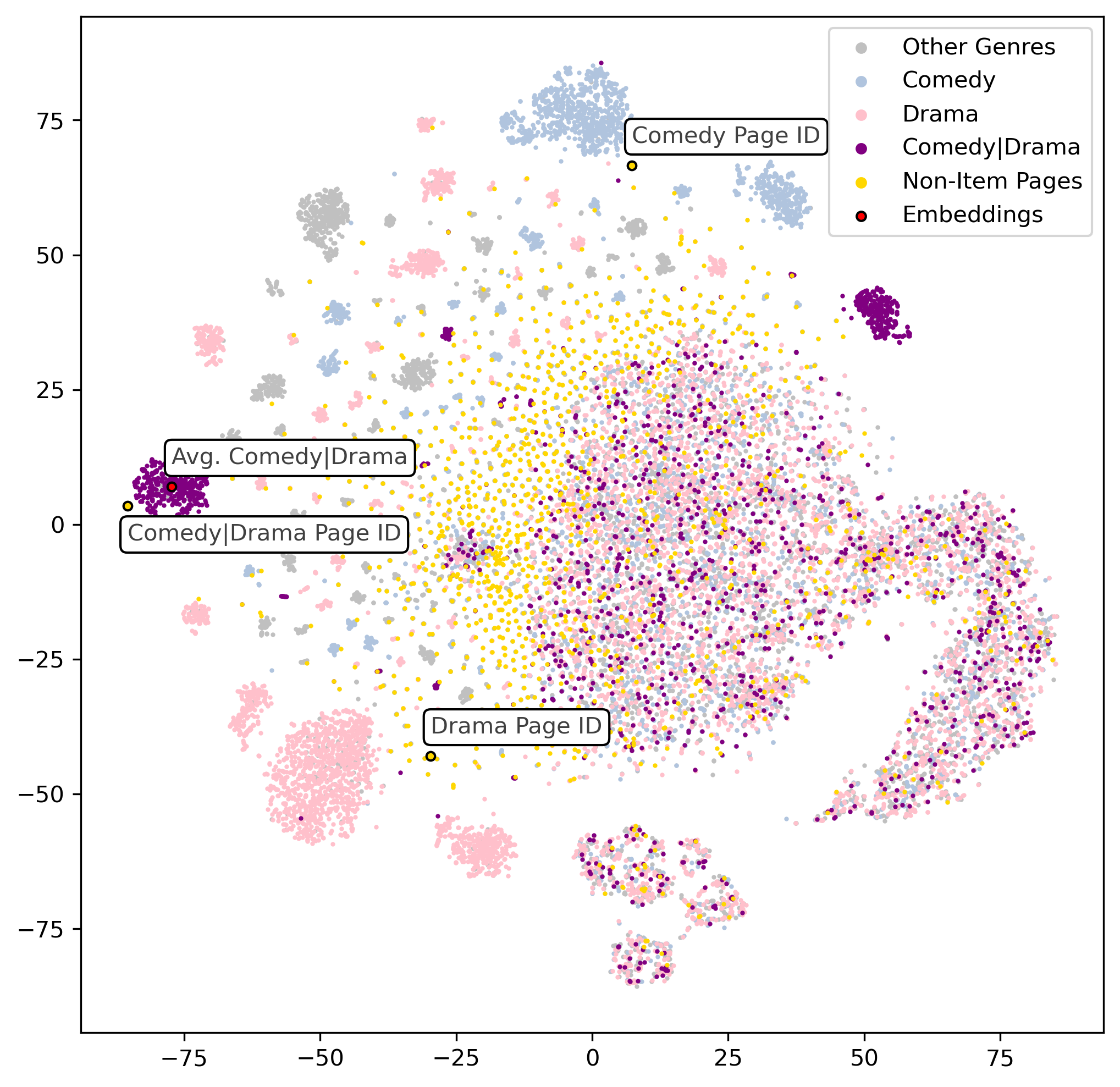}
    \caption{SASRec with content-based page IDs for non-item pages. }
    \label{fig:exp1-tsne-big-b}
    \end{subfigure}
    \medskip
    \begin{subfigure}[t]{0.49\textwidth}
    \centering
    \includegraphics[width=\linewidth]{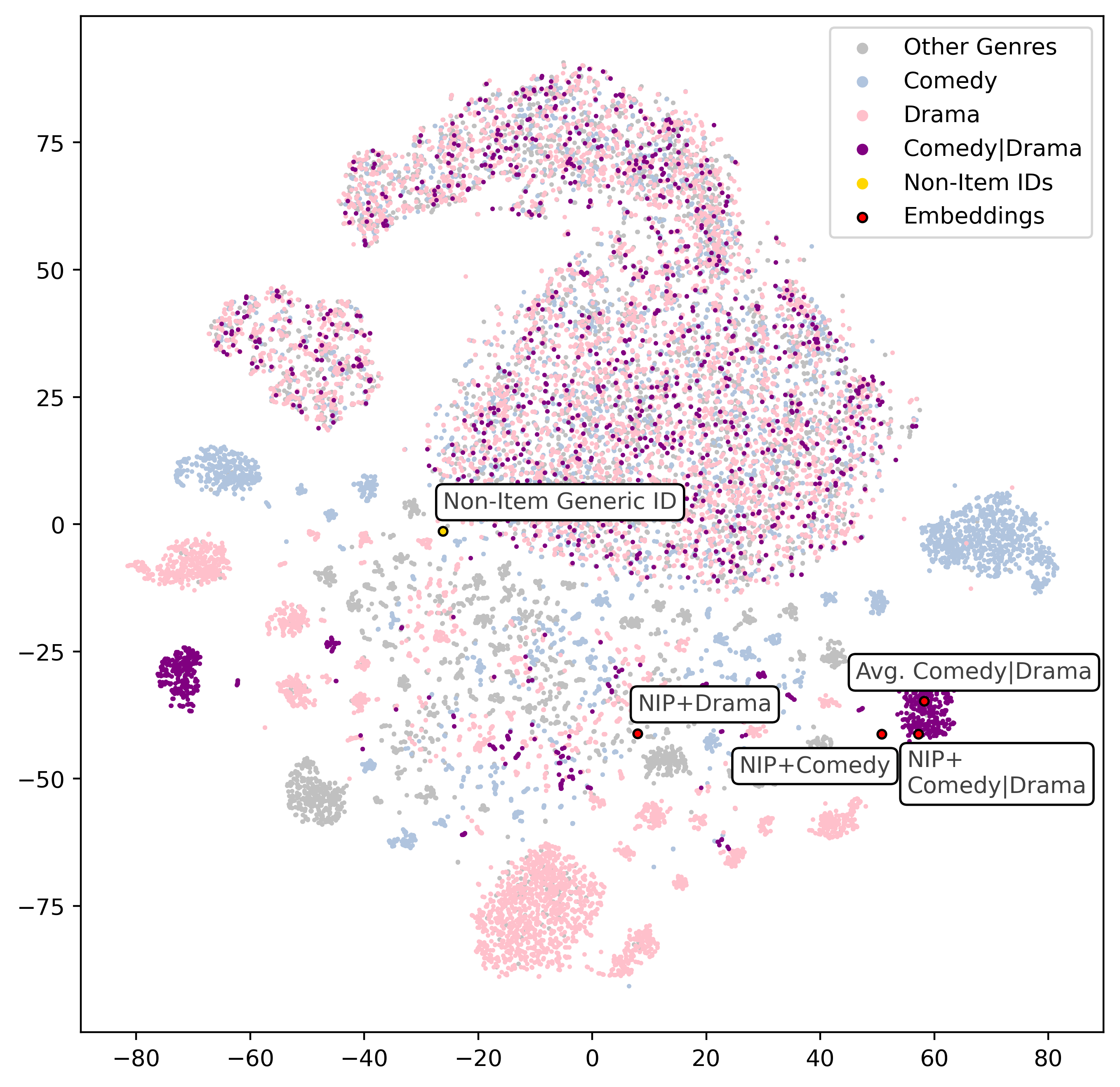}
    \caption{SASRec with genre based page embeddings for non-item pages.}
    \label{fig:exp1-tsne-big-c}
    \end{subfigure}
    \hfill
    \begin{subfigure}[t]{0.49\textwidth}
    \centering
    \includegraphics[width=\linewidth]{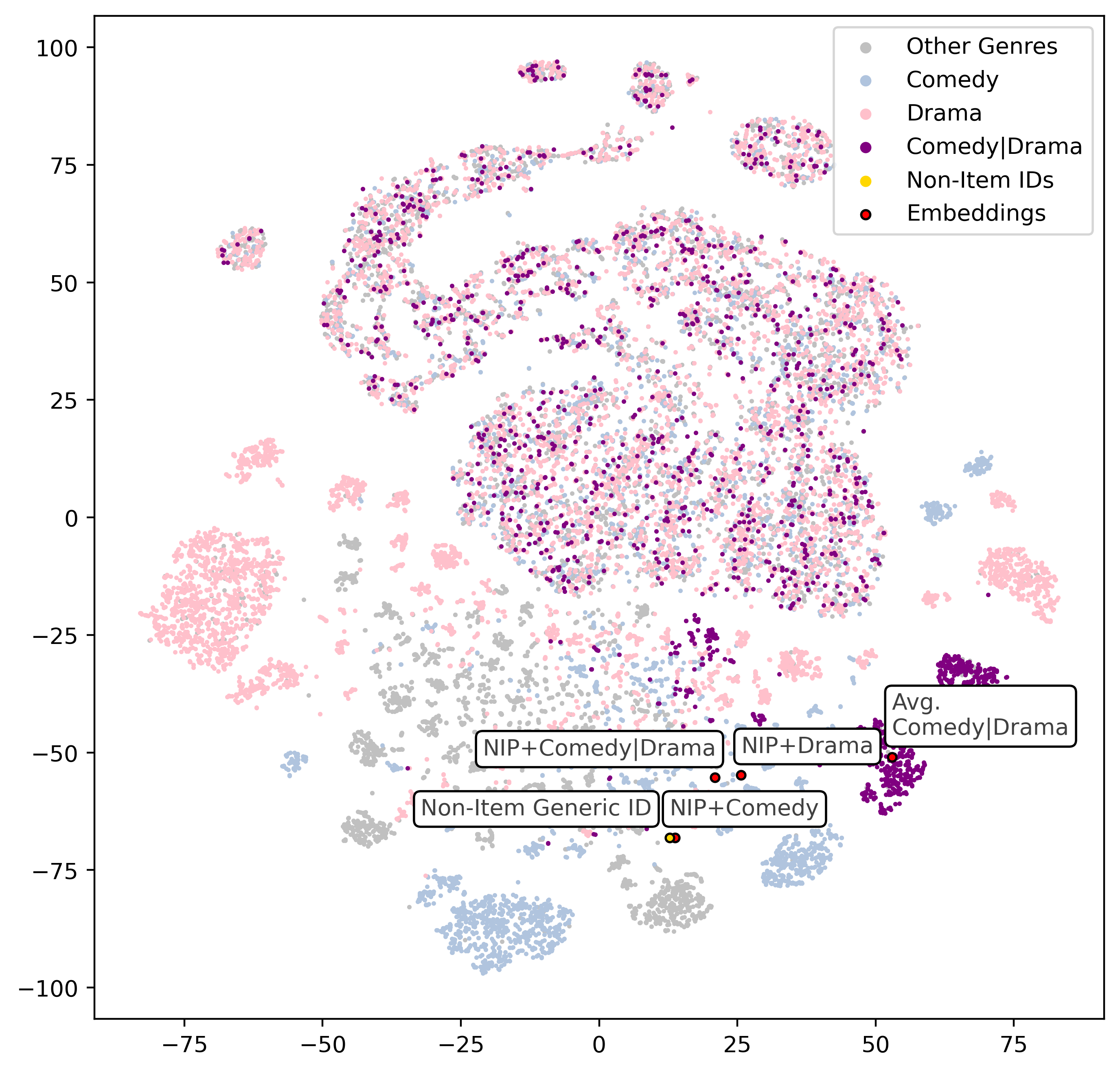}
    \caption{\gruforrec with genre based page embeddings for non-item pages.}
    \label{fig:exp1-tsne-big-d}
    \end{subfigure}
\centering
\caption{t-SNE visualization of item embedding space of different model variants on \syndata and Prev-\syndata with highlighted movies containing the genres ``Comedy'', ``Drama''.}
\label{fig:exp1-tsne-big}
\Description[t-SNE visualization of item embedding space of different model variants on \syndata and Prev-\syndata]{t-SNE visualization of item embedding space of different model variants on \syndata and Prev-\syndata.}
\end{figure}

Using non-item pages in our models can have a high impact on performance. To gain some understanding of the changes happening, we take a look at the item embeddings, as the item embedding layer is where we include non-item pages. We visualize the item embeddings with t-SNE~\cite{maaten_visualizing_2008} exemplary in \Cref{fig:exp1-tsne-big} for one seed. To make patterns more visible, we exemplary highlight movies containing the genres ``Comedy'' and ``Drama''. We also add the average of all ``Comedy|Drama'' movies and highlight the non-item page representations. In case of page embeddings we display the generic non-item page embedding and add the respective genre embeddings.

While the visualization of the baseline version of \sasrec shows only few distinctive structures, e.g. the \textit{hook}-like structure on the top left corner, the item embeddings of \sasrec with both CPID and page embeddings reveal clearer patterns. Both show two bigger clusters containing ``Comedy|Drama'' movies, and we can  find more similar structures for both genres. They also contain the hook-like structure as in the baseline models. Overall, the item embeddings seem to have some similarities to each other, and the non-item pages seem to help distinguishing items of different genres better. Interestingly, the structures seem to be consistent even when looking at different models, like \gruforrec in \Cref{fig:exp1-tsne-big-d}. Here, we can also find some of the structures mentioned above, like the hook and the bigger, genre specific clusters. We can find similar pattern in most visualizations of the other models and variants, too. We include these figures in our git repository for a complete overview.

\begin{figure}
 \begin{subfigure}[b]{0.45\textwidth}
         \centering
         \includegraphics[width=\textwidth]{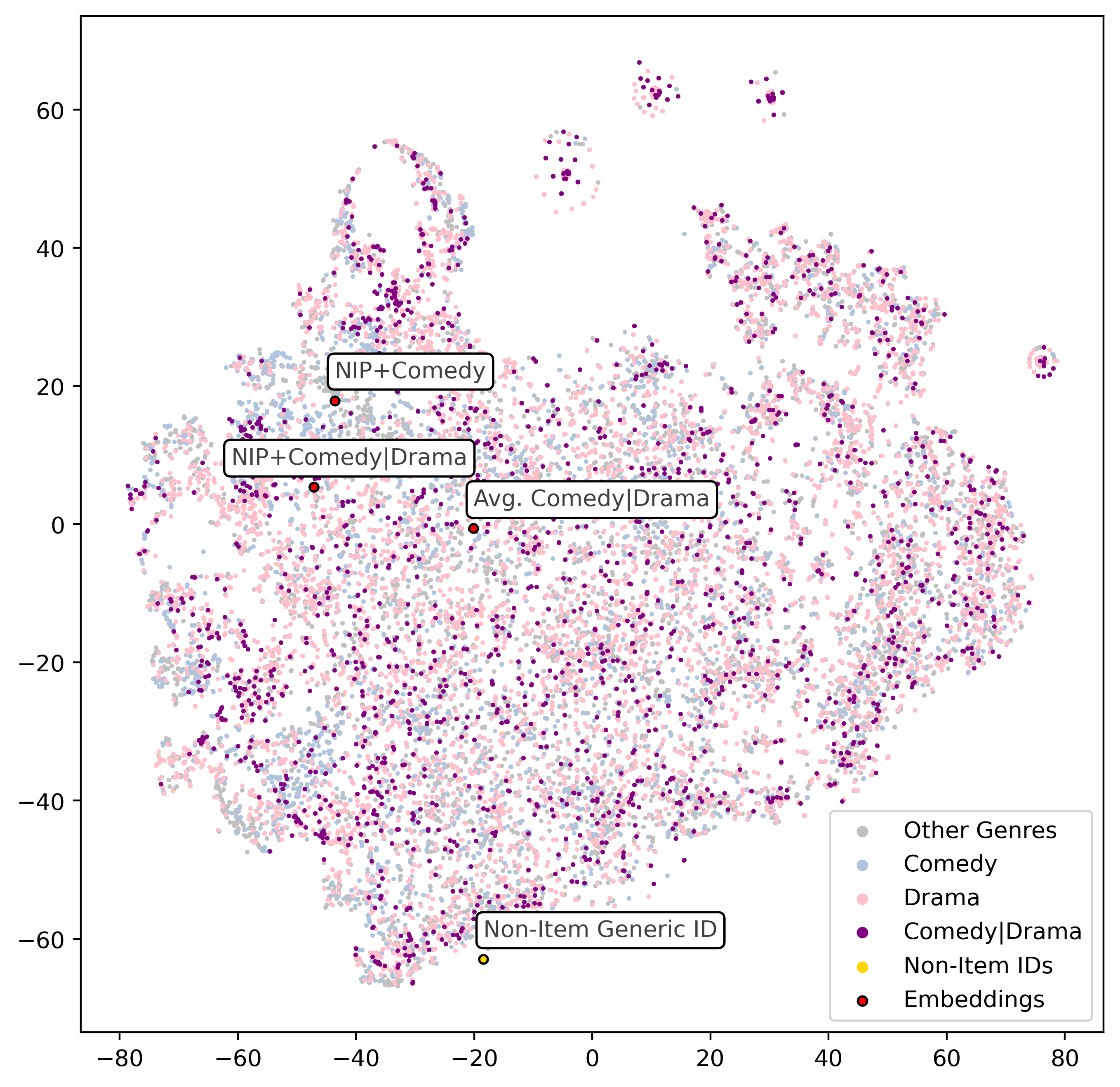}
         \caption{Item embedding space of BERT4Rec on Prev-\syndata.}
         \label{fig:exp1-tsne-broken-bert}
     \end{subfigure}
     \hfill
  \begin{subfigure}[b]{0.45\textwidth}
     \centering
     \includegraphics[width=\textwidth]{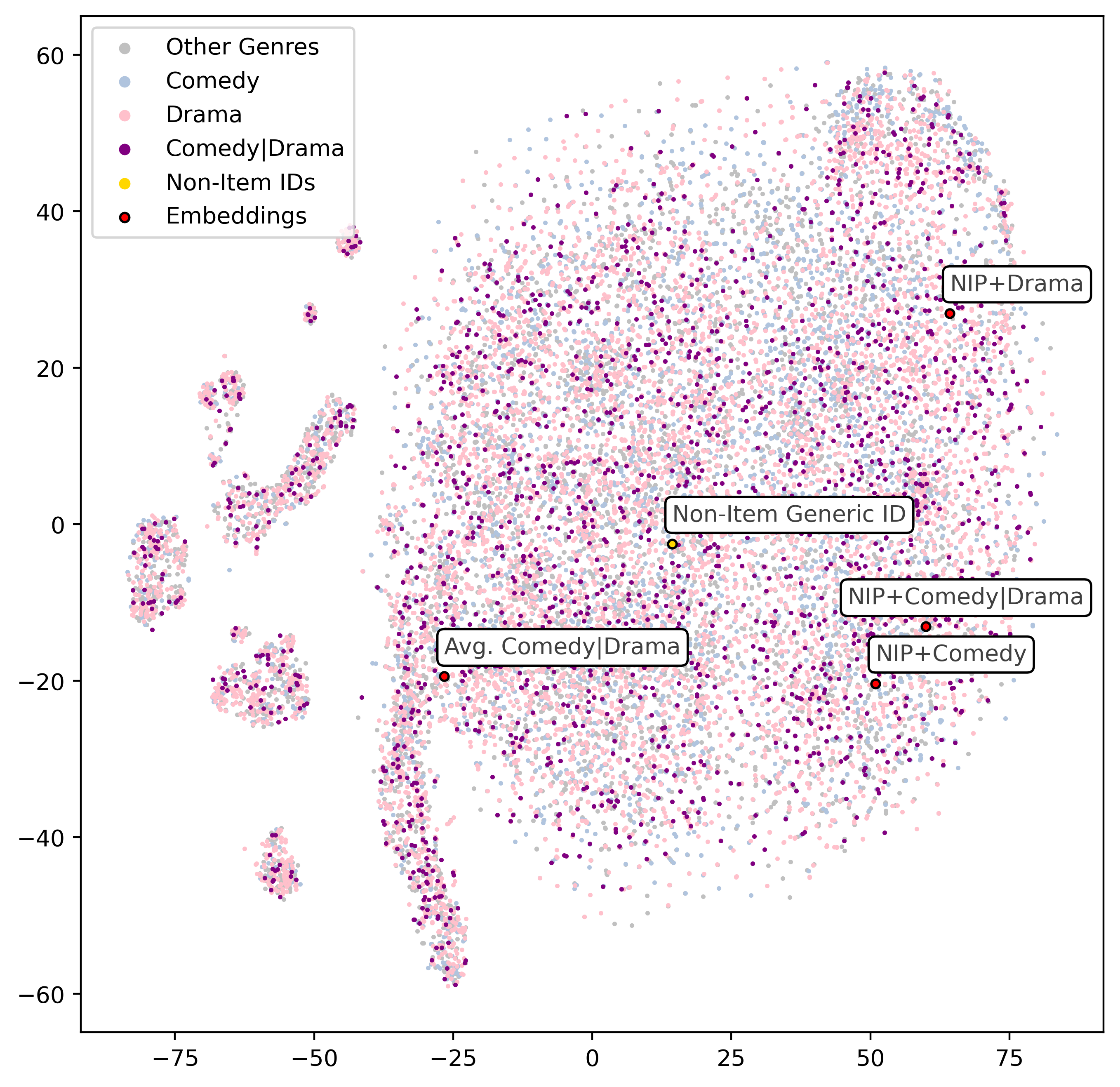}
     \caption{Item embedding space of NextItNet on Prev-\syndata.}
     \label{fig:exp1-tsne-broken-next}
 \end{subfigure}
\centering
\caption{t-SNE visualization of item embedding space of `collapsed' runs of \bertforrec and \nextitnet on Prev-\syndata with page embeddings.}
\label{fig:exp1-tsne-broken}
\Description[t-SNE visualization of item embedding space of `collapsed' runs of \bertforrec and \nextitnet on Prev-\syndata with page embeddings.]{t-SNE visualization of item embedding space of `collapsed' runs of \bertforrec and \nextitnet on Prev-\syndata with page embeddings.}
\end{figure}
Some noticeable exceptions of these patterns can be found when visualizing models which severely underperformed. We visualize both \bertforrec and \nextitnet with page embeddings in \Cref{fig:exp1-tsne-broken}, which both perform similar to their items-only baselines despite added non-item pages. Both figures show no genre specific patterns at all. Some distinct structures are visible, but the difference to other models with non-item pages is evident. 
Overall, for models successfully leveraging non-item pages
the t-SNE visualisations display distinct clusters of movies sharing genres and genre combinations, while the figures for baseline models and `non-successful' runs do not capture the same patterns. 

\subsubsection{Item Embedding Similarity}

As interpreting t-SNE visualizations can be subjective~\cite{wattenberg_how_2016}, we want to verify the similarities and dissimilarities for different models in the item embedding space in a more tangible way. 
While we cannot directly compare item embeddings, we can measure the differences in item similarities between two models. We assume similar item embeddings should keep the semantics between items, for example an item pair closely related in one embedding space should also be closely related in the other embedding space. 
To measure the difference in item similarities between different models, we first compute the cosine similarity between all items for each model. The result is a matrix of item-to-item similarities, which we normalize. To compare two models, we calculate the absolute difference for all item-to-item similarities between the models and report the average as a measure of similarity between the two embeddings.

\begin{figure}
 \begin{subfigure}[b]{0.40\textwidth}
         \centering
         \includegraphics[width=\textwidth]{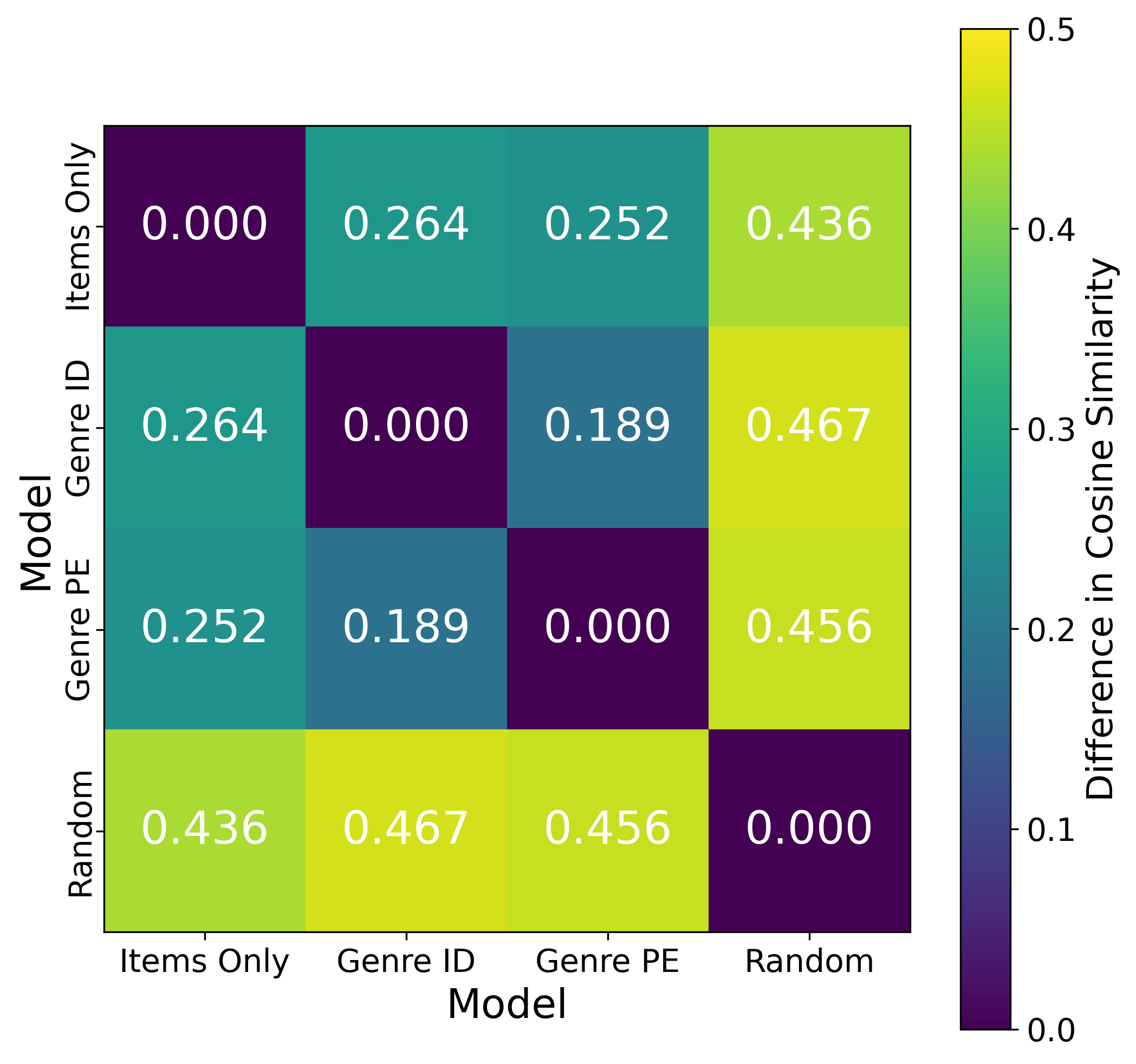}
         \caption{Difference in cosine similarity for \sasrec}
         \label{fig:exp1_sem_emb_sasrec_emb_type}
     \end{subfigure}
     \hfill
  \begin{subfigure}[b]{0.40\textwidth}
     \centering
     \includegraphics[width=\textwidth]{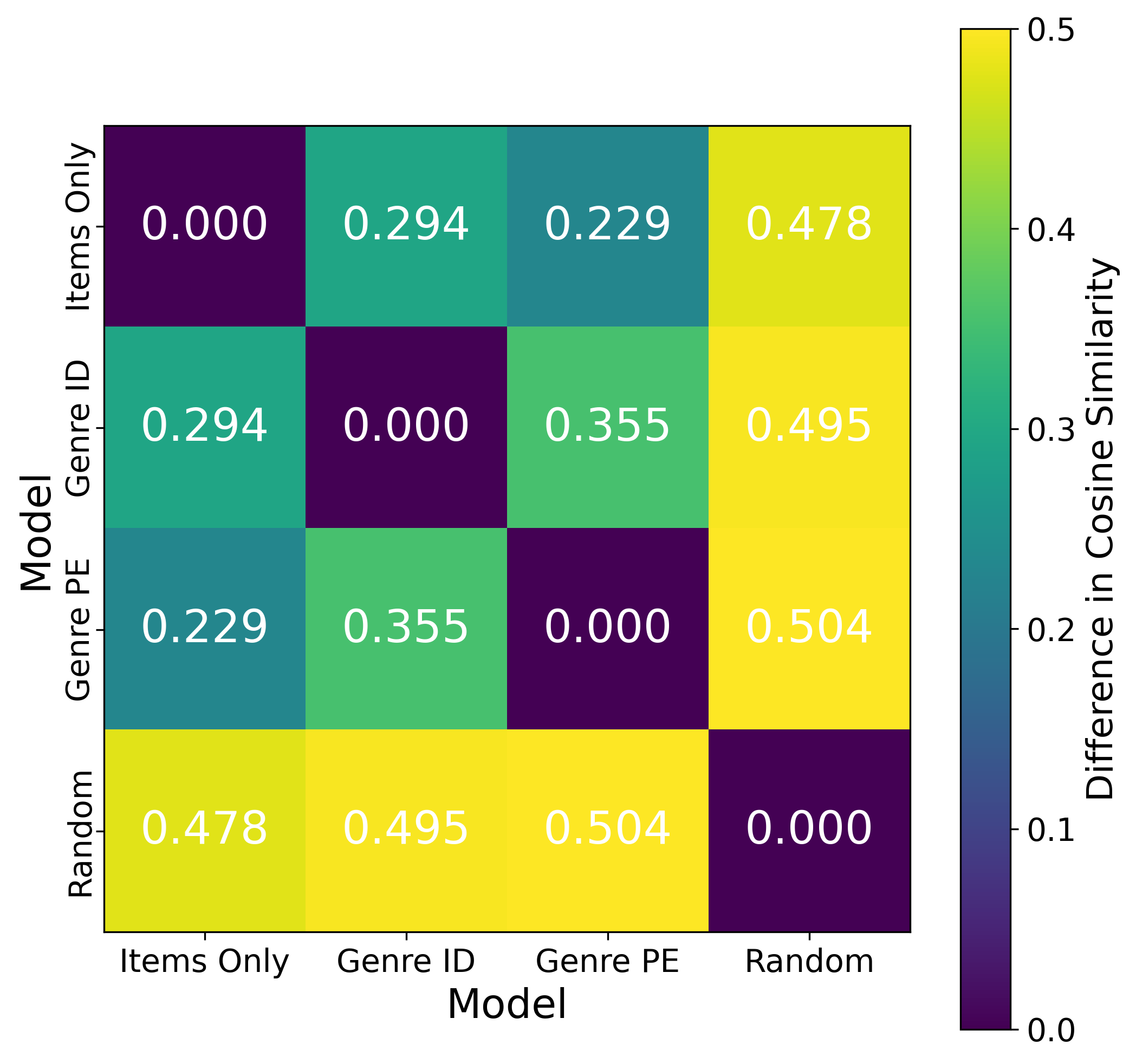}
     \caption{Difference in cosine similarity  for \bertforrec}
     \label{fig:exp1_sem_emb_bert}
 \end{subfigure}
\centering
\Description[Difference in Item-to-Item cosine similarity between different embedding variants for seed 212 on (Prev-)\syndata.]{Difference in Item-to-Item cosine similarity between different embedding variants for seed 212 on (Prev-)\syndata.}
\caption{Difference in Item-to-Item cosine similarity between different embedding variants for seed 212 on (Prev-)\syndata. We report the average of the absolute difference in Item-to-Item similarity between models and a random embedding. A lower score means item-item relations are more similar.}
\label{fig:exp1_sem_emb_type}
\end{figure}

\begin{figure}
 \begin{subfigure}[b]{0.40\textwidth}
         \centering
         \includegraphics[width=\textwidth]{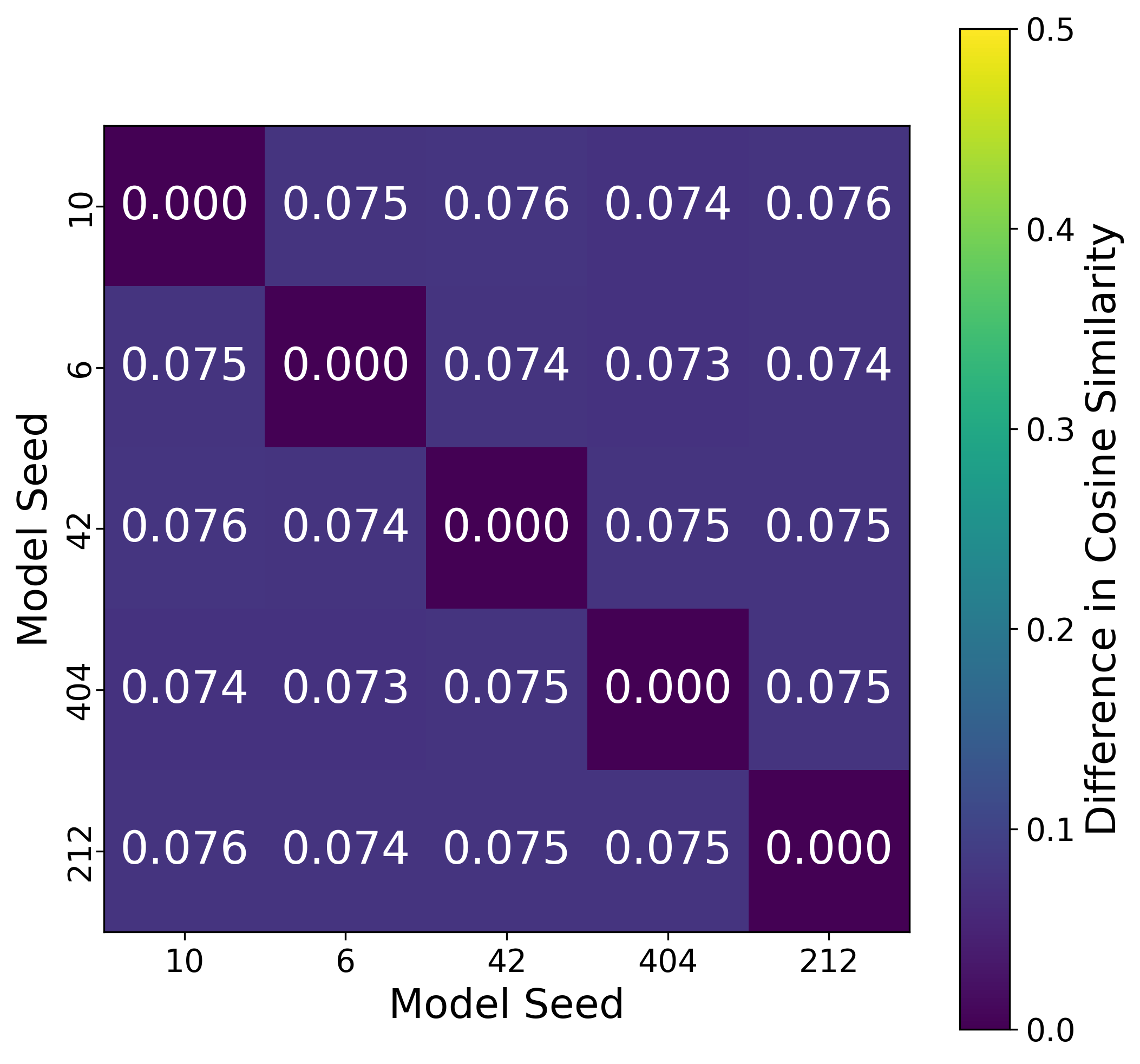}
         \caption{Difference in cosine similarity between different seeds for \sasrec with Items Only.}
         \label{fig:exp1_sem_emb_sasrec_seed}
     \end{subfigure}
     \hfill
  \begin{subfigure}[b]{0.40\textwidth}
     \centering
     \includegraphics[width=\textwidth]{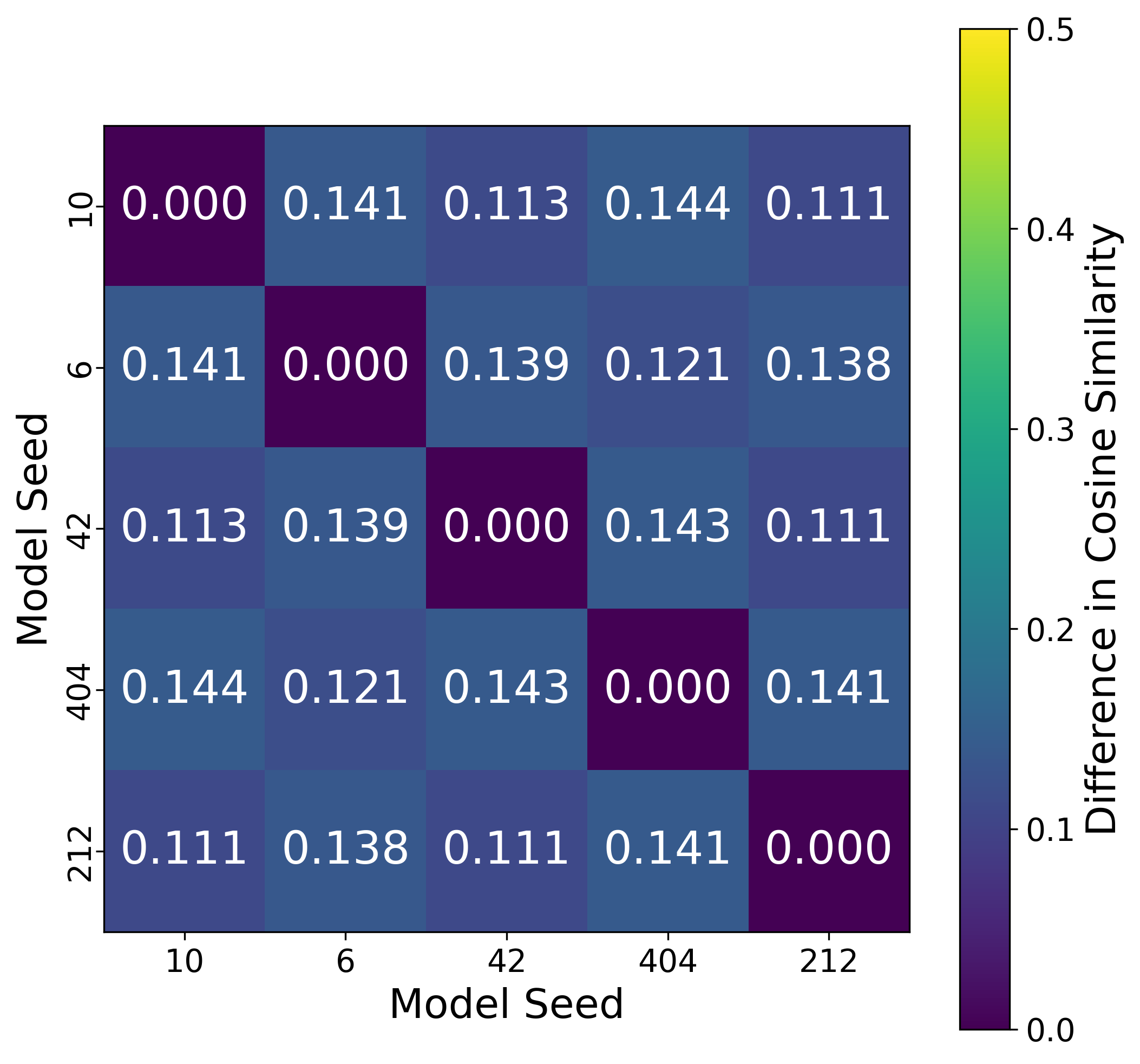}
     \caption{Difference in cosine similarity between different seeds for \bertforrec with page embeddings.}
     \label{fig:exp1_sem_emb_bert_seed}
 \end{subfigure}
\centering
\Description[Difference in Item-to-Item cosine similarity between different embedding variants on (Prev-)\syndata.]{Difference in Item-to-Item cosine similarity between different embedding variants on (Prev-)\syndata.}
\caption{Difference in Item-to-Item cosine similarity between different embedding variants on (Prev-)\syndata. We report the average of the absolute difference in Item-to-Item similarity between models and a random embedding. A lower score means item-item relations are more similar.}
\label{fig:exp1_sem_emb_seed}
\end{figure}

In \Cref{fig:exp1_sem_emb_type} we show the difference in similarity for \sasrec and \bertforrec with and without non-item pages. We show the similarity between the base model and models with CPID and page embeddings as well as a randomly generated embedding. We can confirm our earlier observation: For \sasrec all item embedding variants produce partially similar item embeddings compared to the randomly generated embedding. Therefore, some item properties are shared between all models, but we can also see that the models including non-item pages are a bit closer to each other. For \bertforrec we can also see how the embeddings are in general similar, but that the item embeddings with PE are actually closer to those of the items-only baseline, and dissimilar to the CPID embeddings, which matches the performance of the models for this seed. A comparison between different seeds is shown in \Cref{fig:exp1_sem_emb_seed}.
Similar patterns can be found for the other models, as depicted in \Cref{fig:exp1_sem_emb_appendix}. 

\begin{figure}
 \begin{subfigure}[b]{0.45\textwidth}
         \centering
         \includegraphics[width=\textwidth]{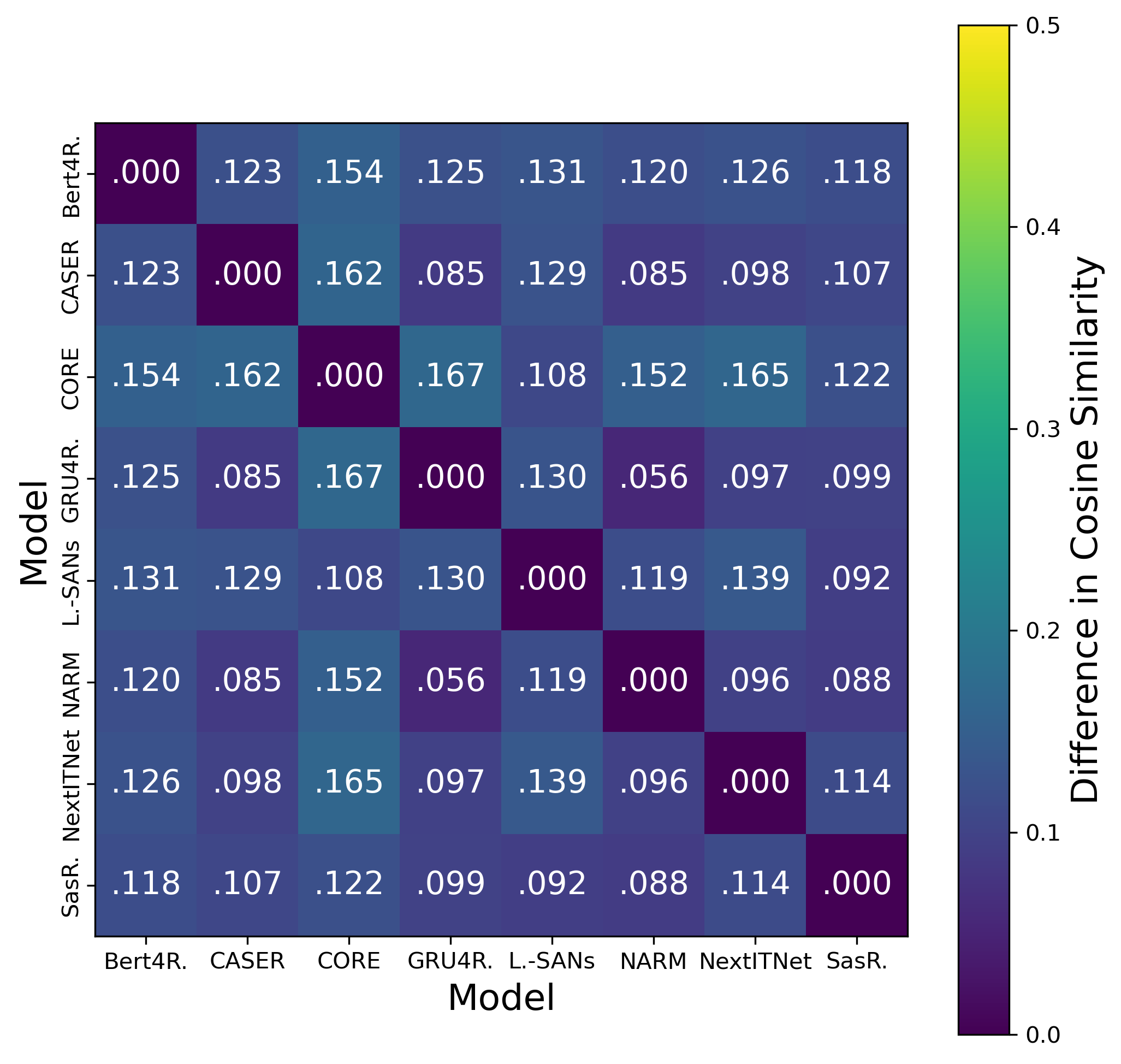}
         \caption{Difference in cosine similarity with Items Only.}
         \label{fig:exp1_sem_emb_models_items_only}
     \end{subfigure}
     \hfill
  \begin{subfigure}[b]{0.45\textwidth}
     \centering
     \includegraphics[width=\textwidth]{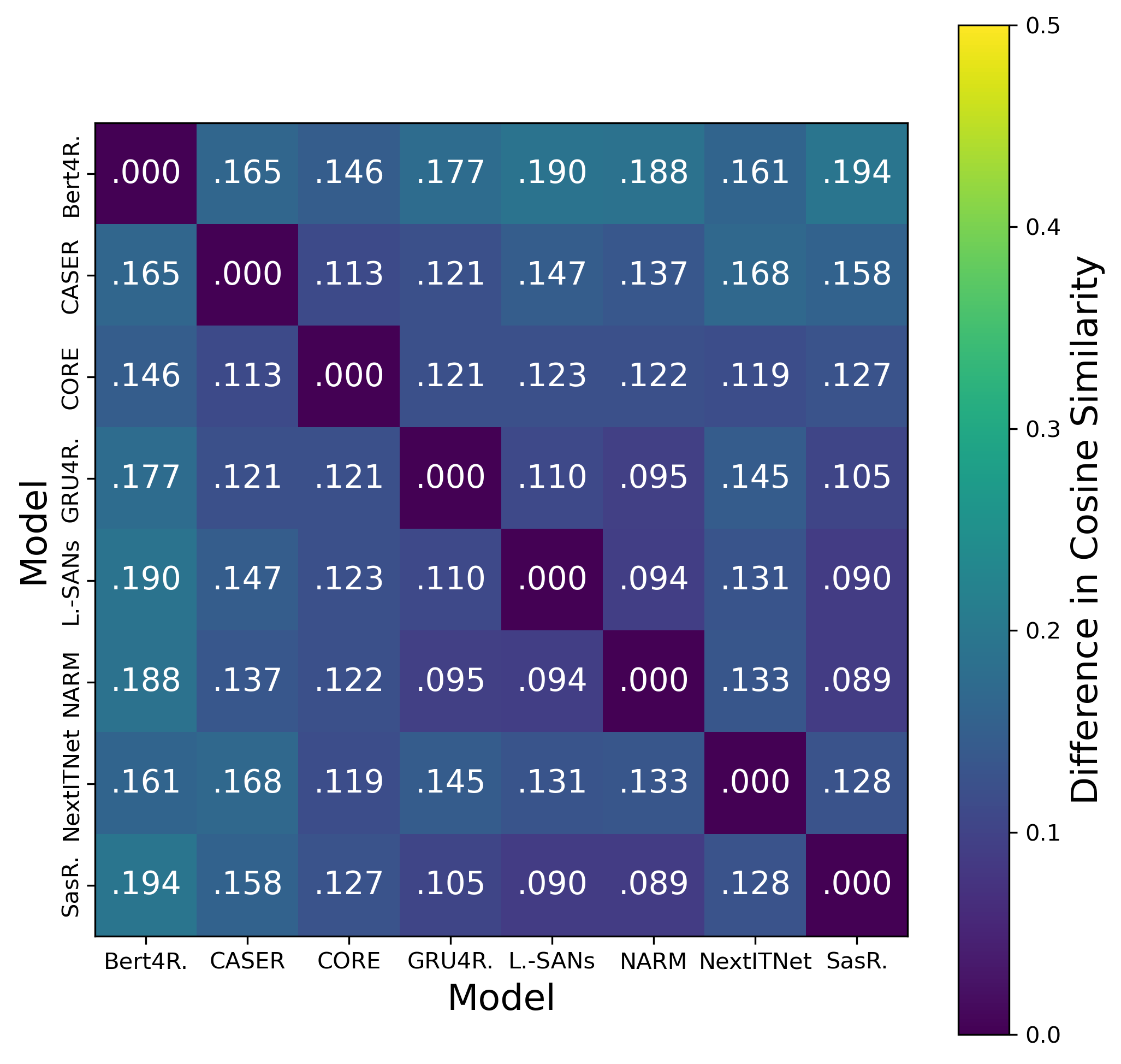}
     \caption{Difference in cosine similarity with page embeddings.}
     \label{fig:exp1_sem_models}
 \end{subfigure}
\centering
\caption{Difference in Item-to-Item cosine similarity between different embedding variants on (Prev-)\syndata. We report the average of the absolute difference in Item-to-Item similarity between models and a random embedding. A lower score means item-item relations are more similar.}
\label{fig:exp1_sem_emb_model}
\Description[Difference in Item-to-Item cosine similarity between different embedding variants on (Prev-)\syndata. ]{Difference in Item-to-Item cosine similarity between different embedding variants on (Prev-)\syndata. }
\end{figure}

We can also compare the similarity of item embeddings for different models, as shown in \Cref{fig:exp1_sem_models} and \Cref{fig:app_exp1_sem_model}.

\FloatBarrier

%% file: data/data_recbole/ml-experiments.tex
\nprounddigits{3}
\npnoaddmissingzero

\begin{tabular}{llrrrrrr}
\toprule
Model & Metrics & Genre-POP & Items-BL & Prev-ID & Prev-PE & Group-ID & Group-PE \\
\midrule
\multirow[t]{5}{*}{BERT4Rec} & HR@1 & \numprint{.010} & \numprint{.01046}\textcolor{lightgray}{\,±\numprint{.0006268971207462989}} & \numprint{.11914}*\textcolor{lightgray}{\,±\numprint{.011573374615901792}} & \numprint{.04008}*\textcolor{lightgray}{\,±\numprint{.04776355305041701}} & \numprint{.29352}*\textcolor{lightgray}{\,±\numprint{.02303045809357686}} & \textbf{\numprint{.30846}}*\textcolor{lightgray}{\,±\numprint{.011797584498531896}} \\
 & HR@5 & \numprint{.031} & \numprint{.04172}\textcolor{lightgray}{\,±\numprint{.0029852973051272473}} & \numprint{.24074}*\textcolor{lightgray}{\,±\numprint{.024835921565345632}} & \numprint{.08688}*\textcolor{lightgray}{\,±\numprint{.09228546472765903}} & \numprint{.53408}*\textcolor{lightgray}{\,±\numprint{.04301548558368253}} & \textbf{\numprint{.5591200000000001}}*\textcolor{lightgray}{\,±\numprint{.008635218584378756}} \\
 & HR@10 & \numprint{.050} & \numprint{.07514000000000001}\textcolor{lightgray}{\,±\numprint{.004413955142499749}} & \numprint{.29791999999999996}*\textcolor{lightgray}{\,±\numprint{.030561855310173816}} & \numprint{.11095999999999999}*\textcolor{lightgray}{\,±\numprint{.108312801644127}} & \numprint{.62436}*\textcolor{lightgray}{\,±\numprint{.046106431221685354}} & \textbf{\numprint{.6527000000000001}}*\textcolor{lightgray}{\,±\numprint{.007340640299047483}} \\
 & NDCG@5 & \numprint{.020} & \numprint{.02594}\textcolor{lightgray}{\,±\numprint{.001804993074778958}} & \numprint{.18288}*\textcolor{lightgray}{\,±\numprint{.018584993946730244}} & \numprint{.06445999999999999}*\textcolor{lightgray}{\,±\numprint{.07131117023300067}} & \numprint{.4210400000000001}*\textcolor{lightgray}{\,±\numprint{.03370672039816392}} & \textbf{\numprint{.441}}*\textcolor{lightgray}{\,±\numprint{.009368030742904286}} \\
 & NDCG@10 & \numprint{.026} & \numprint{.0366}\textcolor{lightgray}{\,±\numprint{.0021943108257491666}} & \numprint{.20136000000000004}*\textcolor{lightgray}{\,±\numprint{.020319522632187993}} & \numprint{.0722}*\textcolor{lightgray}{\,±\numprint{.07648699235817813}} & \numprint{.45034}*\textcolor{lightgray}{\,±\numprint{.03464185329915245}} & \textbf{\numprint{.47138}}*\textcolor{lightgray}{\,±\numprint{.008986768050862336}} \\
\cline{1-8}
\multirow[t]{5}{*}{Caser} & HR@1 & \numprint{.010} & \numprint{.018340000000000002}\textcolor{lightgray}{\,±\numprint{.0014976648490233042}} & \numprint{.31942}*\textcolor{lightgray}{\,±\numprint{.012574259421532544}} & \numprint{.30074}*\textcolor{lightgray}{\,±\numprint{.010747232201827597}} & \numprint{.32192}*\textcolor{lightgray}{\,±\numprint{.010558030119297816}} & \textbf{\numprint{.32288}}*\textcolor{lightgray}{\,±\numprint{.009561746702355175}} \\
 & HR@5 & \numprint{.031} & \numprint{.05382}\textcolor{lightgray}{\,±\numprint{.0016634301909007182}} & \textbf{\numprint{.5762599999999999}}*\textcolor{lightgray}{\,±\numprint{.011684305713220617}} & \numprint{.5565}*\textcolor{lightgray}{\,±\numprint{.01054751155486448}} & \numprint{.57012}*\textcolor{lightgray}{\,±\numprint{.01193616353775365}} & \numprint{.5734999999999999}*\textcolor{lightgray}{\,±\numprint{.00937843270488198}} \\
 & HR@10 & \numprint{.050} & \numprint{.08366}\textcolor{lightgray}{\,±\numprint{.00459597650124541}} & \textbf{\numprint{.6738999999999999}}*\textcolor{lightgray}{\,±\numprint{.005987069399965214}} & \numprint{.6579200000000001}*\textcolor{lightgray}{\,±\numprint{.007217825157206301}} & \numprint{.66614}*\textcolor{lightgray}{\,±\numprint{.010918470588869123}} & \numprint{.6707}*\textcolor{lightgray}{\,±\numprint{.008402975663418272}} \\
 & NDCG@5 & \numprint{.020} & \numprint{.03626}\textcolor{lightgray}{\,±\numprint{.0013849187701811256}} & \numprint{.45548}*\textcolor{lightgray}{\,±\numprint{.011834990494292765}} & \numprint{.4358599999999999}*\textcolor{lightgray}{\,±\numprint{.010839880073137327}} & \numprint{.45336}*\textcolor{lightgray}{\,±\numprint{.011093151040168878}} & \textbf{\numprint{.45553999999999994}}*\textcolor{lightgray}{\,±\numprint{.009063553387055194}} \\
 & NDCG@10 & \numprint{.026} & \numprint{.04586}\textcolor{lightgray}{\,±\numprint{.0023807561823924765}} & \textbf{\numprint{.48717999999999995}}*\textcolor{lightgray}{\,±\numprint{.009981833498911914}} & \numprint{.46882}*\textcolor{lightgray}{\,±\numprint{.00967584621622317}} & \numprint{.48455999999999994}*\textcolor{lightgray}{\,±\numprint{.010845413777260868}} & \numprint{.48714}*\textcolor{lightgray}{\,±\numprint{.008783678045101601}} \\
\cline{1-8}
\multirow[t]{5}{*}{Core} & HR@1 & \numprint{.010} & \numprint{.014600000000000002}\textcolor{lightgray}{\,±\numprint{.0012825755338380658}} & \textbf{\numprint{.34998}}*\textcolor{lightgray}{\,±\numprint{.0016976454282328875}} & \numprint{.30284000000000005}*\textcolor{lightgray}{\,±\numprint{.0072772247457392695}} & \numprint{.34762}*\textcolor{lightgray}{\,±\numprint{.0033796449517663925}} & \numprint{.28687999999999997}*\textcolor{lightgray}{\,±\numprint{.0031188138771013523}} \\
 & HR@5 & \numprint{.031} & \numprint{.058820000000000004}\textcolor{lightgray}{\,±\numprint{.0017683325479106002}} & \textbf{\numprint{.60414}}*\textcolor{lightgray}{\,±\numprint{.002290851370124204}} & \numprint{.54676}*\textcolor{lightgray}{\,±\numprint{.006601742194299942}} & \numprint{.6020800000000001}*\textcolor{lightgray}{\,±\numprint{.0020777391559096154}} & \numprint{.52932}*\textcolor{lightgray}{\,±\numprint{.005546800879786486}} \\
 & HR@10 & \numprint{.050} & \numprint{.10036}\textcolor{lightgray}{\,±\numprint{.0029082640870457446}} & \textbf{\numprint{.69872}}*\textcolor{lightgray}{\,±\numprint{.001788015659886694}} & \numprint{.64236}*\textcolor{lightgray}{\,±\numprint{.005549144078143962}} & \numprint{.6953999999999999}*\textcolor{lightgray}{\,±\numprint{.0013038404810405606}} & \numprint{.6272800000000001}*\textcolor{lightgray}{\,±\numprint{.006979040048602671}} \\
 & NDCG@5 & \numprint{.020} & \numprint{.036359999999999996}\textcolor{lightgray}{\,±\numprint{.0014397916515940783}} & \textbf{\numprint{.4856999999999999}}*\textcolor{lightgray}{\,±\numprint{.0011247221879202023}} & \numprint{.4323}*\textcolor{lightgray}{\,±\numprint{.006419112088131821}} & \numprint{.48302000000000006}*\textcolor{lightgray}{\,±\numprint{.0026395075298244622}} & \numprint{.41524}*\textcolor{lightgray}{\,±\numprint{.0040599261077019526}} \\
 & NDCG@10 & \numprint{.026} & \numprint{.049699999999999994}\textcolor{lightgray}{\,±\numprint{.001650757401921918}} & \textbf{\numprint{.51642}}*\textcolor{lightgray}{\,±\numprint{.000936482781475446}} & \numprint{.46326}*\textcolor{lightgray}{\,±\numprint{.006071078981532029}} & \numprint{.51332}*\textcolor{lightgray}{\,±\numprint{.0020596116138728794}} & \numprint{.44702000000000003}*\textcolor{lightgray}{\,±\numprint{.004460605340085584}} \\
\cline{1-8}
\multirow[t]{5}{*}{GRU4Rec} & HR@1 & \numprint{.010} & \numprint{.02248}\textcolor{lightgray}{\,±\numprint{.001118928058455949}} & \numprint{.35528}*\textcolor{lightgray}{\,±\numprint{.0020753312988532604}} & \numprint{.36410000000000003}*\textcolor{lightgray}{\,±\numprint{.005590617139457866}} & \numprint{.363375}*\textcolor{lightgray}{\,±\numprint{.0027825348155953024}} & \textbf{\numprint{.36838000000000004}}*\textcolor{lightgray}{\,±\numprint{.001207062550160523}} \\
 & HR@5 & \numprint{.031} & \numprint{.07032000000000001}\textcolor{lightgray}{\,±\numprint{.0021545301111843373}} & \numprint{.61214}*\textcolor{lightgray}{\,±\numprint{.0015059880477613296}} & \numprint{.6229}*\textcolor{lightgray}{\,±\numprint{.0023313086453749287}} & \numprint{.616525}*\textcolor{lightgray}{\,±\numprint{.0032014319712694904}} & \textbf{\numprint{.6248400000000001}}*\textcolor{lightgray}{\,±\numprint{.0023839043604976695}} \\
 & HR@10 & \numprint{.050} & \numprint{.10857999999999998}\textcolor{lightgray}{\,±\numprint{.003012806001056157}} & \numprint{.70498}*\textcolor{lightgray}{\,±\numprint{.002710535002541}} & \textbf{\numprint{.71568}}*\textcolor{lightgray}{\,±\numprint{.0035590729129929555}} & \numprint{.7056749999999999}*\textcolor{lightgray}{\,±\numprint{.003163726705432483}} & \numprint{.71486}*\textcolor{lightgray}{\,±\numprint{.0016592166826547602}} \\
 & NDCG@5 & \numprint{.020} & \numprint{.046400000000000004}\textcolor{lightgray}{\,±\numprint{.0015937377450509229}} & \numprint{.49241999999999997}*\textcolor{lightgray}{\,±\numprint{.0012336936410632976}} & \numprint{.50206}*\textcolor{lightgray}{\,±\numprint{.0035697338836389366}} & \numprint{.498275}*\textcolor{lightgray}{\,±\numprint{.0027256497696268346}} & \textbf{\numprint{.5051}}*\textcolor{lightgray}{\,±\numprint{.0015132745950421852}} \\
 & NDCG@10 & \numprint{.026} & \numprint{.058699999999999995}\textcolor{lightgray}{\,±\numprint{.0019196353820452466}} & \numprint{.52256}*\textcolor{lightgray}{\,±\numprint{.0019372661149155812}} & \numprint{.53222}*\textcolor{lightgray}{\,±\numprint{.003949303736103383}} & \numprint{.527225}*\textcolor{lightgray}{\,±\numprint{.0027133927102430543}} & \textbf{\numprint{.5343600000000001}}*\textcolor{lightgray}{\,±\numprint{.0013557285864065902}} \\
\cline{1-8}
\multirow[t]{5}{*}{LightSANs} & HR@1 & \numprint{.010} & \numprint{.02582}\textcolor{lightgray}{\,±\numprint{.0011366617790706257}} & \numprint{.34326}*\textcolor{lightgray}{\,±\numprint{.001985698869416006}} & \numprint{.3452}*\textcolor{lightgray}{\,±\numprint{.003456153931757096}} & \numprint{.34665999999999997}*\textcolor{lightgray}{\,±\numprint{.004066693988979257}} & \textbf{\numprint{.34720000000000006}}*\textcolor{lightgray}{\,±\numprint{.0023653752345029833}} \\
 & HR@5 & \numprint{.031} & \numprint{.0779}\textcolor{lightgray}{\,±\numprint{.0008396427811873341}} & \numprint{.6071600000000001}*\textcolor{lightgray}{\,±\numprint{.001622652150030934}} & \textbf{\numprint{.6087400000000001}}*\textcolor{lightgray}{\,±\numprint{.003113358315388667}} & \numprint{.6031199999999999}*\textcolor{lightgray}{\,±\numprint{.0018143869488066887}} & \numprint{.60256}*\textcolor{lightgray}{\,±\numprint{.0016288032416470568}} \\
 & HR@10 & \numprint{.050} & \numprint{.11750000000000001}\textcolor{lightgray}{\,±\numprint{.0014525839046333946}} & \numprint{.7013199999999999}*\textcolor{lightgray}{\,±\numprint{.0018005554698481087}} & \textbf{\numprint{.7032400000000001}}*\textcolor{lightgray}{\,±\numprint{.002200681712560929}} & \numprint{.69514}*\textcolor{lightgray}{\,±\numprint{.002193855054464629}} & \numprint{.69438}*\textcolor{lightgray}{\,±\numprint{.0023123581037547227}} \\
 & NDCG@5 & \numprint{.020} & \numprint{.05216}\textcolor{lightgray}{\,±\numprint{.0006148170459575736}} & \numprint{.4832000000000001}*\textcolor{lightgray}{\,±\numprint{.0016688319268278653}} & \textbf{\numprint{.48529999999999995}}*\textcolor{lightgray}{\,±\numprint{.0029849623113198578}} & \numprint{.48316}*\textcolor{lightgray}{\,±\numprint{.0009154233993076702}} & \numprint{.48268000000000005}*\textcolor{lightgray}{\,±\numprint{.0011903780911962272}} \\
 & NDCG@10 & \numprint{.026} & \numprint{.06492}\textcolor{lightgray}{\,±\numprint{.0005167204273105539}} & \numprint{.51384}*\textcolor{lightgray}{\,±\numprint{.0013849187701810964}} & \textbf{\numprint{.51596}}*\textcolor{lightgray}{\,±\numprint{.002719007171744864}} & \numprint{.5130399999999999}*\textcolor{lightgray}{\,±\numprint{.0013049904214207761}} & \numprint{.51252}*\textcolor{lightgray}{\,±\numprint{.0011166915420114632}} \\
\cline{1-8}
\multirow[t]{5}{*}{NARM} & HR@1 & \numprint{.010} & \numprint{.02196}\textcolor{lightgray}{\,±\numprint{.001064424727258813}} & \numprint{.35547999999999996}*\textcolor{lightgray}{\,±\numprint{.0043447669672837495}} & \numprint{.3573}*\textcolor{lightgray}{\,±\numprint{.0036338684621213136}} & \numprint{.35638}*\textcolor{lightgray}{\,±\numprint{.002537124356431898}} & \textbf{\numprint{.35801999999999995}}*\textcolor{lightgray}{\,±\numprint{.004967091704408129}} \\
 & HR@5 & \numprint{.031} & \numprint{.07054}\textcolor{lightgray}{\,±\numprint{.0013722244714331538}} & \numprint{.6169399999999999}*\textcolor{lightgray}{\,±\numprint{.0024825390228554295}} & \numprint{.61492}*\textcolor{lightgray}{\,±\numprint{.004774620403759857}} & \numprint{.61338}*\textcolor{lightgray}{\,±\numprint{.0015385057685949643}} & \textbf{\numprint{.6173200000000001}}*\textcolor{lightgray}{\,±\numprint{.0033640749099864095}} \\
 & HR@10 & \numprint{.050} & \numprint{.11072000000000001}\textcolor{lightgray}{\,±\numprint{.0007726577508832756}} & \textbf{\numprint{.71146}}*\textcolor{lightgray}{\,±\numprint{.002255659548779448}} & \numprint{.70808}*\textcolor{lightgray}{\,±\numprint{.0018130085493455502}} & \numprint{.7052599999999999}*\textcolor{lightgray}{\,±\numprint{.0027097970403703775}} & \numprint{.70842}*\textcolor{lightgray}{\,±\numprint{.003784441834669935}} \\
 & NDCG@5 & \numprint{.020} & \numprint{.04636}\textcolor{lightgray}{\,±\numprint{.0010163660757817542}} & \numprint{.49476}*\textcolor{lightgray}{\,±\numprint{.0034238866803677935}} & \numprint{.4946}*\textcolor{lightgray}{\,±\numprint{.003582596823534556}} & \numprint{.49338000000000004}*\textcolor{lightgray}{\,±\numprint{.0019253571097331593}} & \textbf{\numprint{.49622}}*\textcolor{lightgray}{\,±\numprint{.0033010604356782194}} \\
 & NDCG@10 & \numprint{.026} & \numprint{.05926}\textcolor{lightgray}{\,±\numprint{.0008080841540334768}} & \numprint{.5254}*\textcolor{lightgray}{\,±\numprint{.003232645975048928}} & \numprint{.5248800000000001}*\textcolor{lightgray}{\,±\numprint{.0026789923478800453}} & \numprint{.52324}*\textcolor{lightgray}{\,±\numprint{.002302824352832859}} & \textbf{\numprint{.52586}}*\textcolor{lightgray}{\,±\numprint{.0031761612049768548}} \\
\cline{1-8}
\multirow[t]{5}{*}{NextItNet} & HR@1 & \numprint{.010} & \numprint{.006899999999999999}\textcolor{lightgray}{\,±\numprint{.0004898979485566358}} & \textbf{\numprint{.25314000000000003}}*\textcolor{lightgray}{\,±\numprint{.0024306377763870916}} & \numprint{.20871999999999996}*\textcolor{lightgray}{\,±\numprint{.11168004297993442}} & \numprint{.23815}*\textcolor{lightgray}{\,±\numprint{.0037120075430957843}} & \numprint{.1923}*\textcolor{lightgray}{\,±\numprint{.10340766412602113}} \\
 & HR@5 & \numprint{.031} & \numprint{.02196}\textcolor{lightgray}{\,±\numprint{.0010382677881933932}} & \textbf{\numprint{.47051999999999994}}*\textcolor{lightgray}{\,±\numprint{.0028994827124851083}} & \numprint{.38842}*\textcolor{lightgray}{\,±\numprint{.19872012228257105}} & \numprint{.45935}*\textcolor{lightgray}{\,±\numprint{.001839293342563926}} & \numprint{.37532}*\textcolor{lightgray}{\,±\numprint{.19229372584668486}} \\
 & HR@10 & \numprint{.050} & \numprint{.03574}\textcolor{lightgray}{\,±\numprint{.0013813037319865627}} & \textbf{\numprint{.5587799999999999}}*\textcolor{lightgray}{\,±\numprint{.00487052358581703}} & \numprint{.46302000000000004}*\textcolor{lightgray}{\,±\numprint{.22769553574894702}} & \numprint{.5508666666666667}*\textcolor{lightgray}{\,±\numprint{.0020646226451016604}} & \numprint{.45506}*\textcolor{lightgray}{\,±\numprint{.22426723568100626}} \\
 & NDCG@5 & \numprint{.020} & \numprint{.0144}\textcolor{lightgray}{\,±\numprint{.0005049752469181034}} & \textbf{\numprint{.3683}}*\textcolor{lightgray}{\,±\numprint{.0022704625079485517}} & \numprint{.30362}*\textcolor{lightgray}{\,±\numprint{.15785007760530245}} & \numprint{.3545}*\textcolor{lightgray}{\,±\numprint{.002564371267971921}} & \numprint{.2886}*\textcolor{lightgray}{\,±\numprint{.15054872965256133}} \\
 & NDCG@10 & \numprint{.026} & \numprint{.018779999999999998}\textcolor{lightgray}{\,±\numprint{.0005630275304103693}} & \textbf{\numprint{.39693999999999996}}*\textcolor{lightgray}{\,±\numprint{.002822764602300373}} & \numprint{.32780000000000004}*\textcolor{lightgray}{\,±\numprint{.16729726536916256}} & \numprint{.38409999999999994}*\textcolor{lightgray}{\,±\numprint{.0023958297101421967}} & \numprint{.31446}*\textcolor{lightgray}{\,±\numprint{.1609137129022881}} \\
\cline{1-8}
\multirow[t]{5}{*}{SASRec} & HR@1 & \numprint{.010} & \numprint{.025420000000000005}\textcolor{lightgray}{\,±\numprint{.001688786546606764}} & \numprint{.37012}*\textcolor{lightgray}{\,±\numprint{.002653676694701133}} & \numprint{.36766}*\textcolor{lightgray}{\,±\numprint{.0024460171708309683}} & \textbf{\numprint{.37106}}*\textcolor{lightgray}{\,±\numprint{.0020452383724153082}} & \numprint{.36082}*\textcolor{lightgray}{\,±\numprint{.0029149614062625407}} \\
 & HR@5 & \numprint{.031} & \numprint{.08080000000000001}\textcolor{lightgray}{\,±\numprint{.0016867127793432992}} & \textbf{\numprint{.63358}}*\textcolor{lightgray}{\,±\numprint{.0024365959862069834}} & \numprint{.62952}*\textcolor{lightgray}{\,±\numprint{.002552841554033483}} & \numprint{.6286799999999999}*\textcolor{lightgray}{\,±\numprint{.0017512852423291602}} & \numprint{.6175599999999999}*\textcolor{lightgray}{\,±\numprint{.0028360183356247745}} \\
 & HR@10 & \numprint{.050} & \numprint{.12258}\textcolor{lightgray}{\,±\numprint{.0012153188881935484}} & \textbf{\numprint{.72538}}*\textcolor{lightgray}{\,±\numprint{.0024045789652244634}} & \numprint{.72104}*\textcolor{lightgray}{\,±\numprint{.002025586334866977}} & \numprint{.71716}*\textcolor{lightgray}{\,±\numprint{.001145862120850517}} & \numprint{.7070200000000001}*\textcolor{lightgray}{\,±\numprint{.003166543857267729}} \\
 & NDCG@5 & \numprint{.020} & \numprint{.053459999999999994}\textcolor{lightgray}{\,±\numprint{.0012461942063739491}} & \textbf{\numprint{.5107800000000001}}*\textcolor{lightgray}{\,±\numprint{.0020933227175951724}} & \numprint{.5075200000000001}*\textcolor{lightgray}{\,±\numprint{.00238474317275467}} & \numprint{.50838}*\textcolor{lightgray}{\,±\numprint{.0011924764148610952}} & \numprint{.49782000000000004}*\textcolor{lightgray}{\,±\numprint{.001979141227906695}} \\
 & NDCG@10 & \numprint{.026} & \numprint{.06688}\textcolor{lightgray}{\,±\numprint{.0009257429448826492}} & \textbf{\numprint{.5405599999999999}}*\textcolor{lightgray}{\,±\numprint{.0016009372255026105}} & \numprint{.5372399999999999}*\textcolor{lightgray}{\,±\numprint{.0019073541883981512}} & \numprint{.53716}*\textcolor{lightgray}{\,±\numprint{.0012601587201618485}} & \numprint{.52688}*\textcolor{lightgray}{\,±\numprint{.0020608250774871713}} \\
\cline{1-8}
\bottomrule
\end{tabular}

%% file: sections/52_exp_2_pageview.tex
\begin{table}[ht]
  \caption{\revised{Average Hitrate and NDCG with standard deviation over five random seeds for non-item models with Unique Page IDs (UPID) and items-only baseline models (Items-BL) on the \coveopage dataset. Results marked with * show a significantly higher performance compared to the baseline for $p<0.01$.}}
  \centering
  \begin{tabular}{cc}   
  \input{data/data_recbole/pageview-experiments-1}
  &
  \input{data/data_recbole/pageview-experiments-2}
  \end{tabular}
  \label{tab:coveo-page-results-rec}
\end{table}

\subsection{RQ3: Are there gains from including non-item pages without content representations?}
As we have seen that embedding pages via ids is beneficial in an artificial setting, we aim to verify this with the \coveopage dataset on real-life data. In this dataset, the only available representation for non-product interactions is a unique hashed URL, leaving no alternative method for creating a different representation. Using the URL as a unique page id gives us the results shown in \Cref{tab:coveo-page-results-rec}. We train all models for 50 epochs with a batch size of $64$ and a maximum sequence length of $30$.  

\subsubsection{Results}
The influence of non-item pages on the performance varies between models. For \caser , \core and \nextitnet the performance drops with non-item pages. 
The same seems to be the case for \narm, but the model is also very unstable on this dataset: three models with non-item pages and one baseline model collapse and report a HR@10 $\leq .002$. Excluding these runs gives an average HR@10 of $.389$ for the baseline and  UPID at a very similar performance with $.391$. Small differences are also found in \bertforrec and \sasrec, with small decreases but also increases in some metrics. Only for \lightsans and \gruforrec we see a clear improvement in all metrics, e.g., of $.022$ and $.045$ in HR@10. One reason for this might be the number of non-item pages, which is more than twice as many as the number of items in the \coveopage dataset (\Cref{tab:dataset_stats}) and therefore their usage is very sparse.

The growing vocabulary might also hinder the models, as already discussed in \Cref{sec:modnonitems} in more detail. As we do not have further information on the types of non-item pages in the dataset, it is also possible that the non-item pages contain too little useful information and too many random and noisy pages. Interestingly, all models that exhibited a greater performance decline (\caser, \core, \nextitnet ) are models with a steeper drop in performance in our experiment with added noise (see \Cref{sec:exp1:noise}), supporting our finding that they are less effective at handling noisy interactions compared to the others. 
Overall, there is only a small benefit of using the non-item pages here, as only two models can extract useful behavior patterns, but the majority does not improve. 

\subsubsection{Visualization and Item Embedding Similarity}

\begin{figure}
 \begin{subfigure}[b]{0.40\textwidth}
         \centering
         \includegraphics[width=\textwidth]{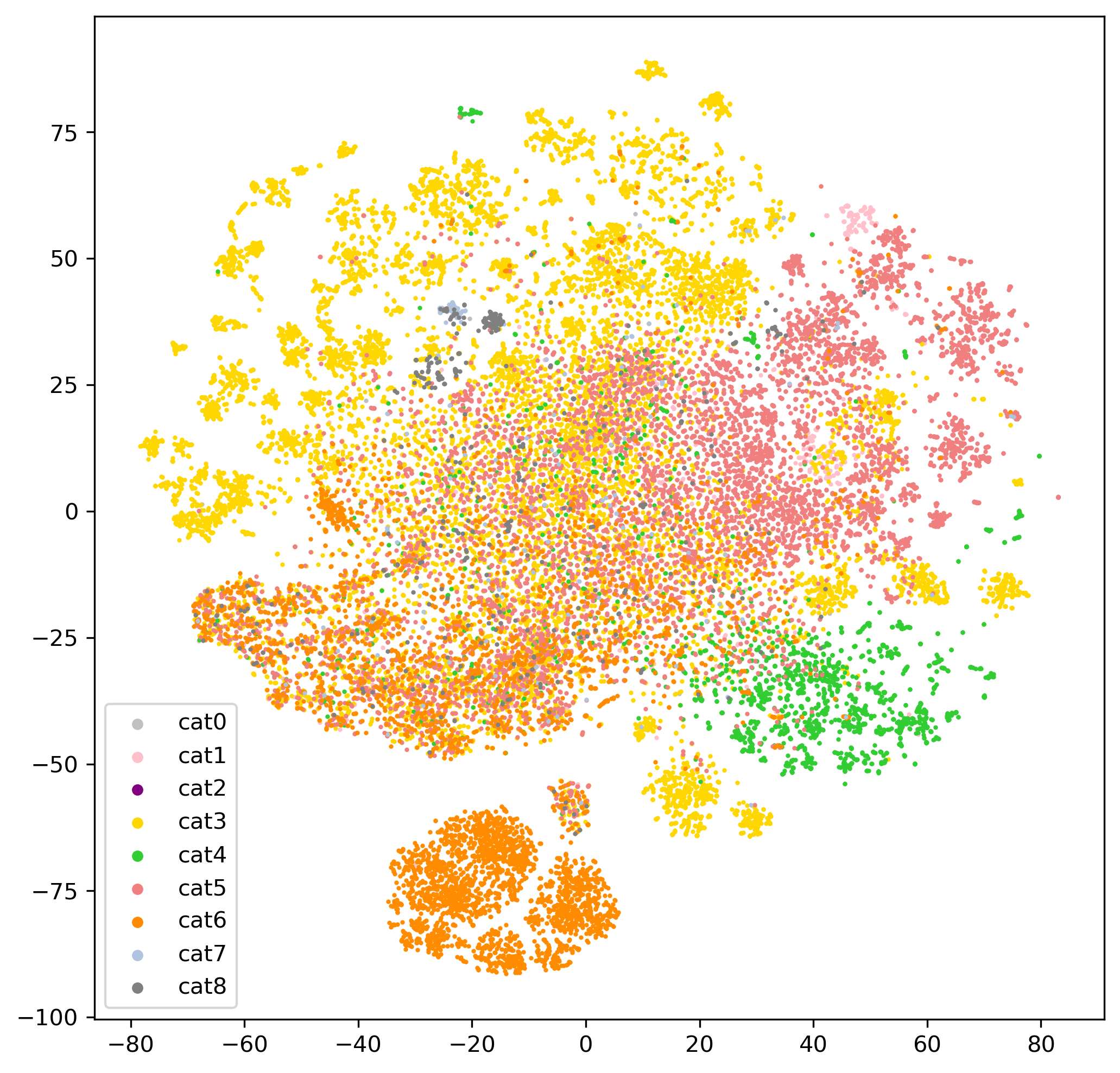}
         \caption{Visualization of the item embedding space of \sasrec on \coveopage with Items Only.}
     \end{subfigure}
     \hfill
  \begin{subfigure}[b]{0.40\textwidth}
     \centering
     \includegraphics[width=\textwidth]{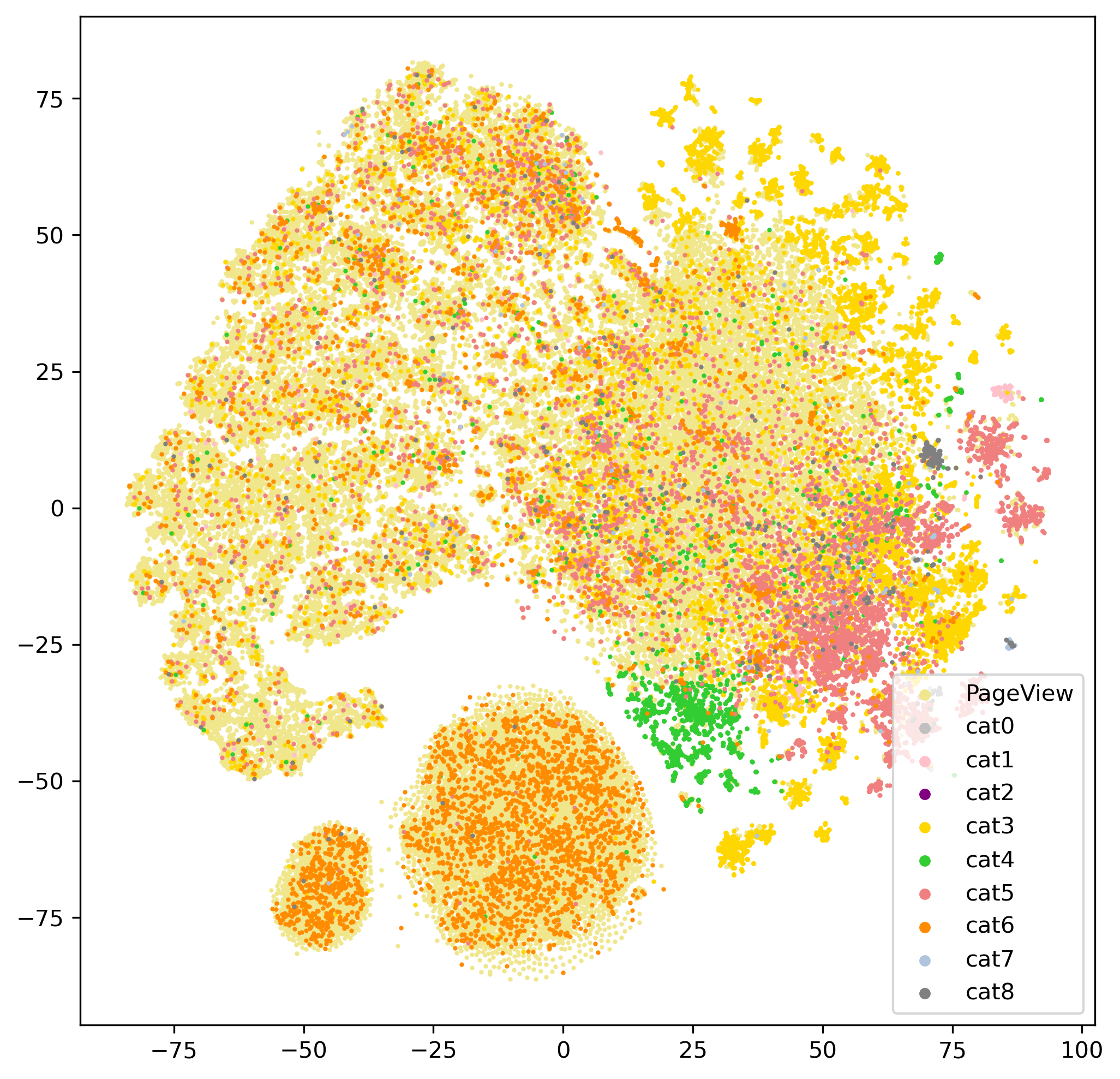}
     \caption{Visualization of the item embedding space of \sasrec on \coveopage with Non-Item Pages.}
 \end{subfigure}
\centering
\caption{t-SNE visualization of item embedding space of \sasrec on \coveopage with CPID embeddings.}
\Description[t-SNE visualization of item embedding space of \sasrec on \coveopage with CPID embeddings.]{t-SNE visualization of item embedding space of \sasrec on \coveopage with CPID embeddings.}
\label{fig:exp2-tsne-sasrec}
\end{figure}

We visualize exemplary the item embedding space for \sasrec
with and without Non-Item Pages in \Cref{fig:exp2-tsne-sasrec} on seed $212$. We can see some recurring patterns in both, based on the top-level categories of the items. We also see the huge amount of non-item pages in this dataset. They also form patterns, for example around the bigger cluster of ``cat6'' on the bottom, which means that they are probably related to these items. In most model visualizations we can identify some similar clusters, but 
the item embedding space looks quite different for models, which have a performance close to zero, for example a run of \narm without pageviews (see \Cref{fig:app-exp2-tsne-narm}). 

\begin{figure}
 \begin{subfigure}[b]{0.40\textwidth}
         \centering
         \includegraphics[width=\textwidth]{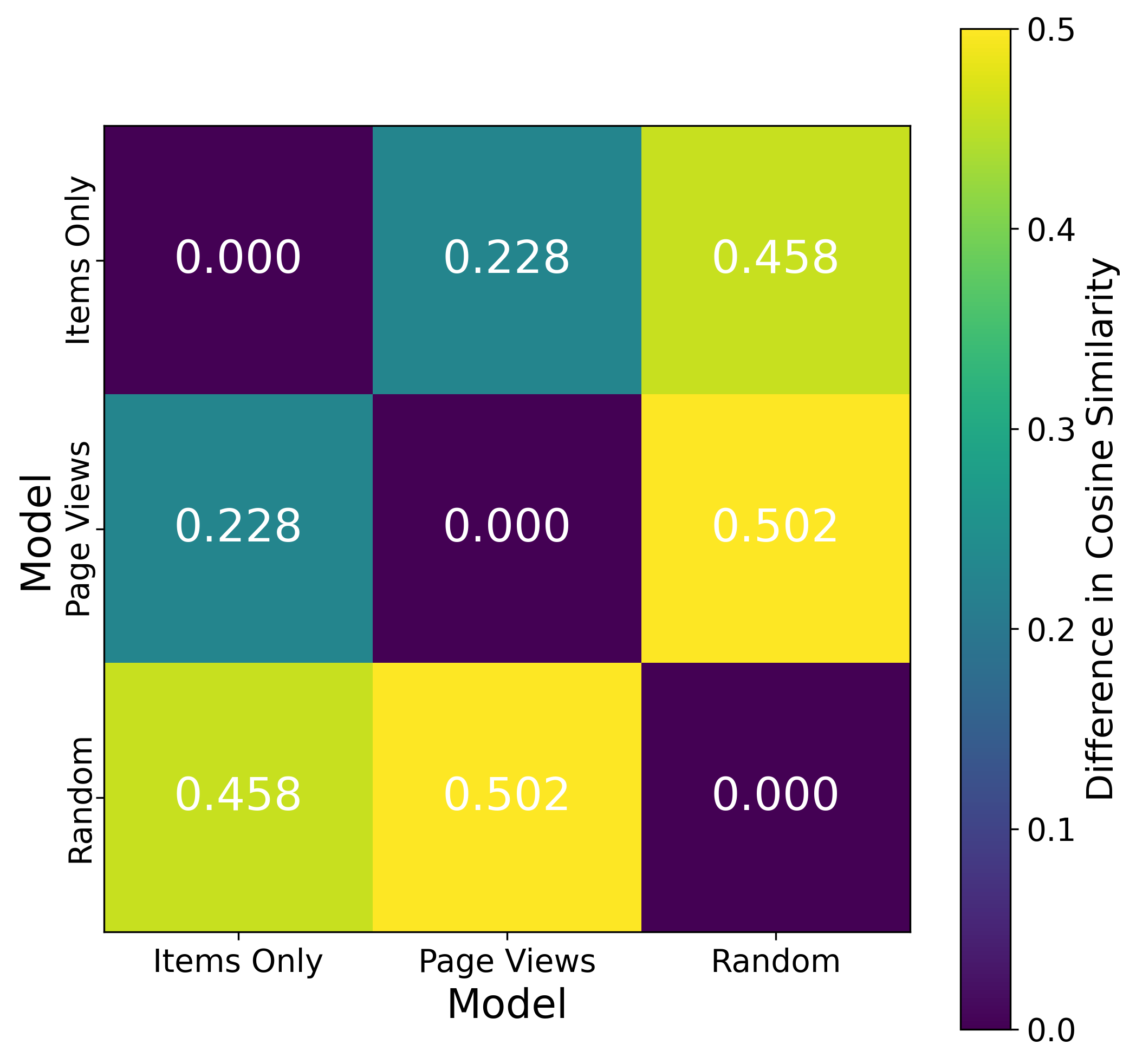}
         \caption{Difference in cosine similarity between embedding variants for \sasrec.}
     \end{subfigure}
     \hfill
  \begin{subfigure}[b]{0.40\textwidth}
     \centering
     \includegraphics[width=\textwidth]{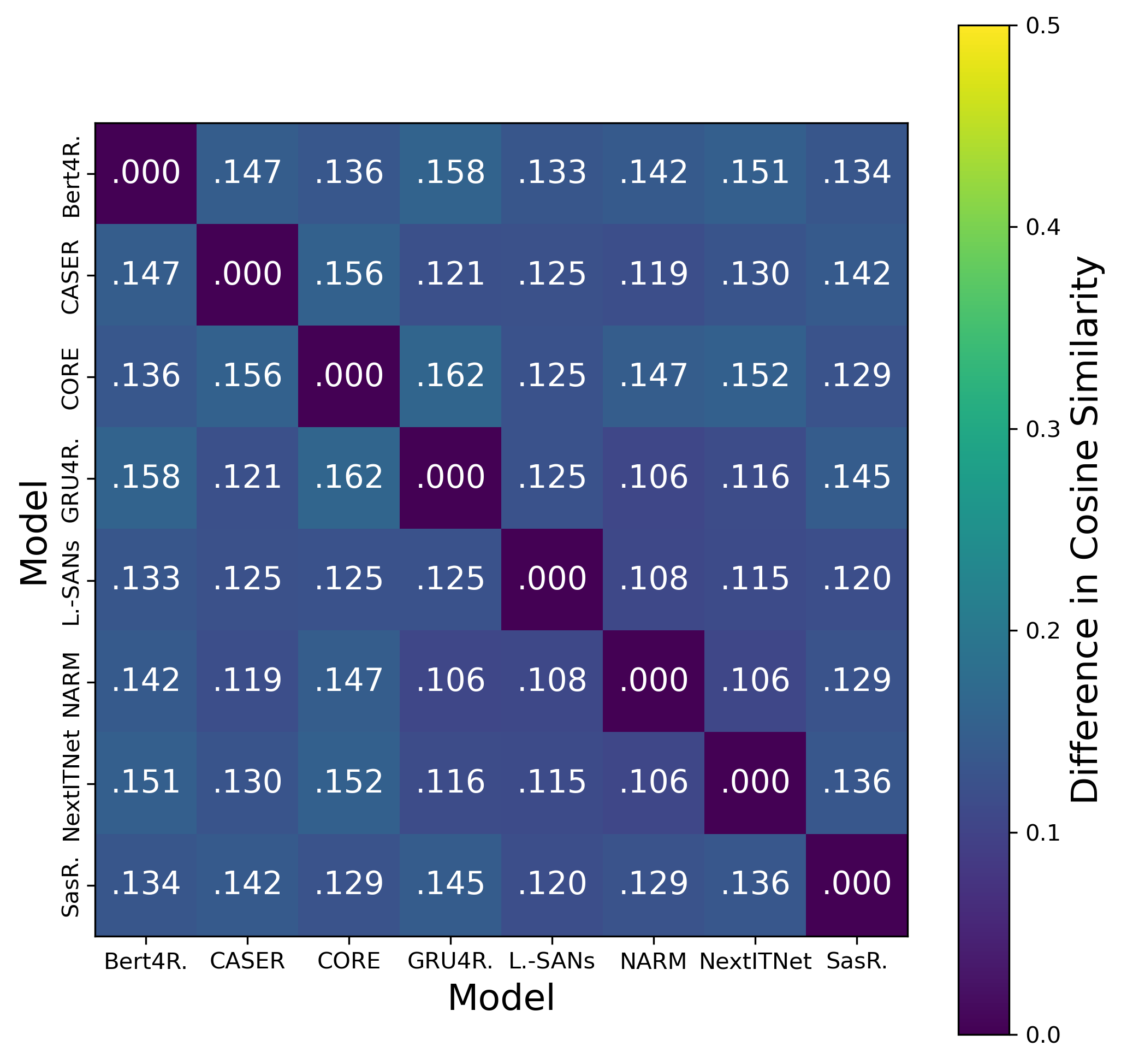}
     \caption{Difference in cosine similarity between different models with page embeddings.}
     \label{fig:exp2_sem_sim}
 \end{subfigure}
\centering
\caption{Difference in Item-to-Item cosine similarity between different  variants on \coveopage. For each model the cosine similarity between all items is computed and normalized. We report the average of the absolute difference in similarity between two models. A lower score means item-item relations are similar in both models. A randomly generated embedding is included as a baseline.}
\Description[Difference in Item-to-Item cosine similarity between different  variants on \coveopage.]{Difference in Item-to-Item cosine similarity between different  variants on \coveopage.}
\label{fig:exp2_sem_emb}
\end{figure}

We also look at the difference in cosine similarity for different models and item embeddings. We observe a similar overall pattern as before in the \syndata dataset: The difference in cosine similarity between our embedding variants is bigger than the difference between different models. We exemplary plot these for \sasrec and models with page embeddings in \Cref{fig:exp2_sem_emb}. Between the models with page embeddings, \core and \gruforrec show bigger differences to most other models. \core is by far the best performing model, while \gruforrec has a huge improvement through non-item pages, which both seems to create more different item embeddings. The \core model also has a high dissimilarity in the items only setting and is also the best performing model there, which suggests that it is able to capture aspects of the data the other models are missing. Additional visualizations are included in our repository.

%% file: data/data_recbole/pageview-experiments-1.tex
\nprounddigits{3}
\npnoaddmissingzero

\begin{tabular}{llrr}
\toprule
 Models & Metrics & Items-BL & UPID \\
\midrule
\multirow[t]{5}{*}{BERT4Rec} & HR@1 & \textbf{\numprint{.08224000000000001}}\textcolor{lightgray}{± \numprint{.0016410362579784748}} & \numprint{.07972}\textcolor{lightgray}{± \numprint{.003783781177605279}} \\
 & HR@5 & \textbf{\numprint{.28578000000000003}}\textcolor{lightgray}{± \numprint{.0033191866473580565}} & \numprint{.28415999999999997}\textcolor{lightgray}{± \numprint{.0028023204670415383}} \\
 & HR@10 & \numprint{.35956}\textcolor{lightgray}{± \numprint{.003423886680367791}} & \textbf{\numprint{.36058}}*\textcolor{lightgray}{± \numprint{.002706843179794504}} \\
 & NDCG@5 & \textbf{\numprint{.187}}\textcolor{lightgray}{± \numprint{.002500999800079962}} & \numprint{.18478}\textcolor{lightgray}{± \numprint{.0026696441710460208}} \\
 & NDCG@10 & \textbf{\numprint{.21112000000000003}}\textcolor{lightgray}{± \numprint{.00253515285535212}} & \numprint{.20974000000000004}\textcolor{lightgray}{± \numprint{.0029636126602509937}} \\
\cline{1-4}
\multirow[t]{5}{*}{Caser} & HR@1 & \textbf{\numprint{.09323999999999999}}\textcolor{lightgray}{± \numprint{.0005412947441089774}} & \numprint{.08348}\textcolor{lightgray}{± \numprint{.004076395466585644}} \\
 & HR@5 & \textbf{\numprint{.2881}}\textcolor{lightgray}{± \numprint{.0019455076458343781}} & \numprint{.2722}\textcolor{lightgray}{± \numprint{.004205353730662862}} \\
 & HR@10 & \textbf{\numprint{.36034}}\textcolor{lightgray}{± \numprint{.0023437149997386623}} & \numprint{.34673999999999994}\textcolor{lightgray}{± \numprint{.0034326374699347427}} \\
 & NDCG@5 & \textbf{\numprint{.19474}}\textcolor{lightgray}{± \numprint{.0015820872289478828}} & \numprint{.18142}\textcolor{lightgray}{± \numprint{.003604441704342017}} \\
 & NDCG@10 & \textbf{\numprint{.21822}}\textcolor{lightgray}{± \numprint{.0017123083834403186}} & \numprint{.2057}\textcolor{lightgray}{± \numprint{.003289376840679707}} \\
\cline{1-4}
\multirow[t]{5}{*}{Core} & HR@1 & \textbf{\numprint{.14222}}\textcolor{lightgray}{± \numprint{.004646719272777307}} & \numprint{.11314000000000002}\textcolor{lightgray}{± \numprint{.006905287828903291}} \\
 & HR@5 & \textbf{\numprint{.40252}}\textcolor{lightgray}{± \numprint{.0008786353054595534}} & \numprint{.38126}\textcolor{lightgray}{± \numprint{.0014152738250953298}} \\
 & HR@10 & \textbf{\numprint{.47648}}\textcolor{lightgray}{± \numprint{.0010986355173577807}} & \numprint{.45996}\textcolor{lightgray}{± \numprint{.001546932448428181}} \\
 & NDCG@5 & \textbf{\numprint{.2808}}\textcolor{lightgray}{± \numprint{.0022715633383201175}} & \numprint{.25624}\textcolor{lightgray}{± \numprint{.003533128924904941}} \\
 & NDCG@10 & \textbf{\numprint{.30494000000000004}}\textcolor{lightgray}{± \numprint{.002532390175308705}} & \numprint{.28188}\textcolor{lightgray}{± \numprint{.003556262082580528}} \\
\cline{1-4}
\multirow[t]{5}{*}{GRU4Rec} & HR@1 & \numprint{.06598}\textcolor{lightgray}{± \numprint{.002537124356431903}} & \textbf{\numprint{.0913}}*\textcolor{lightgray}{± \numprint{.006363568181452915}} \\
 & HR@5 & \numprint{.22834}\textcolor{lightgray}{± \numprint{.0023828554299411436}} & \textbf{\numprint{.26756}}*\textcolor{lightgray}{± \numprint{.002781726082848571}} \\
 & HR@10 & \numprint{.29776}\textcolor{lightgray}{± \numprint{.0024182638400306916}} & \textbf{\numprint{.34308}}*\textcolor{lightgray}{± \numprint{.002267597847943953}} \\
 & NDCG@5 & \numprint{.14977999999999997}\textcolor{lightgray}{± \numprint{.0012557866060760449}} & \textbf{\numprint{.18161999999999998}}*\textcolor{lightgray}{± \numprint{.00427282108214233}} \\
 & NDCG@10 & \numprint{.17229999999999998}\textcolor{lightgray}{± \numprint{.001164044672682272}} & \textbf{\numprint{.20617999999999997}}*\textcolor{lightgray}{± \numprint{.004087419723982351}} \\
\bottomrule
\end{tabular}

%% file: data/data_recbole/pageview-experiments-2.tex
\nprounddigits{3}
\npnoaddmissingzero

\begin{tabular}{llrr}
\toprule
 Models & Metrics & Items-BL & UPID \\
\midrule
\multirow[t]{5}{*}{LightSANs} & HR@1 & \textbf{\numprint{.12702}}\textcolor{lightgray}{± \numprint{.003319939758489603}} & \numprint{.12588000000000002}\textcolor{lightgray}{± \numprint{.0021684095554115185}} \\
 & HR@5 & \numprint{.31108}\textcolor{lightgray}{± \numprint{.0017922053453775883}} & \textbf{\numprint{.32818}}*\textcolor{lightgray}{± \numprint{.0021545301111843434}} \\
 & HR@10 & \numprint{.3861}\textcolor{lightgray}{± \numprint{.0021177818584547405}} & \textbf{\numprint{.40824}}*\textcolor{lightgray}{± \numprint{.001463898903613216}} \\
 & NDCG@5 & \numprint{.22210000000000002}\textcolor{lightgray}{± \numprint{.0010559356040971371}} & \textbf{\numprint{.23003999999999997}}*\textcolor{lightgray}{± \numprint{.0016211107303327605}} \\
 & NDCG@10 & \numprint{.24658000000000002}\textcolor{lightgray}{± \numprint{.0012029131306956382}} & \textbf{\numprint{.25617999999999996}}*\textcolor{lightgray}{± \numprint{.001418449858119773}} \\
\cline{1-4}
\multirow[t]{5}{*}{NARM} & HR@1 & \textbf{\numprint{.0973}}\textcolor{lightgray}{± \numprint{.05430768822183467}} & \numprint{.044219999999999995}\textcolor{lightgray}{± \numprint{.06023173582091089}} \\
 & HR@5 & \textbf{\numprint{.25107999999999997}}\textcolor{lightgray}{± \numprint{.14002832570590854}} & \numprint{.12445999999999999}\textcolor{lightgray}{± \numprint{.16910099940568063}} \\
 & HR@10 & \textbf{\numprint{.31134000000000006}}\textcolor{lightgray}{± \numprint{.17337783306985932}} & \numprint{.15736}\textcolor{lightgray}{± \numprint{.21314704548738178}} \\
 & NDCG@5 & \textbf{\numprint{.17656}}\textcolor{lightgray}{± \numprint{.09847920084972257}} & \numprint{.08529999999999999}\textcolor{lightgray}{± \numprint{.11599778015117358}} \\
 & NDCG@10 & \textbf{\numprint{.1962}}\textcolor{lightgray}{± \numprint{.10934495873153001}} & \numprint{.09603999999999999}\textcolor{lightgray}{± \numprint{.13033766915209125}} \\
\cline{1-4}
\multirow[t]{5}{*}{NextItNet} & HR@1 & \textbf{\numprint{.13757999999999998}}\textcolor{lightgray}{± \numprint{.0017908098726553824}} & \numprint{.11112}\textcolor{lightgray}{± \numprint{.0027887273082895713}} \\
 & HR@5 & \textbf{\numprint{.31672}}\textcolor{lightgray}{± \numprint{.003009485005777552}} & \numprint{.30056}\textcolor{lightgray}{± \numprint{.0024714368290530897}} \\
 & HR@10 & \textbf{\numprint{.39058000000000004}}\textcolor{lightgray}{± \numprint{.002578177650977526}} & \numprint{.3809}\textcolor{lightgray}{± \numprint{.0024072806234421377}} \\
 & NDCG@5 & \textbf{\numprint{.23012000000000002}}\textcolor{lightgray}{± \numprint{.002098094373473226}} & \numprint{.20862000000000003}\textcolor{lightgray}{± \numprint{.002483344518990462}} \\
 & NDCG@10 & \textbf{\numprint{.2542}}\textcolor{lightgray}{± \numprint{.0018828170383762706}} & \numprint{.23481999999999997}\textcolor{lightgray}{± \numprint{.002521309183737684}} \\
\cline{1-4}
\multirow[t]{5}{*}{SASRec} & HR@1 & \textbf{\numprint{.1402}}\textcolor{lightgray}{± \numprint{.005519510847892229}} & \numprint{.12711999999999998}\textcolor{lightgray}{± \numprint{.0021649480363278954}} \\
 & HR@5 & \textbf{\numprint{.3337}}\textcolor{lightgray}{± \numprint{.002118962010041699}} & \numprint{.33282}\textcolor{lightgray}{± \numprint{.0019979989989987418}} \\
 & HR@10 & \numprint{.40998}\textcolor{lightgray}{± \numprint{.0010305338422390738}} & \textbf{\numprint{.41436}}*\textcolor{lightgray}{± \numprint{.0019295077092357003}} \\
 & NDCG@5 & \textbf{\numprint{.24016}}\textcolor{lightgray}{± \numprint{.0035528861507231014}} & \numprint{.23324000000000003}\textcolor{lightgray}{± \numprint{.001352035502492443}} \\
 & NDCG@10 & \textbf{\numprint{.265}}\textcolor{lightgray}{± \numprint{.0029521178838251036}} & \numprint{.25986000000000004}\textcolor{lightgray}{± \numprint{.0013107249902248746}} \\
\bottomrule
\end{tabular}

%% file: sections/53_exp_3_search.tex
\begin{table}[p]
    \centering
    \caption{\revised{Average Hitrate and NDCG with standard deviation over five random seeds for non-item models with Unique Page IDs (UPID) and items-only baseline models (Items-BL) on the \coveosearch dataset. We represent search events by the most frequent categories (FreqC-) or the categories of the first result (FirstC-) as page embeddings (PE) and id embeddings (ID). We also use the id embedding of the first item in the result (1.Item-ID) and the page embedding of the query (Query-PE). Results marked with * show a significantly higher performance compared to the baseline for $p<0.01$ and results marked with $^+$ for $p<0.05$.}}
    \input{data/data_recbole/search-experiments}
    \label{tab:coveo-search-results-rec}
\end{table}

\subsection{RQ4: Can we find good representations for utilizing search requests as non-item interactions?}
\label{sec:rq3}
In this experiment, we investigate content-based page representations. In the \coveosearch dataset, non-item interactions are introduced by search requests. A search request represents a clear expression of the user's intent and should contain valuable information.
We compare three different CPID tokens from the list of retrieved items to represent a search interaction:

(a) the category, (b) the product id of the first item, and (c) the concatenation of the three most frequent categories in the list.
In case of product id, no new ids are added to the vocabulary. 
For both categories, we also test the attribute-based approach with multi-hot embeddings (PE) based on the categories instead of unique ids.
In addition, we also evaluate using the provided query embedding, as it directly represents the search intent and not the feedback from the search engine. 
We use a batch size of $64$, a maximum sequence length of $30$ and train for $100$ epochs.

\Cref{tab:coveo-search-results-rec} shows our results. \revised{Looking at the best-performing setting of each model, we can see that including the search interactions in \bertforrec, \narm, \lightsans and \gruforrec brings significant performance improvements, of at least $1.4$ percentage points and up to $4.2$ percentage points on the HR@10. We also observe small increases in metrics for \sasrec for the embedded query, but they are well within one standard deviation.}

In a similar scope, the baseline performance for \nextitnet is better compared to the query or category id embeddings.
But there are also two models, \caser and \core, where the baseline is clearly outperforming all models containing non-item pages, with the performance dropping visibly for all non-item variants. 

We can also observe common patterns regarding  representations of search interactions. The first item in the results list (FirstItem-ID) hurts the performance of all models and is therefore not a fitting representation for a search request.
Between models using the most frequent or first category we observe similar metrics for the same type of embedding.

But between CPID and PE embeddings there are notable differences: \narm and \bertforrec show slightly higher performance when using CPID, while \lightsans, and \gruforrec seem to work better with page embeddings. 
However, the four models show improvements for all variants of category-based representations. By far the biggest improvement is achieved when using a query embedding to represent the search. \gruforrec has the highest gains in metrics and for \narm, \lightsans and \bertforrec the query embedding is overall the most beneficial way of including the search as well. This can be explained by the fact that the query embedding is a direct representation of the search request. In contrast, other representations are dependent on retrieved items and, therefore, on the quality of the search engine and its interpretation of the user query. Here, the advantage of using more flexible page embeddings becomes evident, as CPID embeddings cannot represent a search request directly.
We can also observe that the three models (\caser, \core and \nextitnet), which do not benefit from the inclusion of the search non-item pages, are all showing lower improvements on highly noisy data in \Cref{sec:exp1:noise}, and the signal from the search could be too weak for them.

\subsubsection{Visualization and Item Embedding Similarity}
\begin{figure}
 \begin{subfigure}[b]{0.40\textwidth}
         \centering
         \includegraphics[width=\textwidth]{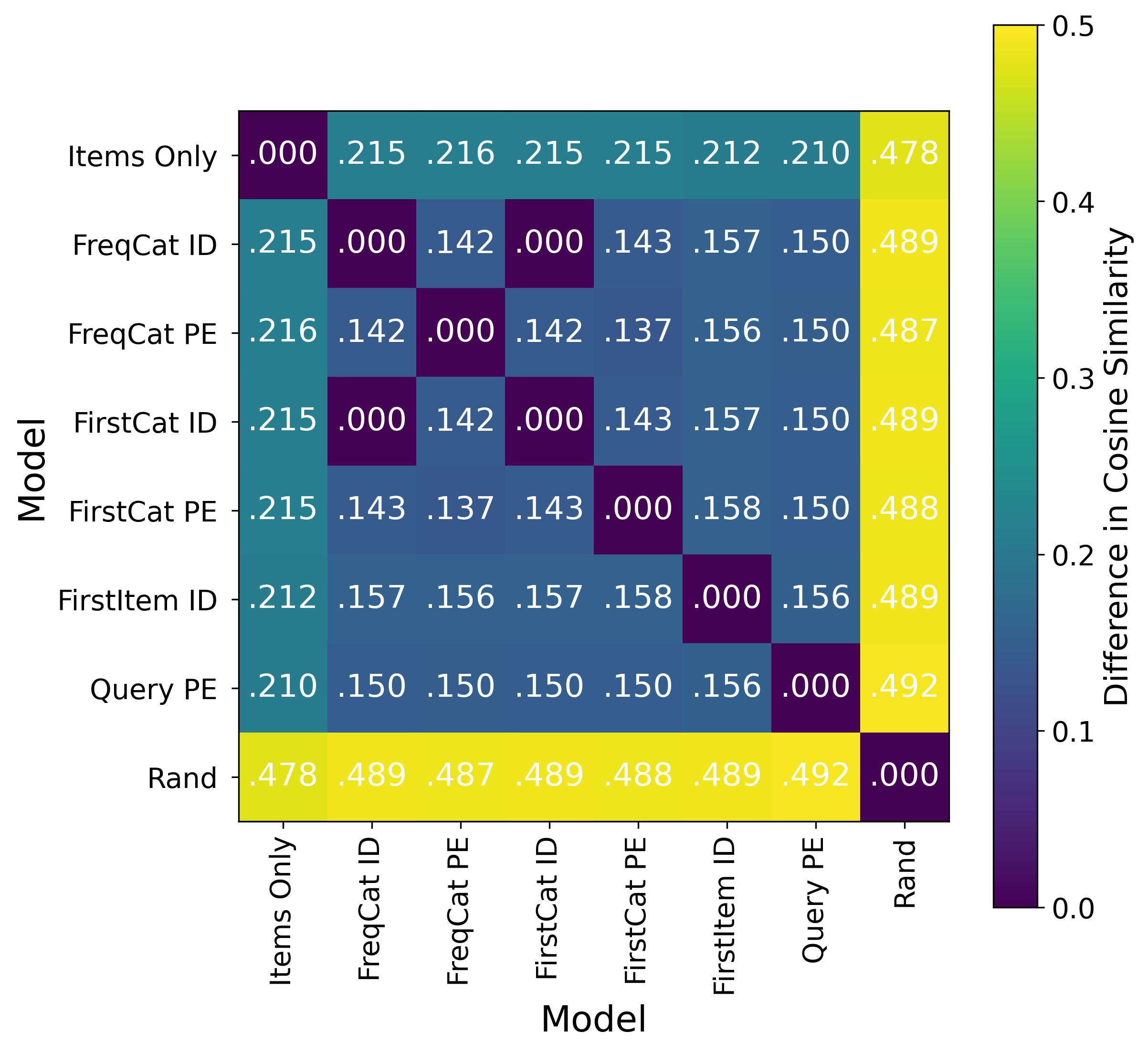}
         \caption{Difference in cosine similarity between embedding variants for SasRec.}
        \label{fig:exp3_sem_sim_sas}
     \end{subfigure}
     \hfill
  \begin{subfigure}[b]{0.40\textwidth}
     \centering
     \includegraphics[width=\textwidth]{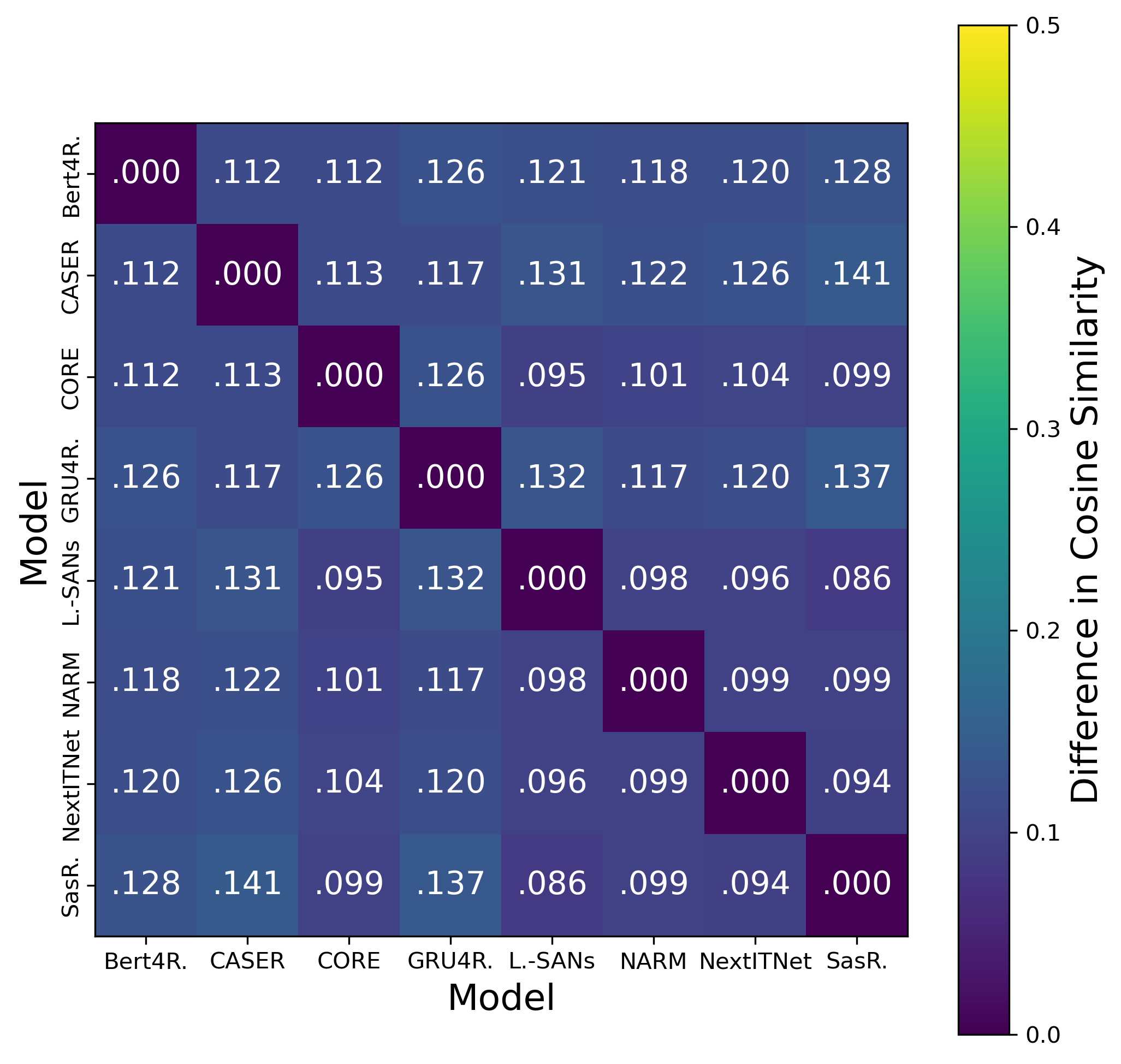}
     \caption{Difference in cosine similarity between different models with the search query embedding.}
     \label{fig:exp3_sem_sim}
 \end{subfigure}
\centering
\caption{Difference in Item-to-Item cosine similarity between different variants on \coveosearch. For each model the cosine similarity between all movies is computed and normalized. We report the average of the absolute difference in similarity between two models. A lower score means item-item relations are similar in both models. A randomly generated embedding is included as a baseline.}
\Description[Difference in Item-to-Item cosine similarity between different variants on \coveosearch. ]{Difference in Item-to-Item cosine similarity between different variants on \coveosearch. }
\label{fig:exp3_sem_emb}
\end{figure}

The item embedding spaces visualized with t-SNE reveal patterns similar to those of the models in the \coveopage dataset, as they are both extracted from the Coveo challenge data (see \Cref{fig:exp3-tsne} and more in our repository). Overall, item embeddings show similar patterns for all models, which we verify by plotting the difference in cosine similarity. \Cref{fig:exp3_sem_sim_sas} shows the difference in cosine similarity for \sasrec. We can observe that the Items-Only model differs the most from all other models. Although not all embedding variants improve performance, each contributes to a distinct change in the item embedding space. We observe this pattern with all model architectures. Meanwhile, the differences between models with any kind of search representation are less pronounced, with the most relevant difference in the item embedding space occurring between including non-items or not including them. 

We also confirm the same pattern as observed earlier: Adding non-items has a greater impact on the item embedding space than changing the model architecture, as shown in \Cref{fig:exp3_sem_sim_sas} for query embeddings. 

%% file: data/data_recbole/search-experiments.tex
\nprounddigits{3}
\npnoaddmissingzero
\addtolength{\tabcolsep}{-0.1em}
\begin{tabular}{llrrrrrrr}
\toprule
Model & Metrics &  Items-BL & FreqC-ID & FreqC-PE & FirstC-ID & FirstC-PE & 1.Item-ID & Query-PE \\
\midrule
\multirow[t]{5}{*}{BERT4R.} 
& HR@1 & \numprint{.07067999999999999}\textcolor{lightgray}{± \numprint{.001167047556871613}} & \textbf{\numprint{.07276}}$^+$\textcolor{lightgray}{± \numprint{.0015175638372075168}} & \numprint{.06808}\textcolor{lightgray}{± \numprint{.0027797481900344893}} & \numprint{.07276}$^+$\textcolor{lightgray}{± \numprint{.0014536161804272806}} & \numprint{.07182}\textcolor{lightgray}{± \numprint{.003417162565638338}} & \numprint{.07006000000000001}\textcolor{lightgray}{± \numprint{.001908664454533587}} & \numprint{.07266}$^+$\textcolor{lightgray}{± \numprint{.0017980545041794476}} \\

 & HR@5 & \numprint{.24634}\textcolor{lightgray}{± \numprint{.0022142718893577644}} & \numprint{.26067999999999997}*\textcolor{lightgray}{± \numprint{.0019279522815671647}} & \numprint{.25228}*\textcolor{lightgray}{± \numprint{.004570776739242462}} & \numprint{.26076}*\textcolor{lightgray}{± \numprint{.001980656456834456}} & \numprint{.2584000000000001}*\textcolor{lightgray}{± \numprint{.0023054283766796965}} & \numprint{.22494}\textcolor{lightgray}{± \numprint{.002876282322721469}} & \textbf{\numprint{.26305999999999996}}*\textcolor{lightgray}{± \numprint{.003973411632338136}} \\
 
 & HR@10 & \numprint{.31832}\textcolor{lightgray}{± \numprint{.003310135948869779}} & \textbf{\numprint{.33368}}*\textcolor{lightgray}{± \numprint{.004082523729263553}} & \numprint{.32324}*\textcolor{lightgray}{± \numprint{.005127670036186018}} & \numprint{.3336}*\textcolor{lightgray}{± \numprint{.003970516339218366}} & \numprint{.32926}*\textcolor{lightgray}{± \numprint{.0037380476187443136}} & \numprint{.28852}\textcolor{lightgray}{± \numprint{.002540078738937045}} & \numprint{.33242000000000005}*\textcolor{lightgray}{± \numprint{.005135854359305774}} \\
 
 & NDCG@5 & \numprint{.16068000000000002}\textcolor{lightgray}{± \numprint{.0009833615815151634}} & \numprint{.1689}*\textcolor{lightgray}{± \numprint{.002216979927739537}} & \numprint{.16232}$^+$\textcolor{lightgray}{± \numprint{.003480229877464986}} & \numprint{.1689}*\textcolor{lightgray}{± \numprint{.0021517434791350092}} & \numprint{.16758}*\textcolor{lightgray}{± \numprint{.002560663976393621}} & \numprint{.14984}\textcolor{lightgray}{± \numprint{.001359411637437312}} & \textbf{\numprint{.17046}}*\textcolor{lightgray}{± \numprint{.002514557615168124}} \\
 
 & NDCG@10 & \numprint{.18414000000000003}\textcolor{lightgray}{± \numprint{.0013520355024924425}} & \numprint{.19274}*\textcolor{lightgray}{± \numprint{.0027627884464793925}} & \numprint{.18548}$^+$\textcolor{lightgray}{± \numprint{.003643075623700387}} & \numprint{.1927}*\textcolor{lightgray}{± \numprint{.002663644120373437}} & \numprint{.19069999999999998}*\textcolor{lightgray}{± \numprint{.0028222331583340164}} & \numprint{.17049999999999998}\textcolor{lightgray}{± \numprint{.0010295630140986986}} & \textbf{\numprint{.19316}}*\textcolor{lightgray}{± \numprint{.0028156704352604897}} \\
\cline{1-9}

\multirow[t]{5}{*}{Caser} & HR@1 & \textbf{\numprint{.07804}}\textcolor{lightgray}{± \numprint{.0022389729788454345}} & \numprint{.07142000000000001}\textcolor{lightgray}{± \numprint{.0022049943310584724}} & \numprint{.07062}\textcolor{lightgray}{± \numprint{.003069527650958698}} & \numprint{.07164000000000001}\textcolor{lightgray}{± \numprint{.0031069277429641}} & \numprint{.07032000000000001}\textcolor{lightgray}{± \numprint{.0006760177512462206}} & \numprint{.07065999999999999}\textcolor{lightgray}{± \numprint{.0016697305171793427}} & \numprint{.07094}\textcolor{lightgray}{± \numprint{.003210607419165417}} \\
 & HR@5 & \textbf{\numprint{.23820000000000002}}\textcolor{lightgray}{± \numprint{.0023398717913595156}} & \numprint{.22444000000000003}\textcolor{lightgray}{± \numprint{.003693643187964961}} & \numprint{.22098}\textcolor{lightgray}{± \numprint{.0011432410069622243}} & \numprint{.22634000000000004}\textcolor{lightgray}{± \numprint{.003066431150376607}} & \numprint{.22206}\textcolor{lightgray}{± \numprint{.0025764316408552327}} & \numprint{.20596}\textcolor{lightgray}{± \numprint{.0018077610461562652}} & \numprint{.22368000000000002}\textcolor{lightgray}{± \numprint{.0024304320603547027}} \\
 & HR@10 & \textbf{\numprint{.30168}}\textcolor{lightgray}{± \numprint{.003084963533009754}} & \numprint{.29048}\textcolor{lightgray}{± \numprint{.0027462701979230064}} & \numprint{.28559999999999997}\textcolor{lightgray}{± \numprint{.0023377339455121905}} & \numprint{.29052}\textcolor{lightgray}{± \numprint{.0019136352839556538}} & \numprint{.2877}\textcolor{lightgray}{± \numprint{.0019170289512680679}} & \numprint{.25993999999999995}\textcolor{lightgray}{± \numprint{.001996997746618652}} & \numprint{.29042}\textcolor{lightgray}{± \numprint{.0034795114599610083}} \\
 & NDCG@5 & \textbf{\numprint{.16146}}\textcolor{lightgray}{± \numprint{.0022853883696212322}} & \numprint{.15052000000000001}\textcolor{lightgray}{± \numprint{.002847279403219852}} & \numprint{.14818}\textcolor{lightgray}{± \numprint{.0013255187663703551}} & \numprint{.15176}\textcolor{lightgray}{± \numprint{.0025559733957926916}} & \numprint{.14878}\textcolor{lightgray}{± \numprint{.001458423806717377}} & \numprint{.14072}\textcolor{lightgray}{± \numprint{.0010802777420645148}} & \numprint{.15008000000000002}\textcolor{lightgray}{± \numprint{.0030954805765825827}} \\
 & NDCG@10 & \textbf{\numprint{.18216000000000002}}\textcolor{lightgray}{± \numprint{.0024541801074900696}} & \numprint{.172}\textcolor{lightgray}{± \numprint{.002487971060924941}} & \numprint{.16920000000000002}\textcolor{lightgray}{± \numprint{.0016140012391569032}} & \numprint{.17265999999999998}\textcolor{lightgray}{± \numprint{.0023394443784796434}} & \numprint{.17018}\textcolor{lightgray}{± \numprint{.0012735776379946353}} & \numprint{.1583}\textcolor{lightgray}{± \numprint{.0013711309200802077}} & \numprint{.1718}\textcolor{lightgray}{± \numprint{.0034561539317570925}} \\
\cline{1-9}

\multirow[t]{5}{*}{Core} & HR@1 & \textbf{\numprint{.11168}}\textcolor{lightgray}{± \numprint{.004202618231531389}} & \numprint{.09283999999999999}\textcolor{lightgray}{± \numprint{.003592770518694451}} & \numprint{.10025999999999999}\textcolor{lightgray}{± \numprint{.001517563837207517}} & \numprint{.09287999999999999}\textcolor{lightgray}{± \numprint{.0018606450494384988}} & \numprint{.10125999999999999}\textcolor{lightgray}{± \numprint{.0019359752064528144}} & \numprint{.09226}\textcolor{lightgray}{± \numprint{.0028156704352604884}} & \numprint{.10266}\textcolor{lightgray}{± \numprint{.003648698398059233}} \\
 & HR@5 & \textbf{\numprint{.35334}}\textcolor{lightgray}{± \numprint{.0014673104647619744}} & \numprint{.34314}\textcolor{lightgray}{± \numprint{.0014223220451079293}} & \numprint{.34102}\textcolor{lightgray}{± \numprint{.0020302709178826386}} & \numprint{.34344}\textcolor{lightgray}{± \numprint{.0009659192512834444}} & \numprint{.3428}\textcolor{lightgray}{± \numprint{.0016232683080747842}} & \numprint{.30918000000000007}\textcolor{lightgray}{± \numprint{.002060825077487174}} & \numprint{.34538}\textcolor{lightgray}{± \numprint{.0031419739018648785}} \\
 & HR@10 & \numprint{.42055999999999993}\textcolor{lightgray}{± \numprint{.0008080841540334779}} & \numprint{.41378000000000004}\textcolor{lightgray}{± \numprint{.0014532721699667993}} & \numprint{.41088}\textcolor{lightgray}{± \numprint{.0024024986992712423}} & \numprint{.41487999999999997}\textcolor{lightgray}{± \numprint{.001512283042290682}} & \numprint{.41356000000000004}\textcolor{lightgray}{± \numprint{.0014170391667134704}} & \numprint{.36974}\textcolor{lightgray}{± \numprint{.000763544366752841}} & \textbf{\numprint{.42119999999999996}}\textcolor{lightgray}{± \numprint{.0017972200755611426}} \\
 & NDCG@5 & \textbf{\numprint{.2404}}\textcolor{lightgray}{± \numprint{.0015842979517754884}} & \numprint{.22670000000000004}\textcolor{lightgray}{± \numprint{.0018641351882307312}} & \numprint{.22846000000000002}\textcolor{lightgray}{± \numprint{.0006580273550544782}} & \numprint{.22690000000000002}\textcolor{lightgray}{± \numprint{.0008124038404636057}} & \numprint{.22972}\textcolor{lightgray}{± \numprint{.001602186006679628}} & \numprint{.20786}\textcolor{lightgray}{± \numprint{.0013575713609236138}} & \numprint{.23175999999999997}\textcolor{lightgray}{± \numprint{.0020452383724153113}} \\
 & NDCG@10 & \textbf{\numprint{.2623}}\textcolor{lightgray}{± \numprint{.0017649362594722828}} & \numprint{.24982}\textcolor{lightgray}{± \numprint{.0018074844397670503}} & \numprint{.25128}\textcolor{lightgray}{± \numprint{.000694262198308378}} & \numprint{.25022}\textcolor{lightgray}{± \numprint{.0008700574693662388}} & \numprint{.25282}\textcolor{lightgray}{± \numprint{.0011966620241321397}} & \numprint{.22762000000000002}\textcolor{lightgray}{± \numprint{.0010568822072492237}} & \numprint{.2565}\textcolor{lightgray}{± \numprint{.0018207141456033117}} \\
\cline{1-9}

\multirow[t]{5}{*}{GRU4R.} & HR@1 & \numprint{.051960000000000006}\textcolor{lightgray}{± \numprint{.0010163660757817508}} & \numprint{.05670000000000001}*\textcolor{lightgray}{± \numprint{.003762977544445356}} & \numprint{.0741}*\textcolor{lightgray}{± \numprint{.003181980515339465}} & \numprint{.05638}*\textcolor{lightgray}{± \numprint{.003944236301237541}} & \textbf{\numprint{.07882}}*\textcolor{lightgray}{± \numprint{.003097095413447897}} & \numprint{.05620000000000001}*\textcolor{lightgray}{± \numprint{.004902550356702111}} & \numprint{.07486000000000001}*\textcolor{lightgray}{± \numprint{.00609696645882196}} \\
 & HR@5 & \numprint{.18480000000000002}\textcolor{lightgray}{± \numprint{.007318811378905732}} & \numprint{.19256}*\textcolor{lightgray}{± \numprint{.006529395071520791}} & \numprint{.21153999999999998}*\textcolor{lightgray}{± \numprint{.008914482598558373}} & \numprint{.19256}*\textcolor{lightgray}{± \numprint{.006221173522736688}} & \numprint{.21888000000000002}*\textcolor{lightgray}{± \numprint{.006557209772456577}} & \numprint{.17786000000000002}\textcolor{lightgray}{± \numprint{.005002799216438734}} & \textbf{\numprint{.22268}}*\textcolor{lightgray}{± \numprint{.005386742986258022}} \\
 & HR@10 & \numprint{.24944000000000002}\textcolor{lightgray}{± \numprint{.008801022667849464}} & \numprint{.25806}*\textcolor{lightgray}{± \numprint{.006019800661151512}} & \numprint{.27606}*\textcolor{lightgray}{± \numprint{.007612358373066778}} & \numprint{.25946}*\textcolor{lightgray}{± \numprint{.005959278479816168}} & \numprint{.28558}*\textcolor{lightgray}{± \numprint{.005047969888975171}} & \numprint{.2366}\textcolor{lightgray}{± \numprint{.005320244355290453}} & \textbf{\numprint{.2913}}*\textcolor{lightgray}{± \numprint{.005117128100800287}} \\
 & NDCG@5 & \numprint{.12002000000000002}\textcolor{lightgray}{± \numprint{.003682662080615055}} & \numprint{.1263}*\textcolor{lightgray}{± \numprint{.004911720676097126}} & \numprint{.14436}*\textcolor{lightgray}{± \numprint{.0059722692504608375}} & \numprint{.12598}*\textcolor{lightgray}{± \numprint{.004979658622837511}} & \numprint{.15036}*\textcolor{lightgray}{± \numprint{.004412822226194933}} & \numprint{.11846000000000001}\textcolor{lightgray}{± \numprint{.0047851854718495524}} & \textbf{\numprint{.15042}}*\textcolor{lightgray}{± \numprint{.005334510286802333}} \\
 & NDCG@10 & \numprint{.14106000000000002}\textcolor{lightgray}{± \numprint{.004105240553244109}} & \numprint{.1476}*\textcolor{lightgray}{± \numprint{.004523825814507007}} & \numprint{.16538}*\textcolor{lightgray}{± \numprint{.005582741262139952}} & \numprint{.14772}*\textcolor{lightgray}{± \numprint{.004822551192055928}} & \numprint{.1721}*\textcolor{lightgray}{± \numprint{.0040718546143004635}} & \numprint{.13752}\textcolor{lightgray}{± \numprint{.004952474129160092}} & \textbf{\numprint{.17276}}*\textcolor{lightgray}{± \numprint{.005269060637343238}} \\
\cline{1-9}

\multirow[t]{5}{*}{L.-SANs} & HR@1 & \numprint{.10462}\textcolor{lightgray}{± \numprint{.010688404932448993}} & \numprint{.09730000000000001}\textcolor{lightgray}{± \numprint{.00410365690573664}} & \numprint{.1056}\textcolor{lightgray}{± \numprint{.005863019699779286}} & \numprint{.09986}\textcolor{lightgray}{± \numprint{.002900517195260186}} & \numprint{.09838}\textcolor{lightgray}{± \numprint{.01038710739330253}} & \numprint{.09206}\textcolor{lightgray}{± \numprint{.005516611278674616}} & \textbf{\numprint{.10588}}\textcolor{lightgray}{± \numprint{.0033003030163910687}} \\
 & HR@5 & \numprint{.27791999999999994}\textcolor{lightgray}{± \numprint{.007668572226953337}} & \numprint{.28195999999999993}*\textcolor{lightgray}{± \numprint{.002909123579361999}} & \numprint{.28624}*\textcolor{lightgray}{± \numprint{.004165093036175783}} & \numprint{.28198}*\textcolor{lightgray}{± \numprint{.001917550520846856}} & \numprint{.28778}*\textcolor{lightgray}{± \numprint{.0037725323060246846}} & \numprint{.25438}\textcolor{lightgray}{± \numprint{.0037144313158274903}} & \textbf{\numprint{.29508}}*\textcolor{lightgray}{± \numprint{.0008757853618324474}} \\
 & HR@10 & \numprint{.34631999999999996}\textcolor{lightgray}{± \numprint{.007019045519157136}} & \numprint{.35188}*\textcolor{lightgray}{± \numprint{.0022851695779525804}} & \numprint{.3577}*\textcolor{lightgray}{± \numprint{.0022416511771459827}} & \numprint{.35253999999999996}*\textcolor{lightgray}{± \numprint{.0024825390228554256}} & \numprint{.35786}*\textcolor{lightgray}{± \numprint{.0024037470748812075}} & \numprint{.31482}\textcolor{lightgray}{± \numprint{.004176362053270767}} & \textbf{\numprint{.3699}}*\textcolor{lightgray}{± \numprint{.0017944358444926271}} \\
 & NDCG@5 & \numprint{.19405999999999998}\textcolor{lightgray}{± \numprint{.009395903362636291}} & \numprint{.19228}\textcolor{lightgray}{± \numprint{.0021347130954767644}} & \numprint{.19848}*\textcolor{lightgray}{± \numprint{.004708715323737469}} & \numprint{.19354}\textcolor{lightgray}{± \numprint{.002159398064276247}} & \numprint{.19622}*\textcolor{lightgray}{± \numprint{.006764022471872791}} & \numprint{.17624}\textcolor{lightgray}{± \numprint{.003280701144572607}} & \textbf{\numprint{.20352}}*\textcolor{lightgray}{± \numprint{.001920156243642688}} \\
 & NDCG@10 & \numprint{.21631999999999998}\textcolor{lightgray}{± \numprint{.009138216456180057}} & \numprint{.2152}\textcolor{lightgray}{± \numprint{.0019313207915827915}} & \numprint{.22182}*\textcolor{lightgray}{± \numprint{.0041020726468457445}} & \numprint{.21661999999999998}\textcolor{lightgray}{± \numprint{.0018619881847100974}} & \numprint{.2191}*\textcolor{lightgray}{± \numprint{.006222941426688834}} & \numprint{.19594}\textcolor{lightgray}{± \numprint{.00293138192666871}} & \textbf{\numprint{.22796}}*\textcolor{lightgray}{± \numprint{.0019424211695716286}} \\
\cline{1-9}

\multirow[t]{5}{*}{NARM} & HR@1 & \numprint{.07202}\textcolor{lightgray}{± \numprint{.003307113545072198}} & \numprint{.07724}*\textcolor{lightgray}{± \numprint{.0016349311912126417}} & \numprint{.08108}*\textcolor{lightgray}{± \numprint{.0030094850057775656}} & \numprint{.07682000000000001}*\textcolor{lightgray}{± \numprint{.001490637447537125}} & \numprint{.07944}*\textcolor{lightgray}{± \numprint{.0028544701785094932}} & \numprint{.0627}\textcolor{lightgray}{± \numprint{.03515835604802932}} & \textbf{\numprint{.08516}}*\textcolor{lightgray}{± \numprint{.0026576305236055676}} \\
 & HR@5 & \numprint{.2238}\textcolor{lightgray}{± \numprint{.0016985287751463038}} & \numprint{.23898000000000003}*\textcolor{lightgray}{± \numprint{.002864786204937465}} & \numprint{.23582}*\textcolor{lightgray}{± \numprint{.00221178660815188}} & \numprint{.23857999999999996}*\textcolor{lightgray}{± \numprint{.002752635101134916}} & \numprint{.23642000000000002}*\textcolor{lightgray}{± \numprint{.002468197723035984}} & \numprint{.17482}\textcolor{lightgray}{± \numprint{.0975658341838986}} & \textbf{\numprint{.24372000000000002}}*\textcolor{lightgray}{± \numprint{.004103900583591172}} \\
 & HR@10 & \numprint{.29222000000000004}\textcolor{lightgray}{± \numprint{.002058397434899306}} & \numprint{.30763999999999997}*\textcolor{lightgray}{± \numprint{.0026063384277564563}} & \numprint{.30579999999999996}*\textcolor{lightgray}{± \numprint{.003012474066278413}} & \numprint{.3074}*\textcolor{lightgray}{± \numprint{.0027322152184628564}} & \numprint{.30402}*\textcolor{lightgray}{± \numprint{.0020191582404556624}} & \numprint{.22361999999999999}\textcolor{lightgray}{± \numprint{.12447249495370453}} & \textbf{\numprint{.31428}}*\textcolor{lightgray}{± \numprint{.0018019433953373823}} \\
 & NDCG@5 & \numprint{.15012}\textcolor{lightgray}{± \numprint{.001853914776897792}} & \numprint{.16046}*\textcolor{lightgray}{± \numprint{.0024429490375363935}} & \numprint{.16038}*\textcolor{lightgray}{± \numprint{.002485357117196642}} & \numprint{.16014}*\textcolor{lightgray}{± \numprint{.0023860008382228215}} & \numprint{.15982}*\textcolor{lightgray}{± \numprint{.0019778776504121706}} & \numprint{.12048}\textcolor{lightgray}{± \numprint{.06729908617507373}} & \textbf{\numprint{.16684}}*\textcolor{lightgray}{± \numprint{.00249559612117025}} \\
 & NDCG@10 & \numprint{.1724}\textcolor{lightgray}{± \numprint{.002115419580130613}} & \numprint{.1828}*\textcolor{lightgray}{± \numprint{.0022293496809607984}} & \numprint{.18317999999999998}*\textcolor{lightgray}{± \numprint{.002597498796919839}} & \numprint{.18253999999999998}*\textcolor{lightgray}{± \numprint{.0022853883696212335}} & \numprint{.18186}*\textcolor{lightgray}{± \numprint{.0019308029417835454}} & \numprint{.13632000000000002}\textcolor{lightgray}{± \numprint{.0760205038131161}} & \textbf{\numprint{.1898}}*\textcolor{lightgray}{± \numprint{.0018841443681416712}} \\
 
\cline{1-9}
\multirow[t]{5}{*}{NexItNet} & HR@1 & \textbf{\numprint{.07136}}\textcolor{lightgray}{± \numprint{.0025967287112827124}} & \numprint{.06888}\textcolor{lightgray}{± \numprint{.0019867058161690677}} & \numprint{.05466}\textcolor{lightgray}{± \numprint{.005232399067349506}} & \numprint{.06856}\textcolor{lightgray}{± \numprint{.0012660963628413114}} & \numprint{.041620000000000004}\textcolor{lightgray}{± \numprint{.030608609899830472}} & \numprint{.06717999999999999}\textcolor{lightgray}{± \numprint{.0019279522815671515}} & \numprint{.06666000000000001}\textcolor{lightgray}{± \numprint{.0022941229260874385}} \\
 & HR@5 & \numprint{.20878000000000002}\textcolor{lightgray}{± \numprint{.0025469589710083677}} & \textbf{\numprint{.20996}}\textcolor{lightgray}{± \numprint{.0034703025804675835}} & \numprint{.17236}\textcolor{lightgray}{± \numprint{.012553206761620719}} & \numprint{.20992000000000002}\textcolor{lightgray}{± \numprint{.0038590154184713985}} & \numprint{.12796000000000002}\textcolor{lightgray}{± \numprint{.08865826526613298}} & \numprint{.19404}\textcolor{lightgray}{± \numprint{.004109501186275532}} & \numprint{.20674}\textcolor{lightgray}{± \numprint{.002127909772523266}} \\
 & HR@10 & \numprint{.27271999999999996}\textcolor{lightgray}{± \numprint{.002185634919193959}} & \numprint{.27408}\textcolor{lightgray}{± \numprint{.004375728510773949}} & \numprint{.23331999999999997}\textcolor{lightgray}{± \numprint{.01325394280959444}} & \textbf{\numprint{.2742}}\textcolor{lightgray}{± \numprint{.0036006943774777667}} & \numprint{.172}\textcolor{lightgray}{± \numprint{.11112787679065951}} & \numprint{.24948}\textcolor{lightgray}{± \numprint{.005071193153489621}} & \numprint{.27278}\textcolor{lightgray}{± \numprint{.0029312113536897927}} \\
 & NDCG@5 & \textbf{\numprint{.14193999999999998}}\textcolor{lightgray}{± \numprint{.0016501515081955322}} & \numprint{.14112}\textcolor{lightgray}{± \numprint{.0020969024774652664}} & \numprint{.11494}\textcolor{lightgray}{± \numprint{.00899377562539782}} & \numprint{.141}\textcolor{lightgray}{± \numprint{.002178302091079196}} & \numprint{.08566}\textcolor{lightgray}{± \numprint{.0604649733316736}} & \numprint{.13224}\textcolor{lightgray}{± \numprint{.003080259729308559}} & \numprint{.13856000000000002}\textcolor{lightgray}{± \numprint{.001993238570768685}} \\
 & NDCG@10 & \textbf{\numprint{.16272000000000003}}\textcolor{lightgray}{± \numprint{.0015530614926653732}} & \numprint{.162}\textcolor{lightgray}{± \numprint{.0025446021299998884}} & \numprint{.13468}\textcolor{lightgray}{± \numprint{.009143959754942055}} & \numprint{.16194}\textcolor{lightgray}{± \numprint{.0020525593779474453}} & \numprint{.09994}\textcolor{lightgray}{± \numprint{.06771826193871192}} & \numprint{.15019999999999997}\textcolor{lightgray}{± \numprint{.0034197953155123213}} & \numprint{.15997999999999998}\textcolor{lightgray}{± \numprint{.0021649480363278954}} \\
\cline{1-9}

\multirow[t]{5}{*}{SASRec} & HR@1 & \numprint{.09716}\textcolor{lightgray}{± \numprint{.0071177946022627}} & \numprint{.09678}\textcolor{lightgray}{± \numprint{.004813210986441379}} & \numprint{.10242}*\textcolor{lightgray}{± \numprint{.004678888756959284}} & \numprint{.09678}\textcolor{lightgray}{± \numprint{.004813210986441379}} & \numprint{.10064}*\textcolor{lightgray}{± \numprint{.008675136886528073}} & \numprint{.09046000000000001}\textcolor{lightgray}{± \numprint{.004063619076635996}} & \textbf{\numprint{.10329999999999999}}*\textcolor{lightgray}{± \numprint{.012092973166264778}} \\

 & HR@5 & \textbf{\numprint{.29308}}\textcolor{lightgray}{± \numprint{.0018019433953373701}} & \numprint{.28447999999999996}\textcolor{lightgray}{± \numprint{.0016037456157383609}} & \numprint{.28966000000000003}\textcolor{lightgray}{± \numprint{.0021847196616499747}} & \numprint{.2845}\textcolor{lightgray}{± \numprint{.001581138830084184}} & \numprint{.2877}\textcolor{lightgray}{± \numprint{.0029664793948382577}} & \numprint{.26064}\textcolor{lightgray}{± \numprint{.0011349008767288994}} & \numprint{.29298}\textcolor{lightgray}{± \numprint{.00347159905519055}} \\
 
 & HR@10 & \numprint{.36426000000000003}\textcolor{lightgray}{± \numprint{.001570986950932433}} & \numprint{.3569}\textcolor{lightgray}{± \numprint{.0017944358444926271}} & \numprint{.36032000000000003}\textcolor{lightgray}{± \numprint{.0016709278859364392}} & \numprint{.3569}\textcolor{lightgray}{± \numprint{.0017944358444926271}} & \numprint{.35900000000000004}\textcolor{lightgray}{± \numprint{.0026466960535732054}} & \numprint{.3214}\textcolor{lightgray}{± \numprint{.0021083168642307966}} & \textbf{\numprint{.36628}}$^+$\textcolor{lightgray}{± \numprint{.00288998269891014}} \\
 
 & NDCG@5 & \numprint{.19856000000000001}\textcolor{lightgray}{± \numprint{.0035648281866031022}} & \numprint{.19342}\textcolor{lightgray}{± \numprint{.002686447468312012}} & \numprint{.19894}\textcolor{lightgray}{± \numprint{.0024673872821265866}} & \numprint{.19342}\textcolor{lightgray}{± \numprint{.002686447468312012}} & \numprint{.19707999999999998}\textcolor{lightgray}{± \numprint{.003965728180296779}} & \numprint{.1787}\textcolor{lightgray}{± \numprint{.0021424285285628563}} & \textbf{\numprint{.20135999999999998}}*\textcolor{lightgray}{± \numprint{.005969757113987141}} \\
 
 & NDCG@10 & \numprint{.22183999999999998}\textcolor{lightgray}{± \numprint{.0031461087075941923}} & \numprint{.21703999999999998}\textcolor{lightgray}{± \numprint{.0027309339061940005}} & \numprint{.22202000000000002}\textcolor{lightgray}{± \numprint{.002144061566280217}} & \numprint{.21703999999999998}\textcolor{lightgray}{± \numprint{.0027309339061940005}} & \numprint{.22033999999999998}\textcolor{lightgray}{± \numprint{.003327611756199936}} & \numprint{.19846000000000003}\textcolor{lightgray}{± \numprint{.0022063544592834643}} & \textbf{\numprint{.22528}}*\textcolor{lightgray}{± \numprint{.006077993089828246}} \\
\bottomrule

\end{tabular}
\addtolength{\tabcolsep}{+0.1em}

%% file: sections/54_exp_4_fashion.tex
\subsection{RQ5: Can we find good representations to integrate list pages as non-item pages?}

\begin{table*}[p]
    \centering
    \caption{\revised{Hitrate and NDCG for non-item models and items-only baseline models (Items-BL) on the \onlineshop dataset for seed 212. We represent list pages based on the categories (Cat-) or the filters (Filter-) as page embeddings (PE) and content-based page id embeddings (CPID). We also use the id embedding of the first item on a list page directly (First-ID). Results marked with * show a significantly higher performance compared to the baseline for $p<0.01$ and results marked with $^+$ for $p<0.05$.}}
    \input{data/data_recbole/fashion-exp}
    \label{tab:fashion-res-rec}
\end{table*}

We further explore the potential benefits of integrating non-item pages with content-based page ids and page embeddings for list pages.
In the \onlineshop dataset, categories and filters set by users are available as direct descriptions of a list page's content. We test both CPID and PE-based approaches for categories and filters. The first item on a list page is also used as a representation. 
We use a batch size of $64$, a maximum sequence length of $50$ and train for $10$ epochs.

We present our results in  \Cref{tab:fashion-res-rec}. We see significant improvements for all models and metrics when adding non-item pages. For example the HR@10 for \nextitnet more than doubles and even models that were already performing well show improvements, e.g., \core increasing on HR@10 from $43.8$\% up to $51.3$\%.
Adding list pages achieves higher metrics across all experiments, regardless of the representation, with three notable exceptions: Using PE embeddings for \bertforrec and Cat-CPID for \narm reduces the performance below the respective baseline. As the performance drop is quite high, we attribute this to the models instabilities, which we have already discussed before.
For \bertforrec the highest improvement is achieved by representing the list pages by the first item on each page. For \gruforrec  and \narm the best representation is a page embedding of the categories, for \core the results are inconclusive between category and filters. All other models improve the most when using representations based on the filters of the list page.
Overall, most models show higher improvements with page embeddings instead of CPID, though the difference is marginal. There are also higher gains from using filter-based representations for most models. 
In the case of the \onlineshop dataset, both category and filter-based representations for list pages perform well. Filters contain more detailed information on the selected categories and therefore seem to be slightly more beneficial. Interestingly, the representation as attributes yields better results than the encoding as an id for this dataset. This might be related to the high number of attributes and their combinations compared to the other datasets.
\FloatBarrier
\subsubsection{Item Embedding Similarity}
\begin{figure}
 \begin{subfigure}[b]{0.40\textwidth}
         \centering
         \includegraphics[width=\textwidth]{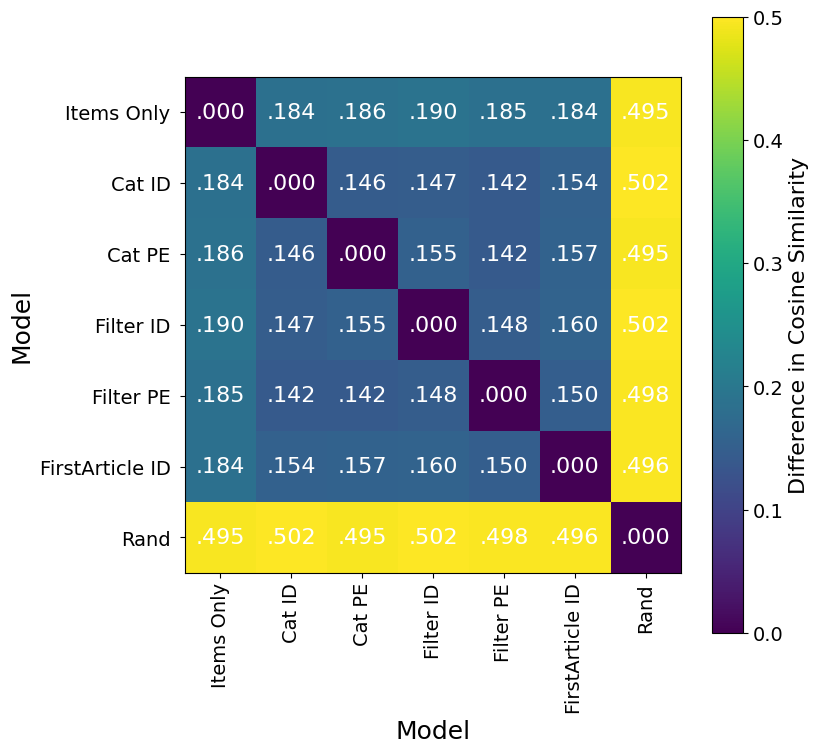}
         \caption{Difference in cosine similarity between embedding variants for \sasrec.}
        \label{fig:exp4_sem_sim_sas}
     \end{subfigure}
     \hfill
  \begin{subfigure}[b]{0.40\textwidth}
     \centering
     \includegraphics[width=\textwidth]{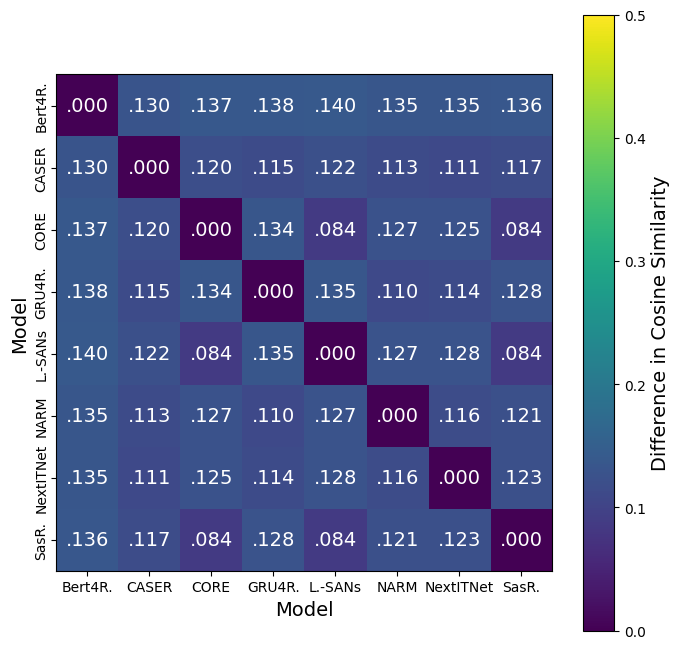}
     \caption{Difference in cosine similarity between different models with page embedding of filters.}
     \label{fig:exp4_sem_sim}
 \end{subfigure}
\centering
\caption{Difference in Item-to-Item cosine similarity between different variants on \onlineshop. For each model the cosine similarity between all items is computed and normalized. We report the average of the absolute difference in similarity between two models. A lower score means item-item relations are similar in both models. A randomly generated embedding is included as a baseline.}
\Description[Difference in Item-to-Item cosine similarity between different variants on \onlineshop.]{Difference in Item-to-Item cosine similarity between different variants on \onlineshop.}
\label{fig:exp4_sem_emb}
\end{figure}

Visualizing the embedding spaces with t-SNE shows similar patterns over all models. We again examine the difference in cosine similarity for the models on the \onlineshop dataset. We observe the same behavior as before: The inclusion of list pages leads to a higher difference in cosine similarity compared to the different embedding variants for nearly all model architectures (\Cref{fig:exp4_sem_sim_sas}). One exception appears for \bertforrec, where most variants show a relatively high distance to each other, presumably due to unstable behavior with page embeddings. 
As noted before, the impact of the model architectures stays relatively low for the different embeddings. We include examples for \sasrec and the page embedding with filters models in \Cref{fig:exp4_sem_sim}.

%% file: data/data_recbole/fashion-exp.tex
\nprounddigits{3}
\npnoaddmissingzero

\begin{tabular}{llllllll}
\toprule
Model & Metrics & Item Only & First-ID & Cat-CPID & Cat-PE & Filter-CPID & Filter-PE \\
\midrule
\multirow[t]{5}{*}{BERT4Rec} & HR@1 & \numprint{.0397} & \textbf{\numprint{.0919}}* & \numprint{.0823}* & \numprint{.0407} & \numprint{.0816}* & \numprint{.0302} \\
 & HR@5 & \numprint{.2786} & \textbf{\numprint{.308}}* & \numprint{.2946}* & \numprint{.195} & \numprint{.3014}* & \numprint{.1698} \\
 & HR@10 & \numprint{.3802} & \textbf{\numprint{.3968}}* & \numprint{.3846}* & \numprint{.2889} & \numprint{.3933}* & \numprint{.2542} \\
 & NDCG@5 & \numprint{.1608} & \textbf{\numprint{.2034}}* & \numprint{.1912}* & \numprint{.1187} & \numprint{.1944}* & \numprint{.1005} \\
 & NDCG@10 & \numprint{.194} & \textbf{\numprint{.2324}}* & \numprint{.2204}* & \numprint{.1492} & \numprint{.2244}* & \numprint{.1279} \\
\cline{1-8}
\multirow[t]{5}{*}{Caser} & HR@1 & \numprint{.02} & \numprint{.0824}* & \numprint{.0762}* & \numprint{.0868}* & \numprint{.0803}* & \textbf{\numprint{.0888}}* \\
 & HR@5 & \numprint{.1279} & \numprint{.26}* & \numprint{.2465}* & \numprint{.274}* & \numprint{.2539}* & \textbf{\numprint{.2816}}* \\
 & HR@10 & \numprint{.1959} & \numprint{.3414}* & \numprint{.3286}* & \numprint{.3584}* & \numprint{.3356}* & \textbf{\numprint{.3678}}* \\
 & NDCG@5 & \numprint{.0746} & \numprint{.1734}* & \numprint{.1637}* & \numprint{.1827}* & \numprint{.1694}* & \textbf{\numprint{.1878}}* \\
 & NDCG@10 & \numprint{.0965} & \numprint{.1999}* & \numprint{.1904}* & \numprint{.2103}* & \numprint{.196}* & \textbf{\numprint{.2158}}* \\
\cline{1-8}
\multirow[t]{5}{*}{Core} & HR@1 & \numprint{.1036}& \numprint{.1014} & \numprint{.1126}* & \textbf{\numprint{.1387}}* & \numprint{.1071}$^+$ & \numprint{.1073}$^+$ \\
 & HR@5 & \numprint{.3521} & \numprint{.4153}* & \numprint{.4173}* & \numprint{.42}* & \numprint{.4228}* & \textbf{\numprint{.427}}* \\
 & HR@10 & \numprint{.4376} & \numprint{.5034}* & \numprint{.5036}* & \numprint{.5065}* & \numprint{.5106}* & \textbf{\numprint{.5132}}* \\
 & NDCG@5 & \numprint{.2346} & \numprint{.2685}* & \numprint{.2751}* & \textbf{\numprint{.2881}}* & \numprint{.2756}* & \numprint{.2778}* \\
 & NDCG@10 & \numprint{.2625} & \numprint{.2973}* & \numprint{.3032}* & \textbf{\numprint{.3163}}* & \numprint{.3042}* & \numprint{.306}* \\
\cline{1-8}
\multirow[t]{5}{*}{GRU4Rec}  & HR@1 & \numprint{.0311} & \numprint{.0846}* & \numprint{.0877}* & \textbf{\numprint{.1022}}* & \numprint{.0916}* & \numprint{.0943}* \\
 & HR@5 & \numprint{.1829} & \numprint{.2703}* & \numprint{.2815}* & \textbf{\numprint{.3055}}* & \numprint{.2905}* & \numprint{.2984}* \\
 & HR@10 & \numprint{.2646} & \numprint{.3543}* & \numprint{.3654}* & \textbf{\numprint{.3881}}* & \numprint{.3745}* & \numprint{.385}* \\
 & NDCG@5 & \numprint{.1077} & \numprint{.1806}* & \numprint{.188}* & \textbf{\numprint{.2071}}* & \numprint{.1945}* & \numprint{.2}* \\
 & NDCG@10 & \numprint{.1341} & \numprint{.2079}* & \numprint{.2152}* & \textbf{\numprint{.234}}* & \numprint{.222}* & \numprint{.2282}* \\
\cline{1-8}
\multirow[t]{5}{*}{LightSANs} & HR@1 & \numprint{.0837} & \numprint{.1525}* & \numprint{.1648}* & \numprint{.1417}* & \textbf{\numprint{.1697}}* & \numprint{.1548}* \\
 & HR@5 & \numprint{.3309} & \numprint{.3953}* & \numprint{.4009}* & \numprint{.3859}* & \textbf{\numprint{.405}}* & \numprint{.399}* \\
 & HR@10 & \numprint{.4298} & \numprint{.4781}* & \numprint{.4833}* & \numprint{.4694}* & \textbf{\numprint{.4901}}* & \numprint{.484}* \\
 & NDCG@5 & \numprint{.2102} & \numprint{.2795}* & \numprint{.2883}* & \numprint{.2692}* & \textbf{\numprint{.2929}}* & \numprint{.2816}* \\
 & NDCG@10 & \numprint{.2424} & \numprint{.3064}* & \numprint{.3151}* & \numprint{.2965}* & \textbf{\numprint{.3206}}* & \numprint{.3094}* \\
\cline{1-8}
\multirow[t]{5}{*}{NARM} & HR@1 & \numprint{.0379} & \numprint{.0844}* & \numprint{.0006} & \textbf{\numprint{.1207}}* & \numprint{.0922}* & \numprint{.0887}* \\
 & HR@5 & \numprint{.1812} & \numprint{.2738}* & \numprint{.0014} & \textbf{\numprint{.3257}}* & \numprint{.289}* & \numprint{.2929}* \\
 & HR@10 & \numprint{.264} & \numprint{.3578}* & \numprint{.0034} & \textbf{\numprint{.4077}}* & \numprint{.3754}* & \numprint{.3777}* \\
 & NDCG@5 & \numprint{.1105} & \numprint{.1815}* & \numprint{.001} & \textbf{\numprint{.2267}}* & \numprint{.1939}* & \numprint{.1942}* \\
 & NDCG@10 & \numprint{.1372} & \numprint{.2089}* & \numprint{.0017} & \textbf{\numprint{.2534}}* & \numprint{.222}* & \numprint{.2218}* \\
\cline{1-8}
\multirow[t]{5}{*}{NextItNet} & HR@1 & \numprint{.0099} & \numprint{.0562}* & \numprint{.0423}* & \numprint{.0521}* & \numprint{.0253}* & \textbf{\numprint{.0693}}* \\
 & HR@5 & \numprint{.0828} & \numprint{.1889}* & \numprint{.1446}* & \numprint{.1737}* & \numprint{.0932}* & \textbf{\numprint{.2208}}* \\
 & HR@10 & \numprint{.1336} & \numprint{.2692}* & \numprint{.2131}* & \numprint{.2511}* & \numprint{.147}* & \textbf{\numprint{.3075}}* \\
 & NDCG@5 & \numprint{.0466} & \numprint{.1237}* & \numprint{.0939}* & \numprint{.114}* & \numprint{.0595}* & \textbf{\numprint{.1464}}* \\
 & NDCG@10 & \numprint{.063} & \numprint{.1498}* & \numprint{.1161}* & \numprint{.1391}* & \numprint{.0768}* & \textbf{\numprint{.1745}}* \\
\cline{1-8}
\multirow[t]{5}{*}{SASRec} & HR@1 & \numprint{.076} & \numprint{.1535}* & \numprint{.1592}* & \numprint{.1228}* & \textbf{\numprint{.1593}}* & \numprint{.1507}* \\
 & HR@5 & \numprint{.3274} & \numprint{.385}* & \numprint{.3939}* & \numprint{.3863}* & \numprint{.3987}* & \textbf{\numprint{.403}}* \\
 & HR@10 & \numprint{.4297} & \numprint{.4724}* & \numprint{.4767}* & \numprint{.4725}* & \numprint{.4835}* & \textbf{\numprint{.4908}}* \\
 & NDCG@5 & \numprint{.2047} & \numprint{.2748}* & \numprint{.2822}* & \numprint{.2603}* & \textbf{\numprint{.2844}}* & \numprint{.2821}* \\
 & NDCG@10 & \numprint{.238} & \numprint{.3032}* & \numprint{.3092}* & \numprint{.2884}* & \textbf{\numprint{.3121}}* & \numprint{.3107}* \\
\bottomrule
\end{tabular}

%% file: sections/60_discussion.tex
\section{Discussion}
\label{sec:discussion}
In this section we consolidate and discuss our overall findings. We discuss performance differences in the models and proposed modeling approaches and compare our results to previous work.

\subsection{Model \revised{D}ifferences}
In the course of our study we find some model architectures able to leverage information from non-item pages better than others. On the \syndata dataset all models increase significantly in their performance, as the signal added through non-item pages is highly indicative for the respective item. 
For \caser and \nextitnet in contrast, we observe a performance drop on both Coveo datasets and for \core there is a similar trend. For \gruforrec, \lightsans and \narm we can report an overall positive outcome, aside from the instabilities of \narm on the \coveopage dataset. 
For both \bertforrec and \sasrec we face partially inconclusive results with a slight positive trend. On the \onlineshop dataset, all architectures can improve their performance significantly, likely due to less abstract non-item interactions. Interestingly, this aligns with our findings in \Cref{sec:exp1:noise}, where lower performance on the randomized data corresponds to lower improvements on real-life data and a higher resistance to noisy data leads to larger performance gains.
\caser and \nextitnet are also the only two CNN-based models in our selection. There might be a common property of CNNs preventing a more effective inclusion of non-item pages. \core explicitly couples the embedding space for encoding and decoding, which might interfere with our additions to the item embedding space, especially with page embeddings.

\revised{\subsection{Sequence Length}
Using non-item pages also means that our sequences will grow longer, as observed in the higher average session length in \Cref{tab:dataset_stats}.
Adding non-item pages to a session can sometimes lead to the session exceeding the maximum sequence length, which means that the oldest interactions are dropped, while the corresponding model without non-item pages would include them.
Adjusting the maximum sequence lengths might lead to higher recommendation performance, as more interactions are considered, but will also increase the computational cost of the model.
For self-attention-based models the influence of the sequence length $n$ on runtime is $O(n^2)$, for recurrent and convolutional architectures $O(n)$~\cite{vaswani_attention_2017}.
For example, by adding a non-item interaction for each item interaction, the cost would increase to $4$ and $2$ times of the original, respectively.
Depending on the proportions of non-item interactions in the dataset, the impact on efficiency can be high and should be considered while optimizing for performance.
In our experimental setup we did not change or optimize the maximum sequence length for our models.
Although this approach results in the loss of some interactions, the advantage of incorporating more recent non-item pages appears to outweigh the impact of missing older interactions.
}

\subsection{Non-Item Representations}
Between the different representations for non-items, we can also observe patterns depending on the models, but also on the datasets. For \core CPIDs work better in all experiments, because of the model's structure. The experiments with randomized data show an overall preference for adding IDs, except for \gruforrec and \narm. On \coveosearch in contrast, we see a clear preference for the query page embedding. Between the category-based embeddings the results are mixed, similar to the results on the \onlineshop dataset.
Overall, the choice between CPIDs and page embeddings seems to depend on how informative and how complex the non-item page representation is. 

\revised{Models utilizing CPIDs work better as long as the number of attributes and therefore added CPIDs stays relatively low, as in \syndata or the category-based representations in \coveosearch.
But once a huge number of Non-Item Pages is added as CPIDs and the vocabulary increases (as in \onlineshop), page embeddings become more efficient.
}
Even if it is more difficult for models to learn from embeddings than explicit IDs, there seems to be a turning-point where adding more IDs becomes a disadvantage.

There is also the factor of the actual informativeness of the content for a representation: The query embedding in \coveosearch contains far more accurate information about users' intents than some average categories and is therefore more helpful. In general, noticeable improvements occur only when the information gain from non-item pages is substantial and clear. For example, the information content of non-item pages in the \coveopage dataset seems to be relatively low and only the models most responsive to non-item pages can leverage it. Overall, selecting useful non-item pages and appropriate means of representation is an important step when including non-item pages. For non-item pages represented by CPIDs, HypTrails~\cite{singer_hyptrails_2015} can be employed in advance to examine the the influence on following items.

\subsection{Comparison with \revised{P}revious \revised{R}esults}
\label{sec:discussion:comparison}
\begin{table}[!ht]
    \centering
    \caption{\revised{Hitrate and NDCG for \bertforrec and \sasrec models on the Prev-\syndata and Group-\syndata datasets for page embeddings (-PE) and id embeddings (-ID), as well as the items-only baseline (Items BL). ``RecB'' marks the results from this paper based on RecBole, averaged over $5$ different seeds, while ``ASME'' shows the respective results reported by \cite{fischer_enhancing_2023} on one seed.}}
    \input{data/implementation_comp_ml}
    \label{tab:impl_comp_ml}
\end{table}

As this paper is a direct extension of the work done in \cite{fischer_enhancing_2023}, we want to consolidate previous and current findings. 
In our previous work, we utilized the recommender framework ASME by Dallmann et al.~\cite{dallmann_case_2021} to conduct experiments on \sasrec and \bertforrec. For the current paper, we used RecBole \cite{zhao_recbole_2021} as basis for our experiments, as more sequential recommender models are readily available and the framework is more commonly used. Our experimental settings are the same as in \cite{fischer_enhancing_2023}. 

We exemplarily compare the results on the \syndata dataset for \sasrec and \bertforrec, the common models in both settings and display the results in detail in \Cref{tab:impl_comp_ml}.

While we can see the benefits of including non-item pages in both previous and current results, the performance of both models differs between the implementations. For \bertforrec we only reach half of the performance for the baseline model. For \sasrec the difference is less prominent but still very noticeable. The difference in models with non-item pages is even higher. One reason for this could be the instability of \bertforrec, especially when comparing models using page embeddings. The fact that our new results are based on multiple seeds can also explain the differences to some extent only, as the standard deviation is not that high (see \Cref{tab:ml_table-rec}).

On the other datasets, the performance for \bertforrec and \sasrec is also lower in RecBole. The overall sentiment, however, is supported by both experiments: On \coveopage, the benefit of non-item pages is ambiguous. However, on \coveosearch and \onlineshop, both models show improvements, with query-based and filter-based representations yielding the greatest gains.

Reproducibility over different implementations has been a known problem especially for \bertforrec, as recently reported in \cite{petrov_systematic_2022}. The study  mentions both, ASME as well as RecBole and reports different performance for both. The implementation of \bertforrec in RecBole has been updated since, necessitating further investigations: When comparing current implementations in both frameworks, we still found some inconsistencies, for example in the masking task. ASME masks items either with the mask token, a random value or keeps the original value, while RecBole only uses the mask token. The generation of target items is also slightly different, as we noticed that the loss can become NaN for some edge cases in RecBole. 
We can also point to differences which can explain the performance gaps in \sasrec: In ASME the loss is computed for each step of the sequences, not only for the last target item, therefore training on all subsequences in one step. It also uses a separate layer for projecting back to the item space, while the implementation of RecBole reuses the item embedding layer. 
\revised{
Overall, independent of the performance of the specific implementations, the results from previous and current experiments confirm that our non-item modeling can improve the performance of several different recommender models in different frameworks.
}

%% file: data/implementation_comp_ml.tex
\nprounddigits{3}
\npnoaddmissingzero

\begin{tabular}{ll|rl|rl|rl|rl|rl}
\toprule
&    & \multicolumn{2}{c}{Items BL}  & \multicolumn{2}{c}{Prev-ID}& \multicolumn{2}{c}{Prev-PE} &\multicolumn{2}{c}{Group-ID} & \multicolumn{2}{c}{Group-PE} \\
Model & Metrics  & RecB & ASME & RecB & ASME & RecB & ASME & RecB & ASME & RecB & ASME \\
\midrule
\multirow[t]{5}{*}{BERT4Rec} & HR@1 & 
\numprint{.01046} &
\numprint{.019} & 
\numprint{.11914} &
\numprint{.393} & 
\numprint{.04008} &
\numprint{.308} & 
\numprint{.29352} &
\numprint{.397} & 
\numprint{.30846} &
\numprint{.016} \\
 & HR@5 & 
 \numprint{.04172} &
 \numprint{.078} & 
 \numprint{.24074} &
 \numprint{.662} & 
 \numprint{.08688} &
 \numprint{.605} & 
 \numprint{.53408} &
 \numprint{.652} & 
 \numprint{.5591200000000001} &
 \numprint{.040} \\
 & HR@10 & 
 \numprint{.07514000000000001} &
 \numprint{.133} & 
 \numprint{.29791999999999996}&
 \numprint{.753} & 
 \numprint{.11095999999999999}&
 \numprint{.719} & 
 \numprint{.62436}&
 \numprint{.745} & 
  \numprint{.6527000000000001}&
 \numprint{.061} \\
 & NDCG@5 & 
 \numprint{.02594} &
 \numprint{.049} & 
 \numprint{.18288}&
 \numprint{.537} & 
 \numprint{.06445999999999999}&
 \numprint{.465} & 
 \numprint{.4210400000000001}&
 \numprint{.533} & 
  \numprint{.441}&
 \numprint{.028} \\
 & NDCG@10 & 
 \numprint{.0366}&\numprint{.066} & 
 \numprint{.20136000000000004}&
 \numprint{.566} & 
 \numprint{.0722}&
 \numprint{.503} & 
 \numprint{.45034}&
 \numprint{.563} & 
  \numprint{.47138}&
  \numprint{.035} \\
 \hline
\multirow[t]{5}{*}{SASRec} & HR@1 & 
\numprint{.025420000000000005}&
\numprint{.027} & 
\numprint{.37012}&\numprint{.426} & 
\numprint{.36766}&\numprint{.384} & 
 \numprint{.37106}&\numprint{.426} & 
\numprint{.36082}&\numprint{.371} \\
 & HR@5 & 
 \numprint{.08080000000000001}&\numprint{.103} & 
  \numprint{.63358}&\numprint{.699} & 
 \numprint{.62952}&\numprint{.681} & 
 \numprint{.6286799999999999}&\numprint{.681} & 
 \numprint{.6175599999999999}&\numprint{.659} \\
 & HR@10 & 
 \numprint{.12258}&\numprint{.164} & 
  \numprint{.72538}&\numprint{.785} & 
 \numprint{.72104}&\numprint{.780} & 
 \numprint{.71716}&\numprint{.768} & 
 \numprint{.7070200000000001}&\numprint{.756} \\
 & NDCG@5 & 
 \numprint{.053459999999999994}&\numprint{.066} & 
  \numprint{.5107800000000001}&\numprint{.572} & 
 \numprint{.5075200000000001}&\numprint{.543} & 
 \numprint{.50838}&\numprint{.563} & 
 \numprint{.49782000000000004}&\numprint{.525} \\
 & NDCG@10 & 
 \numprint{.06688}&\numprint{.085} & 
  \numprint{.5405599999999999}&\numprint{.600} & 
 \numprint{.5372399999999999}&\numprint{.575} & 
 \numprint{.53716}&\numprint{.591} & 
 \numprint{.52688}&\numprint{.556} \\
\bottomrule
\end{tabular}

%% file: sections/70_conclusion.tex
\section{Conclusion}
\label{sec:conclusion}

The effect of the inclusion of item and user properties for sequential item recommendation has been studied, but non-item interactions have been an afterthought, although they are present in most recommendation setups. In our paper, we closed this gap by giving an overview on the various types of non-item interactions and proposing different modeling approaches for them. By analyzing the datasets with HypTrails, we showed that non-item pages are useful indicators for following item interactions. We adapted several sequential recommender baselines in the popular RecBole framework to include non-item representations either by attributes or pre-computed embeddings.
Evaluation on one artificially constructed and two real-life datasets showed that non-item interactions can significantly improve sequential recommendation. While all models can profit from including non-items, the degree of improvement varies between models and is dependent on several factors. One factor is the ability of models to handle noisy data, and another the selection of meaningful non-item pages with content related to the item. Our study also highlights the importance of choosing a suitable representation for non-item interactions for each dataset and scenario individually.
For future work, we plan to investigate alternative fusion functions for incorporating page embeddings within the embedding layer. Additionally, applying our approach to conversational recommendation systems might be a promising direction for future research.

%% file: sections/80_appendix.tex
\begin{table}
    \centering
    \caption{Hitrate and NDCG on the Random-\syndata dataset with Page Embeddings. The datasets with growing percentages of randomization for non-items are denoted as ``X\%R'' }
    \input{data/data_recbole/rand-exp-pe}
    \label{tab:ml_table-rand-pe}
\end{table}

\begin{table}
    \centering
    \caption{Hitrate and NDCG on the Random-\syndata dataset with ID Embeddings. The datasets with growing percentages of randomization for non-items are denoted as ``X\%R'' }
    \input{data/data_recbole/rand-exp-id}
    \label{tab:ml_table-rand-id}
\end{table}

\begin{figure}
 \begin{subfigure}[b]{0.30\textwidth}
     \centering
     \includegraphics[width=\linewidth]{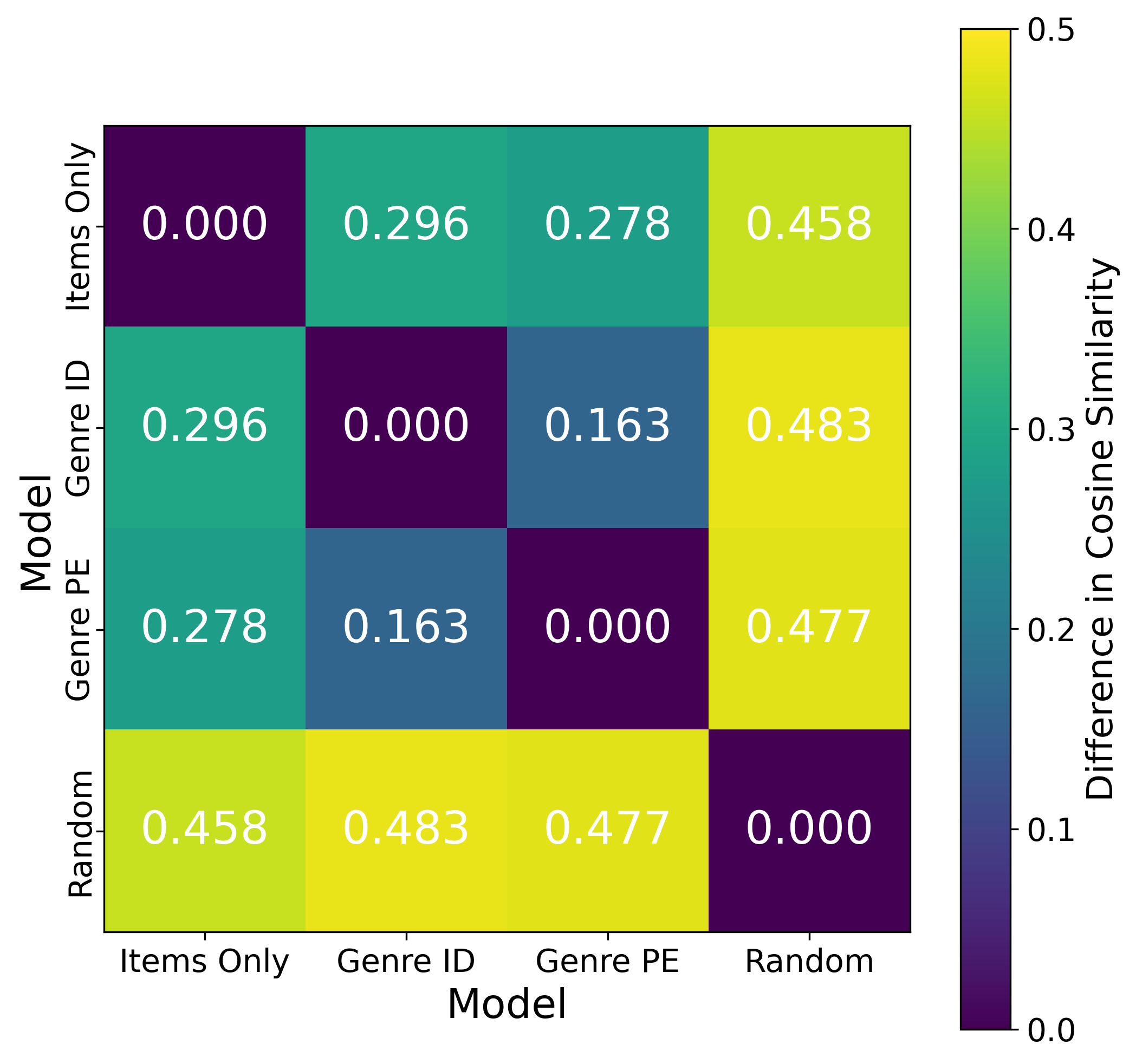}
     \caption{Difference in cosine similarity  for \gruforrec}
 \end{subfigure}
 \hfill
  \begin{subfigure}[b]{0.30\textwidth}
     \centering
     \includegraphics[width=\linewidth]{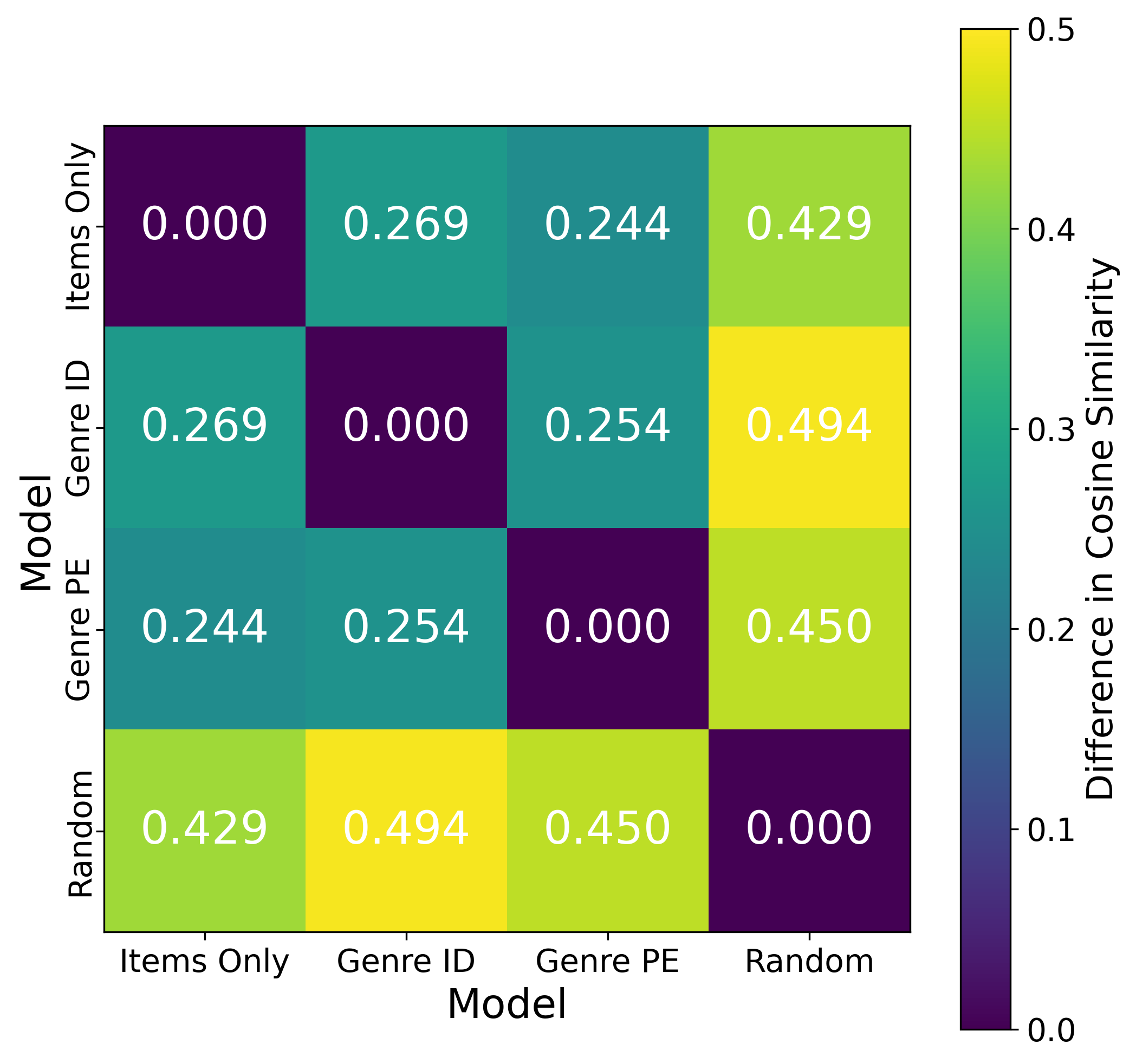}
     \caption{Difference in cosine similarity for \lightsans}
 \end{subfigure}
 \hfill
  \begin{subfigure}[b]{0.30\textwidth}
     \centering
     \includegraphics[width=\linewidth]{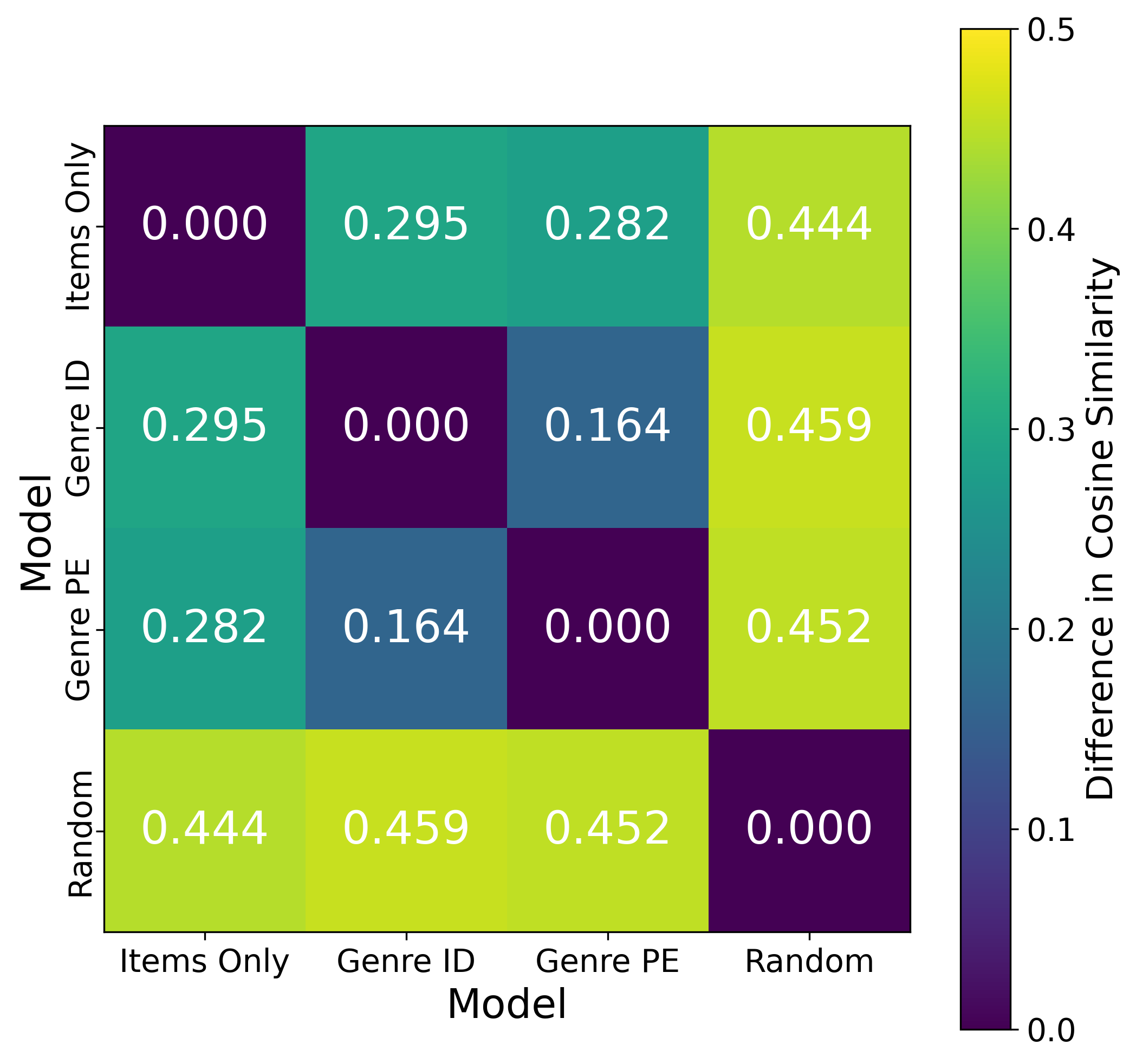}
     \caption{Difference in cosine similarity  for \narm}
 \end{subfigure}
\medskip
  \begin{subfigure}[b]{0.30\textwidth}
     \centering
     \includegraphics[width=\linewidth]{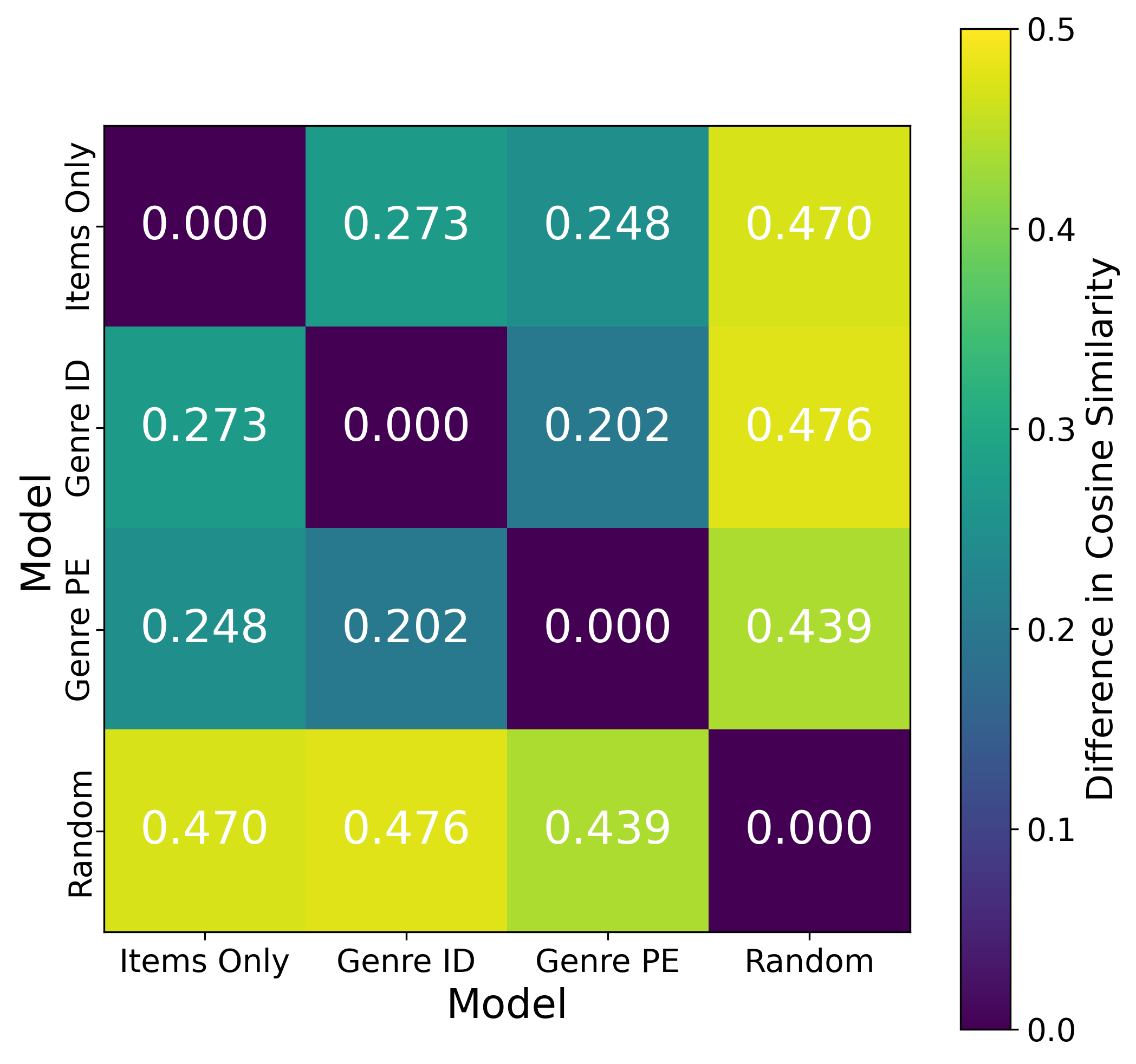}
     \caption{Difference in cosine similarity  for \nextitnet}
 \end{subfigure}
 \hfill
  \begin{subfigure}[b]{0.30\textwidth}
     \centering
     \includegraphics[width=\linewidth]{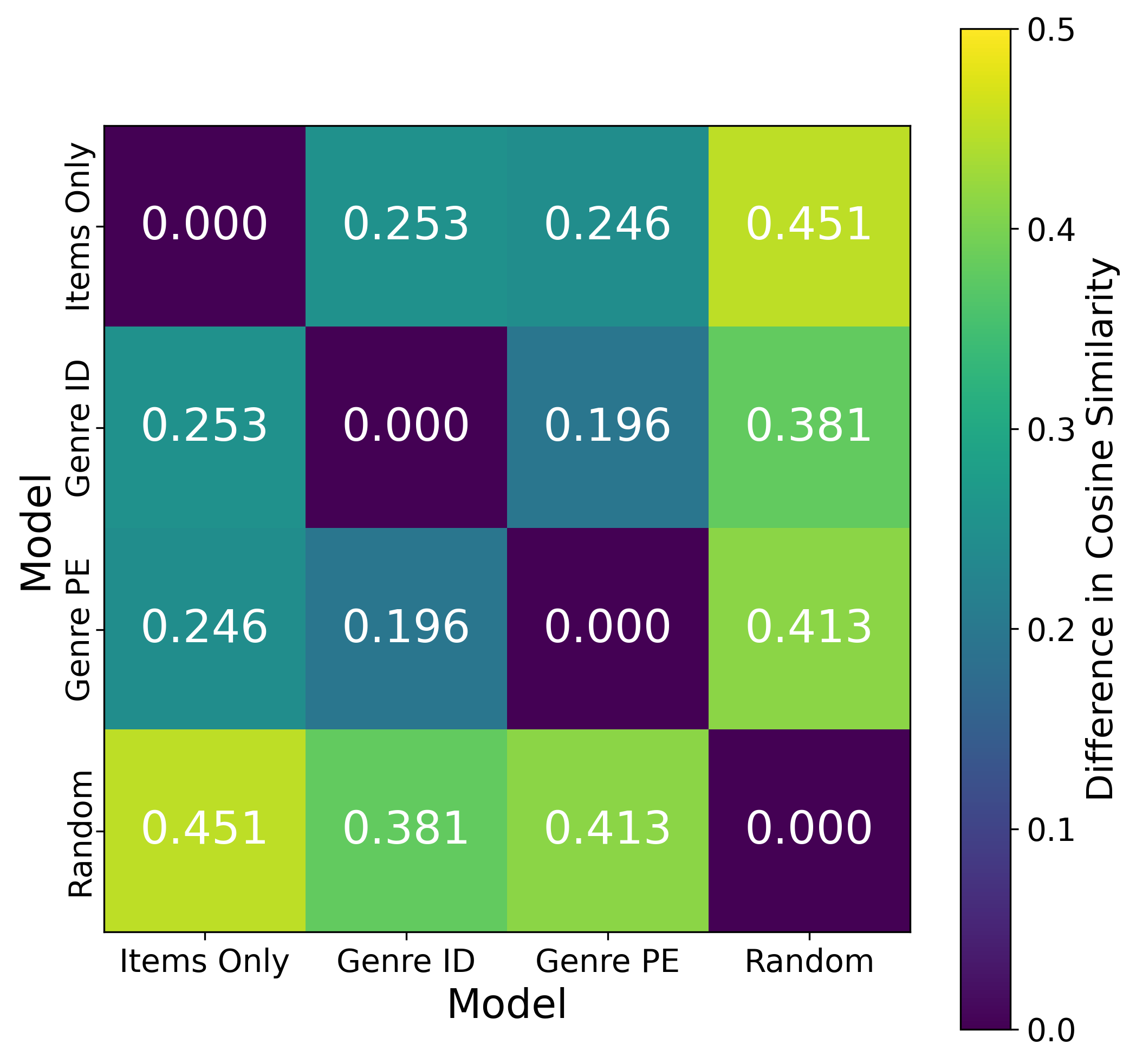}
     \caption{Difference in cosine similarity  for \caser}
 \end{subfigure}
     \hfill
  \begin{subfigure}[b]{0.30\textwidth}
     \centering
     \includegraphics[width=\linewidth]{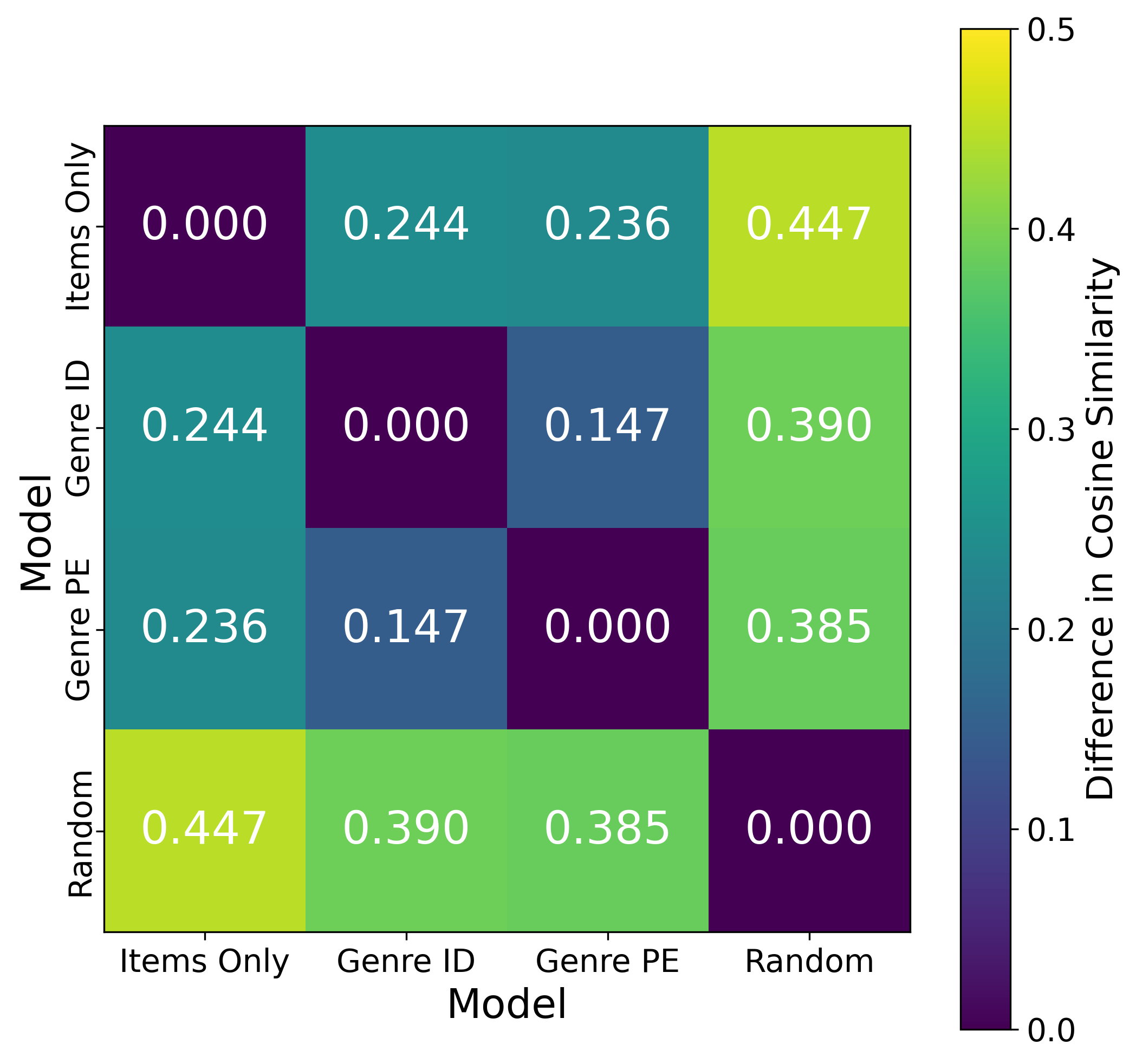}
     \caption{Difference in cosine similarity  for \core}
 \end{subfigure}
\centering
\caption{Difference in Item-to-Item cosine similarity between different model variants. For each model the cosine similarity between all movies is computed and normalized. We report the average of the absolute difference in similarity between two models. A lower score means item-item relations are similar in both models. A randomly generated embedding is included as a baseline.  }
\label{fig:exp1_sem_emb_appendix}
\Description[Difference in Item-to-Item cosine similarity between different model variants.]{Difference in Item-to-Item cosine similarity between different model variants.}
\end{figure}

\begin{figure}
 \includegraphics[width=0.45\textwidth]{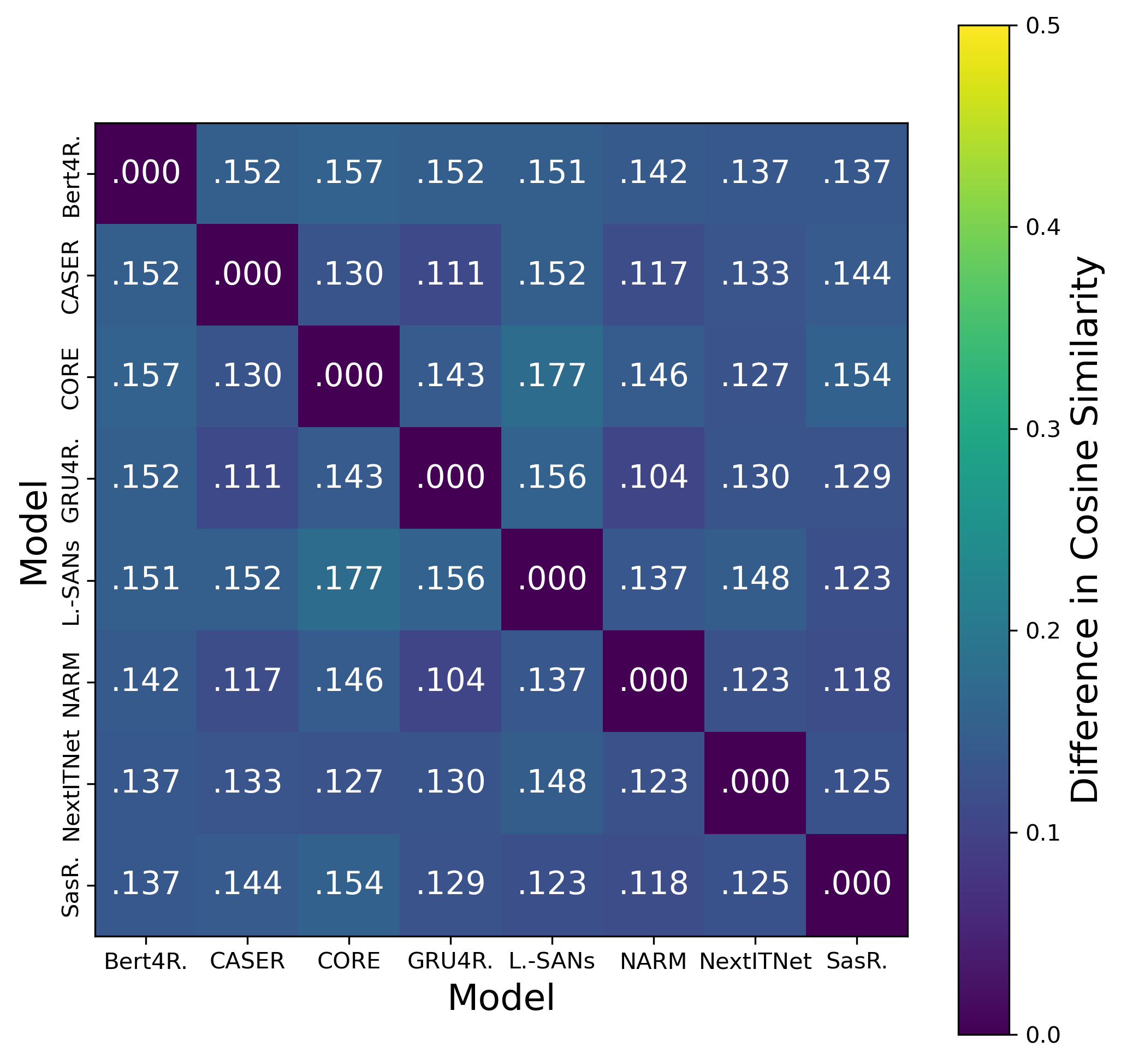}
 \label{fig:exp1_sem_emb_sasrec}
\centering
\caption{Difference in Item-to-Item cosine similarity between different models with CPID on Prev-\syndata. We report the average of the absolute difference in similarity between two models. A lower score means item-item relations are similar in both models. A randomly generated embedding is included as a baseline. }
\label{fig:app_exp1_sem_model}
\Description[Difference in Item-to-Item cosine similarity between different models with CPID on Prev-\syndata.]{Difference in Item-to-Item cosine similarity between different models with CPID on Prev-\syndata.}
\end{figure}

\begin{figure}
 \begin{subfigure}[b]{0.45\textwidth}
         \centering
         \includegraphics[width=\textwidth]{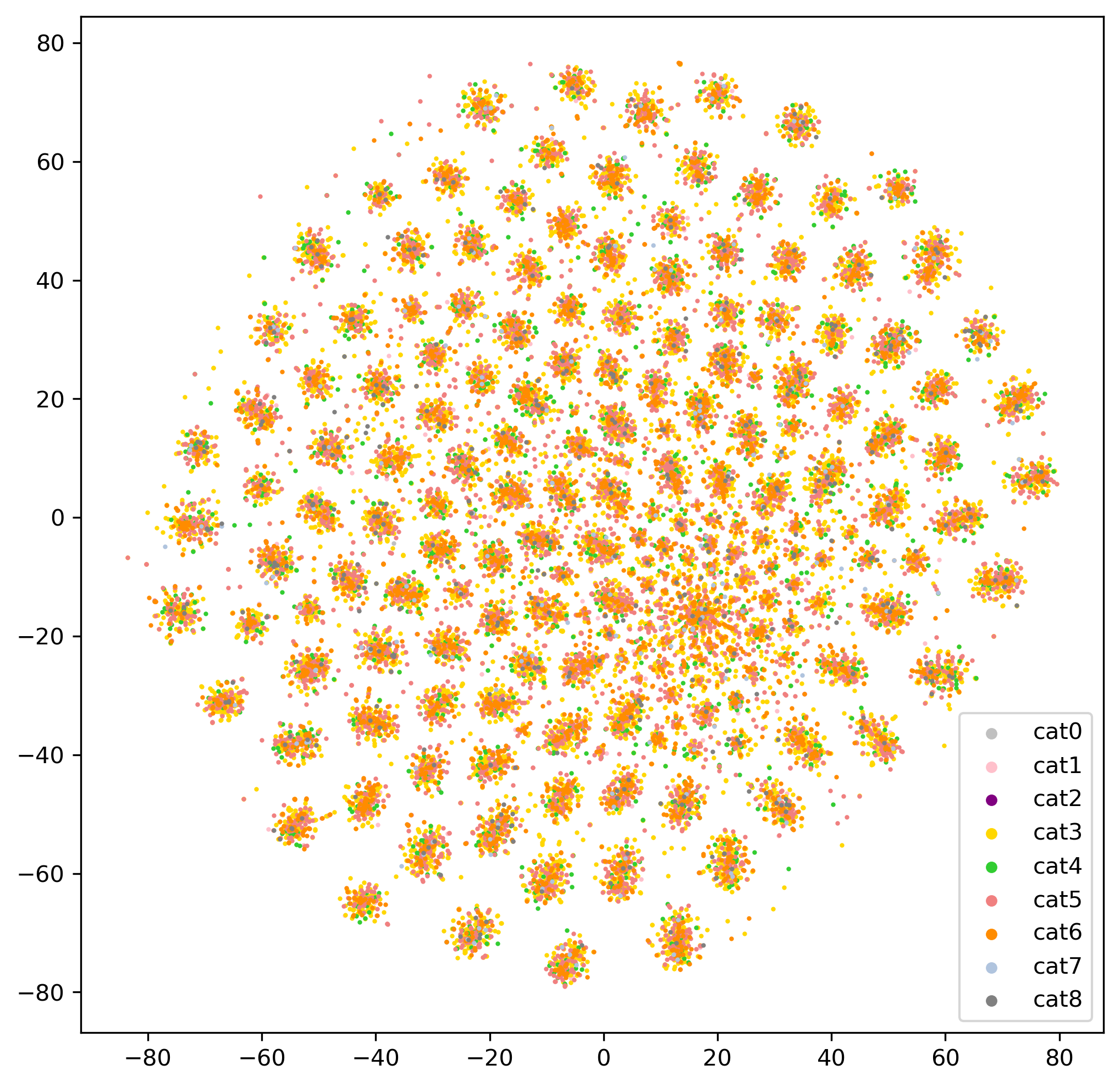}
         \caption{Visualization of the item embedding space of NARM on \coveopage with Items Only.}
     \end{subfigure}
     \hfill
  \begin{subfigure}[b]{0.45\textwidth}
     \centering
     \includegraphics[width=\textwidth]{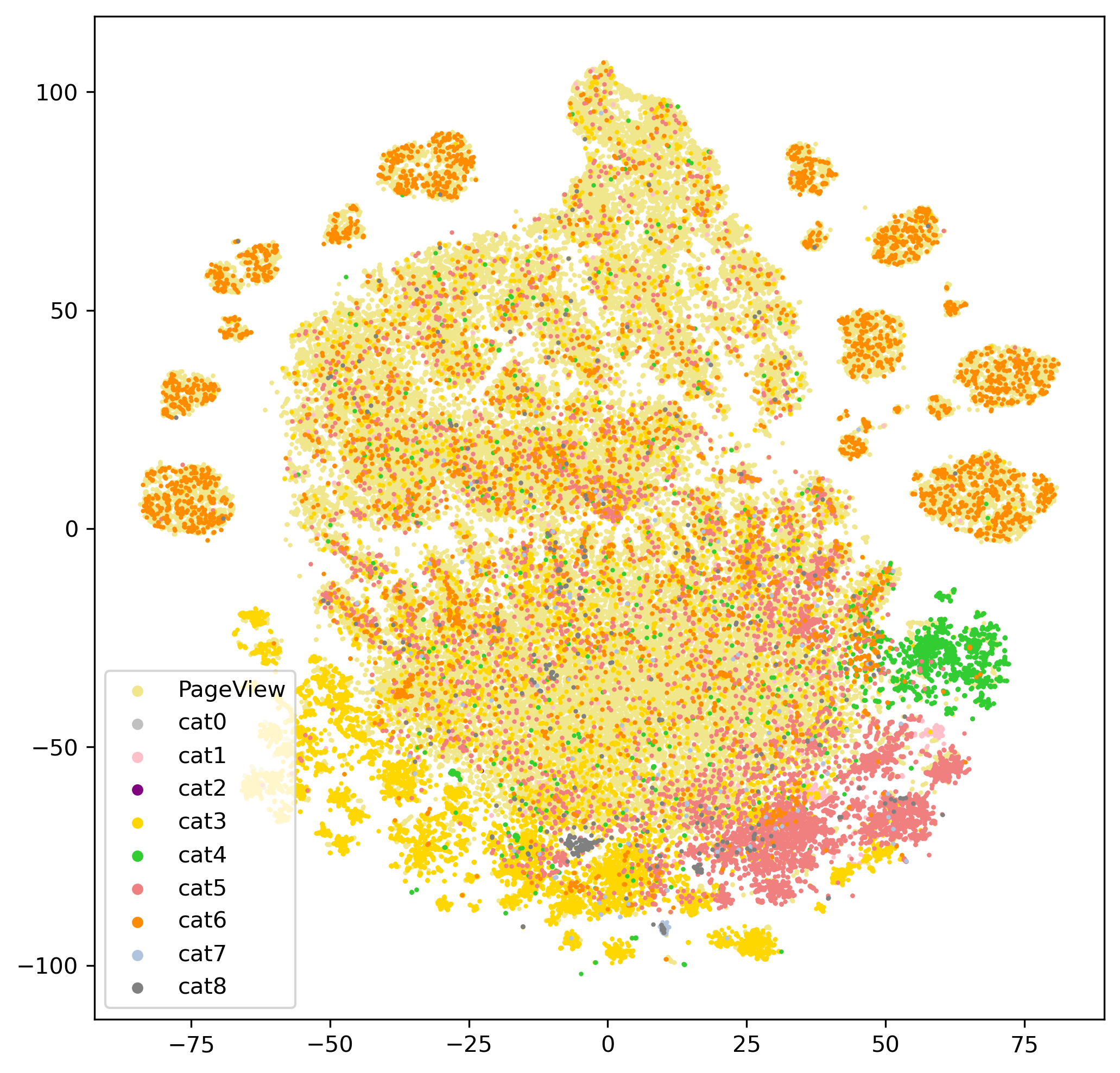}
     \caption{Visualization of the item embedding space of NARM on \coveopage with Non-Item Pages.}
 \end{subfigure}
\centering
\caption{t-SNE visualization of item embedding space of NARM on \coveopage with ID embeddings.}
\label{fig:app-exp2-tsne-narm}
\Description[t-SNE visualization of item embedding space of NARM on \coveopage with ID embeddings.]{t-SNE visualization of item embedding space of NARM on \coveopage with ID embeddings.}
\end{figure}

\begin{figure}
 \begin{subfigure}[b]{0.40\textwidth}
         \centering
         \includegraphics[width=\textwidth]{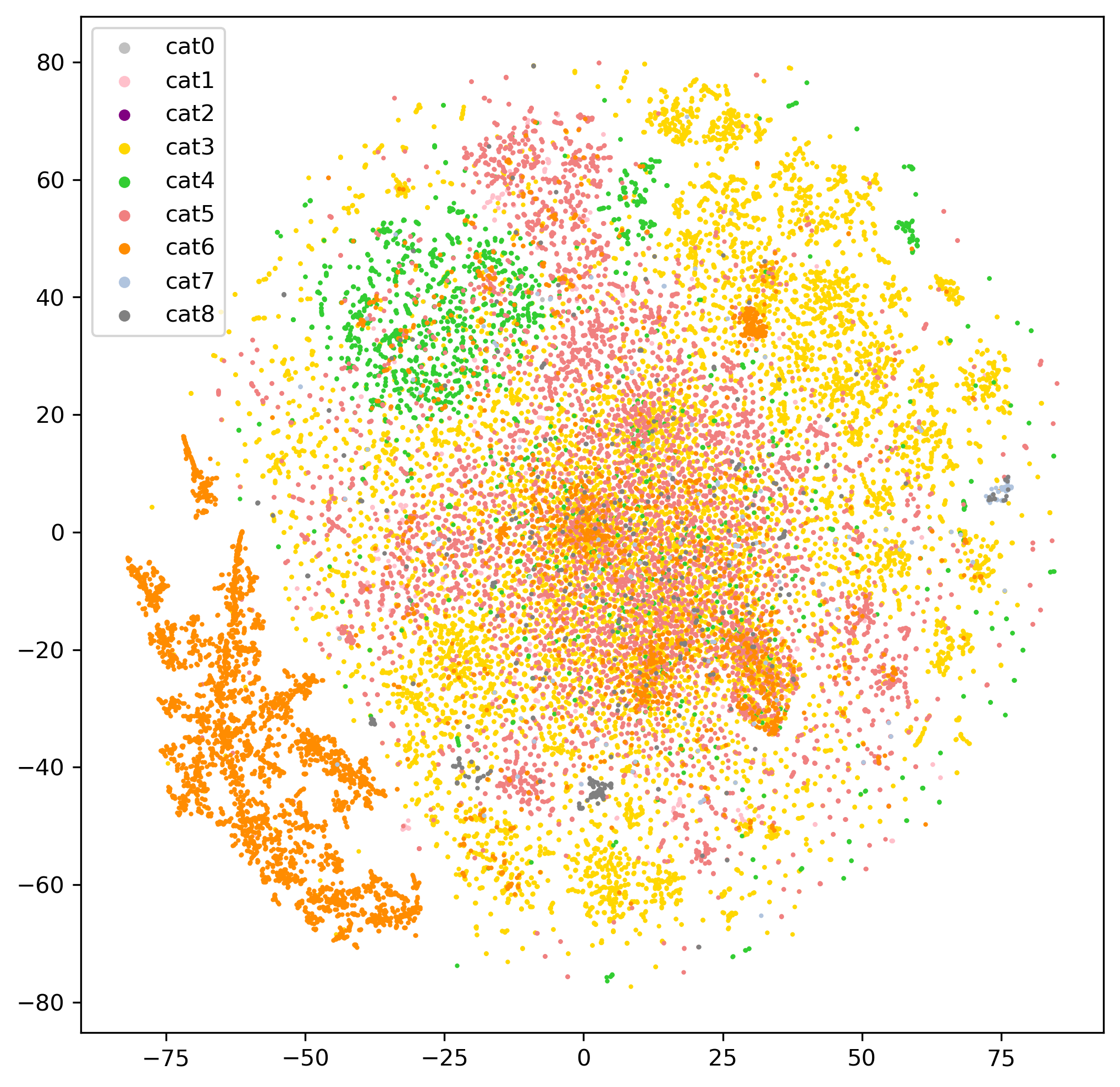}
         \caption{Visualization of the item embedding space of SASRec on \coveosearch with Items Only.}
     \end{subfigure}
     \hfill
  \begin{subfigure}[b]{0.40\textwidth}
     \centering
     \includegraphics[width=\textwidth]{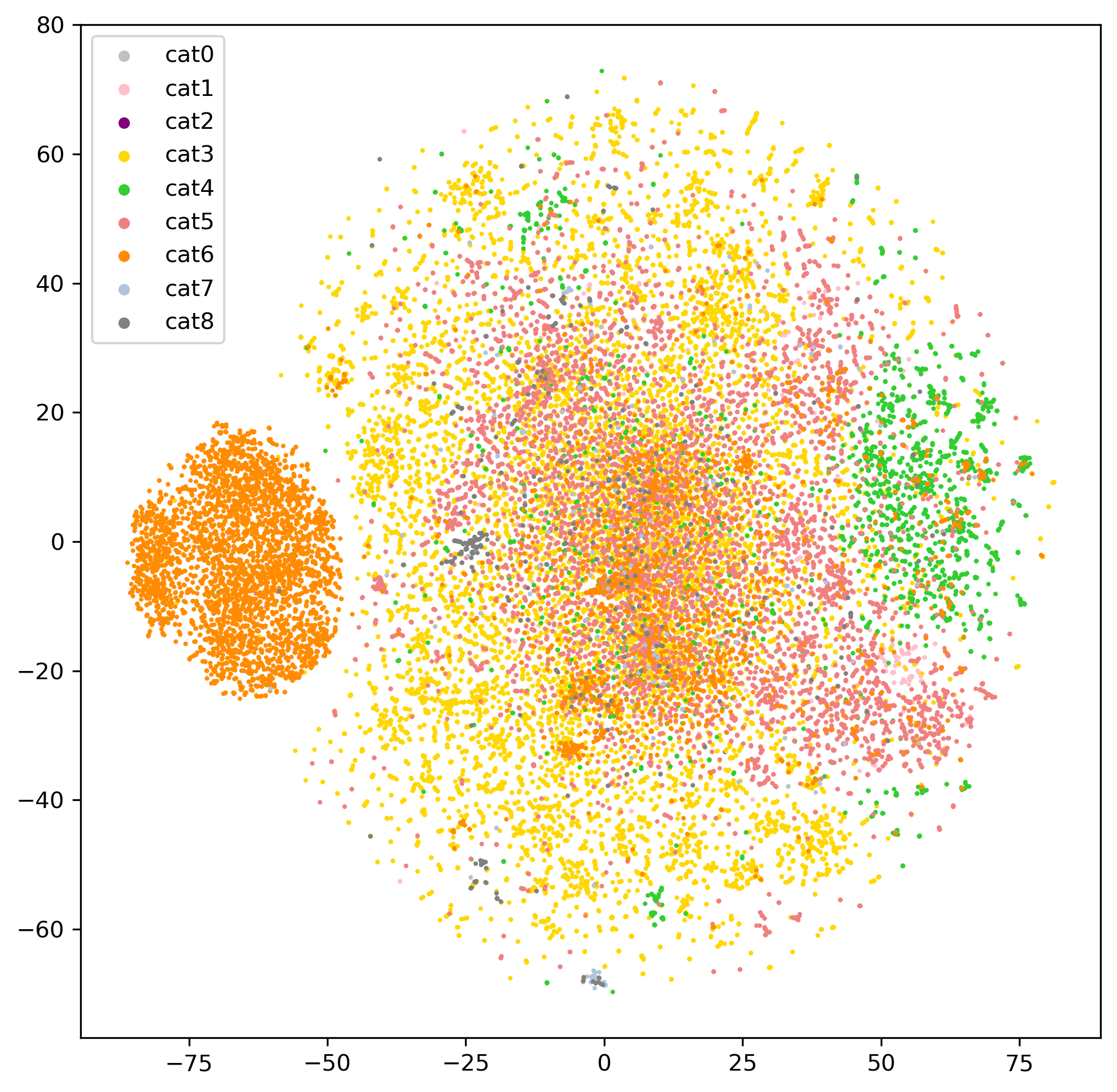}
     \caption{Visualization of the item embedding space of SASRec on \coveosearch trained with the embedded search query.}
 \end{subfigure}
\centering
\caption{t-SNE visualization of item embedding space of SASRec on \coveosearch.}
\label{fig:exp3-tsne}
\Description[t-SNE visualization of item embedding space of SASRec on \coveosearch.]{t-SNE visualization of item embedding space of SASRec on \coveosearch.}
\end{figure}

%% file: data/data_recbole/rand-exp-pe.tex
\nprounddigits{3}
\npnoaddmissingzero

\begin{tabular}{llllllllllllll}
\toprule
Model & Metrics & 0\%R & 10\%R & 20\%R& 30\%R & 40\%R & 50\%R & 60\%R & 70\%R & 80\%R & 90\%R & 100\%R \\
\midrule
\multirow[t]{5}{*}{BERT4Rec} & HR@1 & \numprint{.0922} & \numprint{.0055} & \numprint{.006} & \numprint{.0109} & \numprint{.0052} & \numprint{.0052} & \numprint{.0054} & \numprint{.007} & \numprint{.0066} & \numprint{.0032} & \numprint{.0047} \\
 & HR@5 &  {\numprint{.1891}} & \numprint{.0191} & \numprint{.0178} & \numprint{.0413} & \numprint{.0178} & \numprint{.0183} & \numprint{.0191} & \numprint{.0215} & \numprint{.023} & \numprint{.0149} & \numprint{.0182} \\
 & HR@10 &  {\numprint{.2307}} & \numprint{.0348} & \numprint{.0318} & \numprint{.0706} & \numprint{.0314} & \numprint{.0313} & \numprint{.0355} & \numprint{.0378} & \numprint{.0414} & \numprint{.0296} & \numprint{.0319} \\
 & NDCG@5 &  {\numprint{.1429}} & \numprint{.0122} & \numprint{.0119} & \numprint{.026} & \numprint{.0117} & \numprint{.0118} & \numprint{.0122} & \numprint{.0142} & \numprint{.0146} & \numprint{.009} & \numprint{.0112} \\
 & NDCG@10 &  {\numprint{.1563}} & \numprint{.0171} & \numprint{.0164} & \numprint{.0354} & \numprint{.0161} & \numprint{.0159} & \numprint{.0174} & \numprint{.0194} & \numprint{.0205} & \numprint{.0137} & \numprint{.0156} \\
\cline{1-13}
\multirow[t]{5}{*}{Caser} & HR@1 & \numprint{.2883} & \numprint{.2111} & \numprint{.1749} & \numprint{.1615} & \numprint{.1327} & \numprint{.0428} & \numprint{.0225} & \numprint{.0186} & \numprint{.0168} & \numprint{.016} & \numprint{.018} \\
 & HR@5 & \numprint{.5458} & \numprint{.4481} & \numprint{.3693} & \numprint{.3305} & \numprint{.2751} & \numprint{.1155} & \numprint{.0658} & \numprint{.0601} & \numprint{.0547} & \numprint{.0532} & \numprint{.0586} \\
 & HR@10 & \numprint{.6532} & \numprint{.544} & \numprint{.455} & \numprint{.4067} & \numprint{.3422} & \numprint{.1655} & \numprint{.102} & \numprint{.0952} & \numprint{.0874} & \numprint{.0834} & \numprint{.09} \\
 & NDCG@5 & \numprint{.4239} & \numprint{.3354} & \numprint{.2768} & \numprint{.2499} & \numprint{.2074} & \numprint{.0803} & \numprint{.0444} & \numprint{.0393} & \numprint{.0357} & \numprint{.0349} & \numprint{.0383} \\
 & NDCG@10 & \numprint{.4588} & \numprint{.3665} & \numprint{.3045} & \numprint{.2746} & \numprint{.2292} & \numprint{.0964} & \numprint{.0561} & \numprint{.0506} & \numprint{.0462} & \numprint{.0446} & \numprint{.0484} \\
\cline{1-13}
\multirow[t]{5}{*}{Core} & HR@1 &  {\numprint{.3095}} & \numprint{.2972} & \numprint{.2524} & \numprint{.1906} & \numprint{.1113} & \numprint{.0648} & \numprint{.034} & \numprint{.0212} & \numprint{.0157} & \numprint{.0148} & \numprint{.0136} \\
 & HR@5 & \numprint{.5511} & \numprint{.5143} & \numprint{.4427} & \numprint{.3588} & \numprint{.2588} & \numprint{.1822} & \numprint{.1182} & \numprint{.0806} & \numprint{.0638} & \numprint{.0621} & \numprint{.0615} \\
 & HR@10 & \numprint{.6464} & \numprint{.5952} & \numprint{.5157} & \numprint{.4295} & \numprint{.3277} & \numprint{.2499} & \numprint{.178} & \numprint{.1329} & \numprint{.1101} & \numprint{.1072} & \numprint{.1069} \\
 & NDCG@5 & \numprint{.437} & \numprint{.4127} & \numprint{.3529} & \numprint{.2795} & \numprint{.1879} & \numprint{.1248} & \numprint{.0765} & \numprint{.0507} & \numprint{.0396} & \numprint{.0383} & \numprint{.0371} \\
 & NDCG@10 & \numprint{.4678} & \numprint{.4389} & \numprint{.3766} & \numprint{.3024} & \numprint{.2101} & \numprint{.1466} & \numprint{.0959} & \numprint{.0676} & \numprint{.0544} & \numprint{.0527} & \numprint{.0518} \\
\cline{1-13}
\multirow[t]{5}{*}{GRU4Rec} & HR@1 &  {\numprint{.3683}} & \numprint{.3374} & \numprint{.2959} & \numprint{.2568} & \numprint{.2208} & \numprint{.1732} & \numprint{.1148} & \numprint{.0407} & \numprint{.0238} & \numprint{.0198} & \numprint{.0225} \\
 & HR@5 & \numprint{.6245} & \numprint{.5634} & \numprint{.4965} & \numprint{.4284} & \numprint{.3732} & \numprint{.3009} & \numprint{.2202} & \numprint{.108} & \numprint{.0719} & \numprint{.0681} & \numprint{.0695} \\
 & HR@10 &  {\numprint{.7205}} & \numprint{.6457} & \numprint{.5747} & \numprint{.493} & \numprint{.4312} & \numprint{.3522} & \numprint{.2693} & \numprint{.1545} & \numprint{.1132} & \numprint{.1067} & \numprint{.1098} \\
 & NDCG@5 & \numprint{.5044} & \numprint{.4576} & \numprint{.4024} & \numprint{.3485} & \numprint{.3022} & \numprint{.241} & \numprint{.1695} & \numprint{.0753} & \numprint{.0481} & \numprint{.0442} & \numprint{.0459} \\
 & NDCG@10 &  {\numprint{.5357}} & \numprint{.4843} & \numprint{.4278} & \numprint{.3694} & \numprint{.3209} & \numprint{.2575} & \numprint{.1853} & \numprint{.0902} & \numprint{.0613} & \numprint{.0566} & \numprint{.0589} \\
\cline{1-13}
\multirow[t]{5}{*}{LightSANs} & HR@1 &  {\numprint{.3507}} & \numprint{.3155} & \numprint{.2835} & \numprint{.2423} & \numprint{.208} & \numprint{.177} & \numprint{.1259} & \numprint{.0719} & \numprint{.0324} & \numprint{.0158} & \numprint{.0166} \\
 & HR@5 &  {\numprint{.6133}} & \numprint{.5465} & \numprint{.4884} & \numprint{.4191} & \numprint{.363} & \numprint{.3065} & \numprint{.229} & \numprint{.148} & \numprint{.083} & \numprint{.0475} & \numprint{.0488} \\
 & HR@10 &  {\numprint{.7063}} & \numprint{.6325} & \numprint{.5684} & \numprint{.4846} & \numprint{.4216} & \numprint{.3569} & \numprint{.2754} & \numprint{.1867} & \numprint{.1166} & \numprint{.0734} & \numprint{.0768} \\
 & NDCG@5 &  {\numprint{.4905}} & \numprint{.4387} & \numprint{.3925} & \numprint{.3363} & \numprint{.2912} & \numprint{.2456} & \numprint{.1801} & \numprint{.1111} & \numprint{.0582} & \numprint{.0318} & \numprint{.0328} \\
 & NDCG@10 &  {\numprint{.5206}} & \numprint{.4667} & \numprint{.4184} & \numprint{.3575} & \numprint{.3102} & \numprint{.2619} & \numprint{.1951} & \numprint{.1235} & \numprint{.069} & \numprint{.0401} & \numprint{.0418} \\
\cline{1-13}
\multirow[t]{5}{*}{NARM} & HR@1 &  {\numprint{.3615}} & \numprint{.3323} & \numprint{.2968} & \numprint{.2523} & \numprint{.2159} & \numprint{.1672} & \numprint{.1072} & \numprint{.0352} & \numprint{.0218} & \numprint{.0219} & \numprint{.0247} \\
 & HR@5 &  {\numprint{.6211}} & \numprint{.5591} & \numprint{.5019} & \numprint{.4302} & \numprint{.3741} & \numprint{.2988} & \numprint{.2086} & \numprint{.1036} & \numprint{.072} & \numprint{.0752} & \numprint{.0734} \\
 & HR@10 &  {\numprint{.7112}} & \numprint{.6405} & \numprint{.5725} & \numprint{.4957} & \numprint{.4339} & \numprint{.3553} & \numprint{.2608} & \numprint{.1536} & \numprint{.1138} & \numprint{.1168} & \numprint{.1142} \\
 & NDCG@5 &  {\numprint{.4992}} & \numprint{.4536} & \numprint{.4052} & \numprint{.3471} & \numprint{.3002} & \numprint{.237} & \numprint{.1607} & \numprint{.0701} & \numprint{.0474} & \numprint{.0487} & \numprint{.0489} \\
 & NDCG@10 &  {\numprint{.5285}} & \numprint{.4799} & \numprint{.428} & \numprint{.3683} & \numprint{.3195} & \numprint{.2553} & \numprint{.1776} & \numprint{.0862} & \numprint{.0609} & \numprint{.062} & \numprint{.062} \\
\cline{1-13}
\multirow[t]{5}{*}{NextItNet} & HR@1 & \numprint{.2642} & \numprint{.2306} & \numprint{.1308} & \numprint{.0833} & \numprint{.0345} & \numprint{.0183} & \numprint{.0112} & \numprint{.0107} & \numprint{.008} & \numprint{.0105} & \numprint{.0113} \\
 & HR@5 & \numprint{.4834} & \numprint{.4275} & \numprint{.2928} & \numprint{.1941} & \numprint{.09} & \numprint{.0568} & \numprint{.0376} & \numprint{.033} & \numprint{.0316} & \numprint{.0349} & \numprint{.0397} \\
 & HR@10 &  {\numprint{.5749}} & \numprint{.5049} & \numprint{.3679} & \numprint{.2582} & \numprint{.1318} & \numprint{.089} & \numprint{.062} & \numprint{.0541} & \numprint{.0534} & \numprint{.0559} & \numprint{.063} \\
 & NDCG@5 & \numprint{.3799} & \numprint{.3347} & \numprint{.2154} & \numprint{.1407} & \numprint{.0623} & \numprint{.038} & \numprint{.0241} & \numprint{.0217} & \numprint{.0199} & \numprint{.0225} & \numprint{.0255} \\
 & NDCG@10 & \numprint{.4096} & \numprint{.3598} & \numprint{.2397} & \numprint{.1614} & \numprint{.0757} & \numprint{.0484} & \numprint{.0318} & \numprint{.0285} & \numprint{.0269} & \numprint{.0292} & \numprint{.033} \\
\cline{1-13}
\multirow[t]{5}{*}{SASRec} & HR@1 & \numprint{.3666} & \numprint{.3471} & \numprint{.3018} & \numprint{.2618} & \numprint{.2289} & \numprint{.182} & \numprint{.1177} & \numprint{.0706} & \numprint{.0308} & \numprint{.0211} & \numprint{.0222} \\
 & HR@5 & \numprint{.6279} & \numprint{.5755} & \numprint{.51} & \numprint{.4352} & \numprint{.3822} & \numprint{.3167} & \numprint{.2272} & \numprint{.1532} & \numprint{.0897} & \numprint{.0705} & \numprint{.0689} \\
 & HR@10 & \numprint{.7193} & \numprint{.6571} & \numprint{.5841} & \numprint{.5027} & \numprint{.4411} & \numprint{.372} & \numprint{.2784} & \numprint{.2011} & \numprint{.1371} & \numprint{.1105} & \numprint{.1124} \\
 & NDCG@5 & \numprint{.5061} & \numprint{.468} & \numprint{.4124} & \numprint{.3542} & \numprint{.3105} & \numprint{.2529} & \numprint{.1752} & \numprint{.1132} & \numprint{.0605} & \numprint{.0461} & \numprint{.0456} \\
 & NDCG@10 & \numprint{.5358} & \numprint{.4945} & \numprint{.4364} & \numprint{.3762} & \numprint{.3296} & \numprint{.2709} & \numprint{.1917} & \numprint{.1286} & \numprint{.0757} & \numprint{.0589} & \numprint{.0595} \\
\cline{1-13}
\bottomrule
\end{tabular}

%% file: data/data_recbole/rand-exp-id.tex
\nprounddigits{3}
\npnoaddmissingzero

\begin{tabular}{lllllllllllll}
\toprule
Model & Metrics & 0\%R & 10\%R & 20\%R& 30\%R & 40\%R & 50\%R & 60\%R & 70\%R & 80\%R & 90\%R & 100\%R \\
\midrule
\multirow[t]{5}{*}{BERT4Rec} & HR@1 & \numprint{.1264} & \numprint{.0937} & \numprint{.084} & \numprint{.0648} & \numprint{.0634} & \numprint{.0044} & \numprint{.004} & \numprint{.0036} & \numprint{.0039} & \numprint{.0019} & \numprint{.0044} \\
 & HR@5 & \numprint{.2488} & \numprint{.1813} & \numprint{.1696} & \numprint{.1381} & \numprint{.1326} & \numprint{.0157} & \numprint{.0141} & \numprint{.0123} & \numprint{.0121} & \numprint{.0068} & \numprint{.011} \\
 & HR@10 & \numprint{.3033} & \numprint{.2253} & \numprint{.2137} & \numprint{.1733} & \numprint{.1617} & \numprint{.0274} & \numprint{.0229} & \numprint{.0196} & \numprint{.0189} & \numprint{.0138} & \numprint{.0175} \\
 & NDCG@5 & \numprint{.1911} & \numprint{.1395} & \numprint{.129} & \numprint{.103} & \numprint{.0997} & \numprint{.01} & \numprint{.0089} & \numprint{.0079} & \numprint{.008} & \numprint{.0045} & \numprint{.0078} \\
 & NDCG@10 & \numprint{.2087} & \numprint{.1537} & \numprint{.1433} & \numprint{.1144} & \numprint{.1092} & \numprint{.0138} & \numprint{.0118} & \numprint{.0102} & \numprint{.0102} & \numprint{.0067} & \numprint{.0099} \\
\cline{1-13}
\multirow[t]{5}{*}{Caser} & HR@1 & \numprint{.3177} & \numprint{.275} & \numprint{.2369} & \numprint{.18} & \numprint{.099} & \numprint{.0518} & \numprint{.0224} & \numprint{.0175} & \numprint{.0172} & \numprint{.018} & \numprint{.0182} \\
 & HR@5 & \numprint{.5722} & \numprint{.4931} & \numprint{.4331} & \numprint{.3384} & \numprint{.2124} & \numprint{.1331} & \numprint{.0685} & \numprint{.0556} & \numprint{.0568} & \numprint{.0607} & \numprint{.0574} \\
 & HR@10 & \numprint{.6735} & \numprint{.5805} & \numprint{.5095} & \numprint{.4081} & \numprint{.2725} & \numprint{.1804} & \numprint{.1043} & \numprint{.0897} & \numprint{.0886} & \numprint{.0924} & \numprint{.0909} \\
 & NDCG@5 & \numprint{.452} & \numprint{.3908} & \numprint{.3401} & \numprint{.2631} & \numprint{.1583} & \numprint{.0936} & \numprint{.046} & \numprint{.0366} & \numprint{.0371} & \numprint{.0393} & \numprint{.0382} \\
 & NDCG@10 & \numprint{.4848} & \numprint{.4191} & \numprint{.3648} & \numprint{.2858} & \numprint{.1778} & \numprint{.1089} & \numprint{.0575} & \numprint{.0476} & \numprint{.0473} & \numprint{.0495} & \numprint{.049} \\
\cline{1-13}
\multirow[t]{5}{*}{Core} & HR@1 & \numprint{.3482} & \numprint{.3241} & \numprint{.2818} & \numprint{.2239} & \numprint{.1753} & \numprint{.1303} & \numprint{.0887} & \numprint{.051} & \numprint{.03} & \numprint{.0141} & \numprint{.0164} \\
 & HR@5 & \numprint{.602} & \numprint{.5524} & \numprint{.4907} & \numprint{.4166} & \numprint{.3502} & \numprint{.2817} & \numprint{.2088} & \numprint{.1371} & \numprint{.0943} & \numprint{.0625} & \numprint{.0594} \\
 & HR@10 & \numprint{.6974} & \numprint{.6328} & \numprint{.5648} & \numprint{.4854} & \numprint{.4144} & \numprint{.3423} & \numprint{.2676} & \numprint{.1937} & \numprint{.1425} & \numprint{.1065} & \numprint{.102} \\
 & NDCG@5 & \numprint{.4839} & \numprint{.4451} & \numprint{.3931} & \numprint{.3267} & \numprint{.2684} & \numprint{.2106} & \numprint{.1516} & \numprint{.0951} & \numprint{.0623} & \numprint{.0378} & \numprint{.0373} \\
 & NDCG@10 & \numprint{.5149} & \numprint{.4713} & \numprint{.4172} & \numprint{.3489} & \numprint{.2892} & \numprint{.2303} & \numprint{.1705} & \numprint{.1134} & \numprint{.0779} & \numprint{.0519} & \numprint{.051} \\
\cline{1-13}
\multirow[t]{5}{*}{GRU4Rec} & HR@1 & \numprint{.3559} & \numprint{.3267} & \numprint{.294} & \numprint{.2512} & \numprint{.2161} & \numprint{.1774} & \numprint{.121} & \numprint{.0367} & \numprint{.0206} & \numprint{.0176} & \numprint{.0192} \\
 & HR@5 & \numprint{.6134} & \numprint{.5579} & \numprint{.5007} & \numprint{.4216} & \numprint{.3681} & \numprint{.3038} & \numprint{.2203} & \numprint{.0991} & \numprint{.0687} & \numprint{.0589} & \numprint{.0635} \\
 & HR@10 & \numprint{.7059} & \numprint{.6347} & \numprint{.5719} & \numprint{.4842} & \numprint{.423} & \numprint{.3533} & \numprint{.2697} & \numprint{.1418} & \numprint{.1049} & \numprint{.0958} & \numprint{.1001} \\
 & NDCG@5 &  {\numprint{.4937}} & \numprint{.45} & \numprint{.4044} & \numprint{.3424} & \numprint{.2973} & \numprint{.244} & \numprint{.1736} & \numprint{.0685} & \numprint{.0448} & \numprint{.0385} & \numprint{.0414} \\
 & NDCG@10 & \numprint{.5238} & \numprint{.475} & \numprint{.4275} & \numprint{.3628} & \numprint{.3152} & \numprint{.26} & \numprint{.1895} & \numprint{.0822} & \numprint{.0565} & \numprint{.0503} & \numprint{.0532} \\
\cline{1-13}
\multirow[t]{5}{*}{LightSANs} & HR@1 & \numprint{.3439} & \numprint{.3133} & \numprint{.2798} & \numprint{.2391} & \numprint{.2142} & \numprint{.182} & \numprint{.1421} & \numprint{.1028} & \numprint{.0604} & \numprint{.0179} & \numprint{.0139} \\
 & HR@5 & \numprint{.6072} & \numprint{.5498} & \numprint{.4855} & \numprint{.4209} & \numprint{.3734} & \numprint{.3127} & \numprint{.2485} & \numprint{.1825} & \numprint{.1173} & \numprint{.0546} & \numprint{.0446} \\
 & HR@10 & \numprint{.7015} & \numprint{.633} & \numprint{.5638} & \numprint{.49} & \numprint{.4319} & \numprint{.3664} & \numprint{.2928} & \numprint{.2209} & \numprint{.1501} & \numprint{.083} & \numprint{.0727} \\
 & NDCG@5 & \numprint{.4834} & \numprint{.4386} & \numprint{.3889} & \numprint{.3359} & \numprint{.2986} & \numprint{.2517} & \numprint{.1985} & \numprint{.1445} & \numprint{.0902} & \numprint{.0362} & \numprint{.0293} \\
 & NDCG@10 & \numprint{.5141} & \numprint{.4656} & \numprint{.4143} & \numprint{.3582} & \numprint{.3175} & \numprint{.269} & \numprint{.2129} & \numprint{.157} & \numprint{.1008} & \numprint{.0454} & \numprint{.0383} \\
\cline{1-13}
\multirow[t]{5}{*}{NARM} & HR@1 & \numprint{.3546} & \numprint{.3258} & \numprint{.2894} & \numprint{.2511} & \numprint{.2127} & \numprint{.1638} & \numprint{.1121} & \numprint{.0368} & \numprint{.0188} & \numprint{.019} & \numprint{.0189} \\
 & HR@5 & \numprint{.615} & \numprint{.556} & \numprint{.4917} & \numprint{.426} & \numprint{.3678} & \numprint{.2939} & \numprint{.2121} & \numprint{.0994} & \numprint{.0688} & \numprint{.066} & \numprint{.0631} \\
 & HR@10 & \numprint{.7094} & \numprint{.6403} & \numprint{.5698} & \numprint{.4934} & \numprint{.4257} & \numprint{.3463} & \numprint{.2643} & \numprint{.1423} & \numprint{.1088} & \numprint{.1} & \numprint{.1043} \\
 & NDCG@5 & \numprint{.4934} & \numprint{.4486} & \numprint{.397} & \numprint{.3439} & \numprint{.2948} & \numprint{.2321} & \numprint{.1642} & \numprint{.0684} & \numprint{.0438} & \numprint{.0429} & \numprint{.041} \\
 & NDCG@10 & \numprint{.5241} & \numprint{.4761} & \numprint{.4223} & \numprint{.3659} & \numprint{.3135} & \numprint{.2492} & \numprint{.1811} & \numprint{.0822} & \numprint{.0567} & \numprint{.0539} & \numprint{.0543} \\
\cline{1-13}
\multirow[t]{5}{*}{NextItNet} & HR@1 &  {\numprint{.2562}} & \numprint{.201} & \numprint{.1502} & \numprint{.0901} & \numprint{.0433} & \numprint{.0304} & \numprint{.0118} & \numprint{.01} & \numprint{.0099} & \numprint{.0084} & \numprint{.0105} \\
 & HR@5 &  {\numprint{.4747}} & \numprint{.3837} & \numprint{.3089} & \numprint{.2075} & \numprint{.1022} & \numprint{.0814} & \numprint{.0402} & \numprint{.0354} & \numprint{.0361} & \numprint{.0337} & \numprint{.035} \\
 & HR@10 &  {\numprint{.5656}} & \numprint{.465} & \numprint{.3837} & \numprint{.2692} & \numprint{.1394} & \numprint{.1146} & \numprint{.0636} & \numprint{.0548} & \numprint{.0604} & \numprint{.0552} & \numprint{.062} \\
 & NDCG@5 &\numprint{.3715} & \numprint{.2974} & \numprint{.2332} & \numprint{.1511} & \numprint{.0736} & \numprint{.0565} & \numprint{.0258} & \numprint{.0225} & \numprint{.0229} & \numprint{.0209} & \numprint{.0228} \\
 & NDCG@10 &  {\numprint{.4009}} & \numprint{.3236} & \numprint{.2574} & \numprint{.171} & \numprint{.0856} & \numprint{.0672} & \numprint{.0333} & \numprint{.0288} & \numprint{.0306} & \numprint{.0278} & \numprint{.0315} \\
\cline{1-13}
\multirow[t]{5}{*}{SASRec} & HR@1 & \numprint{.3667} & \numprint{.3443} & \numprint{.3032} & \numprint{.2545} & \numprint{.2255} & \numprint{.1957} & \numprint{.1489} & \numprint{.1121} & \numprint{.0638} & \numprint{.0261} & \numprint{.0214} \\
 & HR@5 & \numprint{.6313} & \numprint{.5779} & \numprint{.5145} & \numprint{.4362} & \numprint{.3886} & \numprint{.3299} & \numprint{.2656} & \numprint{.1974} & \numprint{.1334} & \numprint{.0799} & \numprint{.0665} \\
 & HR@10 & \numprint{.7249} & \numprint{.6624} & \numprint{.5887} & \numprint{.5053} & \numprint{.4481} & \numprint{.3799} & \numprint{.3133} & \numprint{.2385} & \numprint{.1711} & \numprint{.1178} & \numprint{.1053} \\
 & NDCG@5 & \numprint{.5082} & \numprint{.4685} & \numprint{.4156} & \numprint{.3512} & \numprint{.3124} & \numprint{.2671} & \numprint{.2109} & \numprint{.157} & \numprint{.1} & \numprint{.0534} & \numprint{.0435} \\
 & NDCG@10 & \numprint{.5386} & \numprint{.4959} & \numprint{.4398} & \numprint{.3736} & \numprint{.3318} & \numprint{.2832} & \numprint{.2263} & \numprint{.1702} & \numprint{.1122} & \numprint{.0656} & \numprint{.0559} \\
\cline{1-13}
\bottomrule
\end{tabular}

%% file: main.bbl

\begin{thebibliography}{41}


\ifx \showCODEN    \undefined \def \showCODEN     #1{\unskip}     \fi
\ifx \showDOI      \undefined \def \showDOI       #1{#1}\fi
\ifx \showISBNx    \undefined \def \showISBNx     #1{\unskip}     \fi
\ifx \showISBNxiii \undefined \def \showISBNxiii  #1{\unskip}     \fi
\ifx \showISSN     \undefined \def \showISSN      #1{\unskip}     \fi
\ifx \showLCCN     \undefined \def \showLCCN      #1{\unskip}     \fi
\ifx \shownote     \undefined \def \shownote      #1{#1}          \fi
\ifx \showarticletitle \undefined \def \showarticletitle #1{#1}   \fi
\ifx \showURL      \undefined \def \showURL       {\relax}        \fi
\providecommand\bibfield[2]{#2}
\providecommand\bibinfo[2]{#2}
\providecommand\natexlab[1]{#1}
\providecommand\showeprint[2][]{arXiv:#2}

\bibitem[Chen et~al\mbox{.}(2019)]%
        {chen_behavior_2019}
\bibfield{author}{\bibinfo{person}{Qiwei Chen}, \bibinfo{person}{Huan Zhao}, \bibinfo{person}{Wei Li}, \bibinfo{person}{Pipei Huang}, {and} \bibinfo{person}{Wenwu Ou}.} \bibinfo{year}{2019}\natexlab{}.
\newblock \showarticletitle{Behavior {Sequence} {Transformer} for {E}-{Commerce} {Recommendation} in {Alibaba}}. In \bibinfo{booktitle}{\emph{Proceedings of the 1st {International} {Workshop} on {Deep} {Learning} {Practice} for {High}-{Dimensional} {Sparse} {Data}}}. \bibinfo{publisher}{ACM}.
\newblock
\urldef\tempurl%
\url{https://doi.org/10.1145/3326937.3341261}
\showDOI{\tempurl}


\bibitem[Clark et~al\mbox{.}(2020)]%
        {clark_electra_2020}
\bibfield{author}{\bibinfo{person}{Kevin Clark}, \bibinfo{person}{Minh-Thang Luong}, \bibinfo{person}{Quoc~V. Le}, {and} \bibinfo{person}{Christopher~D. Manning}.} \bibinfo{year}{2020}\natexlab{}.
\newblock \showarticletitle{{ELECTRA}: {Pre}-training {Text} {Encoders} as {Discriminators} {Rather} {Than} {Generators}}. In \bibinfo{booktitle}{\emph{{CoRR}}}, Vol.~\bibinfo{volume}{abs/2003.10555}.
\newblock
\urldef\tempurl%
\url{https://arxiv.org/abs/2003.10555}
\showURL{%
\tempurl}


\bibitem[Dallmann et~al\mbox{.}(2021)]%
        {dallmann_case_2021}
\bibfield{author}{\bibinfo{person}{Alexander Dallmann}, \bibinfo{person}{Daniel Zoller}, {and} \bibinfo{person}{Andreas Hotho}.} \bibinfo{year}{2021}\natexlab{}.
\newblock \showarticletitle{A {Case} {Study} on {Sampling} {Strategies} for {Evaluating} {Neural} {Sequential} {Item} {Recommendation} {Models}}. In \bibinfo{booktitle}{\emph{Proceedings of the 15th {ACM} {Conference} on {Recommender} {Systems}}} \emph{(\bibinfo{series}{{RecSys} '21})}. \bibinfo{publisher}{Association for Computing Machinery}, \bibinfo{address}{New York, NY, USA}, \bibinfo{pages}{505--514}.
\newblock
\showISBNx{978-1-4503-8458-2}
\urldef\tempurl%
\url{https://doi.org/10.1145/3460231.3475943}
\showDOI{\tempurl}


\bibitem[de~Souza Pereira~Moreira et~al\mbox{.}(2021)]%
        {de_souza_pereira_moreira_transformers4rec_2021}
\bibfield{author}{\bibinfo{person}{Gabriel de Souza Pereira~Moreira}, \bibinfo{person}{Sara Rabhi}, \bibinfo{person}{Jeong~Min Lee}, \bibinfo{person}{Ronay Ak}, {and} \bibinfo{person}{Even Oldridge}.} \bibinfo{year}{2021}\natexlab{}.
\newblock \showarticletitle{{Transformers4Rec}: {Bridging} the {Gap} between {NLP} and {Sequential} / {Session}-{Based} {Recommendation}}. In \bibinfo{booktitle}{\emph{Proceedings of the 15th {ACM} {Conference} on {Recommender} {Systems}}} \emph{(\bibinfo{series}{{RecSys} '21})}. \bibinfo{publisher}{Association for Computing Machinery}, \bibinfo{address}{New York, NY, USA}, \bibinfo{pages}{143--153}.
\newblock
\showISBNx{978-1-4503-8458-2}
\urldef\tempurl%
\url{https://doi.org/10.1145/3460231.3474255}
\showDOI{\tempurl}
\newblock
\shownote{event-place: Amsterdam, Netherlands}.


\bibitem[Devlin et~al\mbox{.}(2019)]%
        {devlin_bert_2019}
\bibfield{author}{\bibinfo{person}{Jacob Devlin}, \bibinfo{person}{Ming-Wei Chang}, \bibinfo{person}{Kenton Lee}, {and} \bibinfo{person}{Kristina Toutanova}.} \bibinfo{year}{2019}\natexlab{}.
\newblock \showarticletitle{{BERT}: {Pre}-training of {Deep} {Bidirectional} {Transformers} for {Language} {Understanding}}. In \bibinfo{booktitle}{\emph{Proceedings of the 2019 {Conference} of the {North} {American} {Chapter} of the {Association} for {Computational} {Linguistics}: {Human} {Language} {Technologies}, {Volume} 1 ({Long} and {Short} {Papers})}}. \bibinfo{publisher}{Association for Computational Linguistics}, \bibinfo{address}{Minneapolis, Minnesota}, \bibinfo{pages}{4171--4186}.
\newblock
\urldef\tempurl%
\url{https://doi.org/10.18653/v1/N19-1423}
\showDOI{\tempurl}


\bibitem[Fan et~al\mbox{.}(2021)]%
        {fan_lighter_2021}
\bibfield{author}{\bibinfo{person}{Xinyan Fan}, \bibinfo{person}{Zheng Liu}, \bibinfo{person}{Jianxun Lian}, \bibinfo{person}{Wayne Zhao}, \bibinfo{person}{Xing Xie}, {and} \bibinfo{person}{Ji-Rong Wen}.} \bibinfo{year}{2021}\natexlab{}.
\newblock \showarticletitle{Lighter and {Better}: {Low}-{Rank} {Decomposed} {Self}-{Attention} {Networks} for {Next}-{Item} {Recommendation}}. In \bibinfo{booktitle}{\emph{Proceedings of the 44th {International} {ACM} {SIGIR} {Conference} on {Research} and {Development} in {Information} {Retrieval}}}. \bibinfo{publisher}{ACM}, \bibinfo{address}{Virtual Event Canada}, \bibinfo{pages}{1733--1737}.
\newblock
\showISBNx{978-1-4503-8037-9}
\urldef\tempurl%
\url{https://doi.org/10.1145/3404835.3462978}
\showDOI{\tempurl}


\bibitem[Fischer et~al\mbox{.}(2023)]%
        {fischer_enhancing_2023}
\bibfield{author}{\bibinfo{person}{Elisabeth Fischer}, \bibinfo{person}{Daniel Schlör}, \bibinfo{person}{Albin Zehe}, {and} \bibinfo{person}{Andreas Hotho}.} \bibinfo{year}{2023}\natexlab{}.
\newblock \showarticletitle{Enhancing {Sequential} {Next}-{Item} {Prediction} through {Modelling} {Non}-{Item} {Pages}}. In \bibinfo{booktitle}{\emph{2023 {IEEE} {International} {Conference} on {Data} {Mining} {Workshops} ({ICDMW})}}. \bibinfo{pages}{128--136}.
\newblock
\urldef\tempurl%
\url{https://doi.org/10.1109/ICDMW60847.2023.00024}
\showDOI{\tempurl}


\bibitem[Fischer et~al\mbox{.}(2020)]%
        {fischer_integrating_2020}
\bibfield{author}{\bibinfo{person}{Elisabeth Fischer}, \bibinfo{person}{Daniel Zoller}, \bibinfo{person}{Alexander Dallmann}, {and} \bibinfo{person}{Andreas Hotho}.} \bibinfo{year}{2020}\natexlab{}.
\newblock \showarticletitle{Integrating {Keywords} into {BERT4Rec} for {Sequential} {Recommendation}}. In \bibinfo{booktitle}{\emph{{KI} 2020: {Advances} in {Artificial} {Intelligence}}}, \bibfield{editor}{\bibinfo{person}{Ute Schmid}, \bibinfo{person}{Franziska Klügl}, {and} \bibinfo{person}{Diedrich Wolter}} (Eds.). \bibinfo{publisher}{Springer International Publishing}, \bibinfo{address}{Cham}, \bibinfo{pages}{275--282}.
\newblock
\showISBNx{978-3-030-58285-2}
\urldef\tempurl%
\url{https://doi.org/10.1007/978-3-030-58285-2_23}
\showDOI{\tempurl}


\bibitem[Fischer et~al\mbox{.}(2021)]%
        {fischer_comparison_2021}
\bibfield{author}{\bibinfo{person}{Elisabeth Fischer}, \bibinfo{person}{Daniel Zoller}, {and} \bibinfo{person}{Andreas Hotho}.} \bibinfo{year}{2021}\natexlab{}.
\newblock \showarticletitle{Comparison of {Transformer}-{Based} {Sequential} {Product} {Recommendation} {Models} for the {Coveo} {Data} {Challenge}}. In \bibinfo{booktitle}{\emph{{SIGIR} 2021 {Workshop} on {eCommerce}}}.
\newblock


\bibitem[Hidasi and Karatzoglou(2018)]%
        {hidasi_recurrent_2018}
\bibfield{author}{\bibinfo{person}{Balázs Hidasi} {and} \bibinfo{person}{Alexandros Karatzoglou}.} \bibinfo{year}{2018}\natexlab{}.
\newblock \showarticletitle{Recurrent neural networks with top-k gains for session-based recommendations}. In \bibinfo{booktitle}{\emph{Proceedings of the 27th {ACM} international conference on information and knowledge management}}. \bibinfo{pages}{843--852}.
\newblock


\bibitem[Hidasi et~al\mbox{.}(2016a)]%
        {hidasi_session-based_2016}
\bibfield{author}{\bibinfo{person}{Balázs Hidasi}, \bibinfo{person}{Alexandros Karatzoglou}, \bibinfo{person}{Linas Baltrunas}, {and} \bibinfo{person}{Domonkos Tikk}.} \bibinfo{year}{2016}\natexlab{a}.
\newblock \showarticletitle{Session-based {Recommendations} with {Recurrent} {Neural} {Networks}}. In \bibinfo{booktitle}{\emph{{ICLR} ({Poster})}}, \bibfield{editor}{\bibinfo{person}{Yoshua Bengio} {and} \bibinfo{person}{Yann LeCun}} (Eds.).
\newblock


\bibitem[Hidasi et~al\mbox{.}(2016b)]%
        {hidasi_parallel_2016}
\bibfield{author}{\bibinfo{person}{Balázs Hidasi}, \bibinfo{person}{Massimo Quadrana}, \bibinfo{person}{Alexandros Karatzoglou}, {and} \bibinfo{person}{Domonkos Tikk}.} \bibinfo{year}{2016}\natexlab{b}.
\newblock \showarticletitle{Parallel {Recurrent} {Neural} {Network} {Architectures} for {Feature}-rich {Session}-based {Recommendations}}. In \bibinfo{booktitle}{\emph{Proceedings of the 10th {ACM} {Conference} on {Recommender} {Systems}}} \emph{(\bibinfo{series}{{RecSys} '16})}. \bibinfo{publisher}{ACM}, \bibinfo{address}{New York, NY, USA}, \bibinfo{pages}{241--248}.
\newblock
\urldef\tempurl%
\url{https://doi.org/10.1145/2959100.2959167}
\showDOI{\tempurl}
\newblock
\shownote{event-place: Boston, Massachusetts, USA}.


\bibitem[Hou et~al\mbox{.}(2022)]%
        {hou_core_2022}
\bibfield{author}{\bibinfo{person}{Yupeng Hou}, \bibinfo{person}{Binbin Hu}, \bibinfo{person}{Zhiqiang Zhang}, {and} \bibinfo{person}{Wayne~Xin Zhao}.} \bibinfo{year}{2022}\natexlab{}.
\newblock \showarticletitle{{CORE}: {Simple} and {Effective} {Session}-based {Recommendation} within {Consistent} {Representation} {Space}}. In \bibinfo{booktitle}{\emph{Proceedings of the 45th {International} {ACM} {SIGIR} {Conference} on {Research} and {Development} in {Information} {Retrieval}}} \emph{(\bibinfo{series}{{SIGIR} '22})}. \bibinfo{publisher}{Association for Computing Machinery}, \bibinfo{address}{New York, NY, USA}, \bibinfo{pages}{1796--1801}.
\newblock
\showISBNx{978-1-4503-8732-3}
\urldef\tempurl%
\url{https://doi.org/10.1145/3477495.3531955}
\showDOI{\tempurl}
\newblock
\shownote{event-place: Madrid, Spain}.


\bibitem[Ishihara et~al\mbox{.}(2021)]%
        {ishihara_adversarial_2021}
\bibfield{author}{\bibinfo{person}{Shotaro Ishihara}, \bibinfo{person}{Shuhei Goda}, {and} \bibinfo{person}{Hidehisa Arai}.} \bibinfo{year}{2021}\natexlab{}.
\newblock \showarticletitle{Adversarial validation to select validation data for evaluating performance in e-commerce purchase intent prediction}. In \bibinfo{booktitle}{\emph{{SIGIR} 2021 {Workshop} on {eCommerce}}}.
\newblock


\bibitem[Jagatap et~al\mbox{.}(2023)]%
        {jagatap_attribert_2023}
\bibfield{author}{\bibinfo{person}{Akshay Jagatap}, \bibinfo{person}{Nikki Gupta}, \bibinfo{person}{Sachin Farfade}, {and} \bibinfo{person}{Prakash~Mandayam Comar}.} \bibinfo{year}{2023}\natexlab{}.
\newblock \showarticletitle{{AttriBERT} - {Session}-based {Product} {Attribute} {Recommendation} with {BERT}}. In \bibinfo{booktitle}{\emph{Proceedings of the 46th {International} {ACM} {SIGIR} {Conference} on {Research} and {Development} in {Information} {Retrieval}}} \emph{(\bibinfo{series}{{SIGIR} '23})}. \bibinfo{publisher}{Association for Computing Machinery}, \bibinfo{address}{New York, NY, USA}, \bibinfo{pages}{3421--3425}.
\newblock
\showISBNx{978-1-4503-9408-6}
\urldef\tempurl%
\url{https://doi.org/10.1145/3539618.3594714}
\showDOI{\tempurl}
\newblock
\shownote{event-place: Taipei, Taiwan}.


\bibitem[Kang and McAuley(2018)]%
        {kang_self-attentive_2018}
\bibfield{author}{\bibinfo{person}{W. Kang} {and} \bibinfo{person}{J. McAuley}.} \bibinfo{year}{2018}\natexlab{}.
\newblock \showarticletitle{Self-{Attentive} {Sequential} {Recommendation}}. In \bibinfo{booktitle}{\emph{2018 {IEEE} {International} {Conference} on {Data} {Mining} ({ICDM})}}. \bibinfo{publisher}{IEEE Computer Society}, \bibinfo{address}{Los Alamitos, CA, USA}, \bibinfo{pages}{197--206}.
\newblock
\urldef\tempurl%
\url{https://doi.org/10.1109/ICDM.2018.00035}
\showDOI{\tempurl}


\bibitem[Koopmann et~al\mbox{.}(2024)]%
        {koopmann_comptrails_2024}
\bibfield{author}{\bibinfo{person}{Tobias Koopmann}, \bibinfo{person}{Martin Becker}, \bibinfo{person}{Florian Lemmerich}, {and} \bibinfo{person}{Andreas Hotho}.} \bibinfo{year}{2024}\natexlab{}.
\newblock \showarticletitle{{CompTrails}: comparing hypotheses across behavioral networks}.
\newblock \bibinfo{journal}{\emph{Data Min. Knowl. Discov.}} \bibinfo{volume}{38}, \bibinfo{number}{3} (\bibinfo{date}{Jan.} \bibinfo{year}{2024}), \bibinfo{pages}{1258--1288}.
\newblock
\showISSN{1384-5810}
\urldef\tempurl%
\url{https://doi.org/10.1007/s10618-023-00996-8}
\showDOI{\tempurl}
\newblock
\shownote{Place: USA Publisher: Kluwer Academic Publishers}.


\bibitem[Li et~al\mbox{.}(2017)]%
        {li_neural_2017}
\bibfield{author}{\bibinfo{person}{Jing Li}, \bibinfo{person}{Pengjie Ren}, \bibinfo{person}{Zhumin Chen}, \bibinfo{person}{Zhaochun Ren}, \bibinfo{person}{Tao Lian}, {and} \bibinfo{person}{Jun Ma}.} \bibinfo{year}{2017}\natexlab{}.
\newblock \showarticletitle{Neural {Attentive} {Session}-based {Recommendation}}. In \bibinfo{booktitle}{\emph{Proceedings of the 2017 {ACM} on {Conference} on {Information} and {Knowledge} {Management}}} \emph{(\bibinfo{series}{{CIKM} '17})}. \bibinfo{publisher}{Association for Computing Machinery}, \bibinfo{address}{New York, NY, USA}, \bibinfo{pages}{1419--1428}.
\newblock
\showISBNx{978-1-4503-4918-5}
\urldef\tempurl%
\url{https://doi.org/10.1145/3132847.3132926}
\showDOI{\tempurl}
\newblock
\shownote{event-place: Singapore, Singapore}.


\bibitem[Liu et~al\mbox{.}(2021)]%
        {liu_non-invasive_2021}
\bibfield{author}{\bibinfo{person}{Chang Liu}, \bibinfo{person}{Xiaoguang Li}, \bibinfo{person}{Guohao Cai}, \bibinfo{person}{Zhenhua Dong}, \bibinfo{person}{Hong Zhu}, {and} \bibinfo{person}{Lifeng Shang}.} \bibinfo{year}{2021}\natexlab{}.
\newblock \showarticletitle{Noninvasive {Self}-attention for {Side} {Information} {Fusion} in {Sequential} {Recommendation}}.
\newblock \bibinfo{journal}{\emph{Proceedings of the AAAI Conference on Artificial Intelligence}} \bibinfo{volume}{35}, \bibinfo{number}{5} (\bibinfo{date}{May} \bibinfo{year}{2021}), \bibinfo{pages}{4249--4256}.
\newblock
\showISSN{2374-3468}
\urldef\tempurl%
\url{https://doi.org/10.1609/aaai.v35i5.16549}
\showDOI{\tempurl}
\newblock
\shownote{Number: 5}.


\bibitem[Liu et~al\mbox{.}(2016)]%
        {liu_context-aware_2016}
\bibfield{author}{\bibinfo{person}{Qiang Liu}, \bibinfo{person}{Shu Wu}, \bibinfo{person}{Diyi Wang}, \bibinfo{person}{Zhaokang Li}, {and} \bibinfo{person}{Liang Wang}.} \bibinfo{year}{2016}\natexlab{}.
\newblock \showarticletitle{Context-{Aware} {Sequential} {Recommendation}}. In \bibinfo{booktitle}{\emph{2016 {IEEE} 16th {International} {Conference} on {Data} {Mining} ({ICDM})}}. \bibinfo{pages}{1053--1058}.
\newblock
\urldef\tempurl%
\url{https://doi.org/10.1109/ICDM.2016.0135}
\showDOI{\tempurl}


\bibitem[Moreira et~al\mbox{.}(2021)]%
        {moreira_transformers_2021}
\bibfield{author}{\bibinfo{person}{Gabriel de Souza~P. Moreira}, \bibinfo{person}{Sara Rabhi}, \bibinfo{person}{Ronay Ak}, \bibinfo{person}{Md~Yasin Kabir}, {and} \bibinfo{person}{Even Oldridge}.} \bibinfo{year}{2021}\natexlab{}.
\newblock \showarticletitle{Transformers with multi-modal features and post-fusion context for e-commerce session-based recommendation}. In \bibinfo{booktitle}{\emph{{SIGIR} 2021 {Workshop} on {eCommerce}}}.
\newblock
\newblock
\shownote{\_eprint: 2107.05124}.


\bibitem[Petrov and Macdonald(2022)]%
        {petrov_systematic_2022}
\bibfield{author}{\bibinfo{person}{Aleksandr Petrov} {and} \bibinfo{person}{Craig Macdonald}.} \bibinfo{year}{2022}\natexlab{}.
\newblock \showarticletitle{A {Systematic} {Review} and {Replicability} {Study} of {BERT4Rec} for {Sequential} {Recommendation}}. In \bibinfo{booktitle}{\emph{Proceedings of the 16th {ACM} {Conference} on {Recommender} {Systems}}} \emph{(\bibinfo{series}{{RecSys} '22})}. \bibinfo{publisher}{Association for Computing Machinery}, \bibinfo{address}{New York, NY, USA}, \bibinfo{pages}{436--447}.
\newblock
\showISBNx{978-1-4503-9278-5}
\urldef\tempurl%
\url{https://doi.org/10.1145/3523227.3548487}
\showDOI{\tempurl}
\newblock
\shownote{event-place: Seattle, WA, USA}.


\bibitem[Sakatani(2021)]%
        {sakatani_session-based_2021}
\bibfield{author}{\bibinfo{person}{Yoshihiro Sakatani}.} \bibinfo{year}{2021}\natexlab{}.
\newblock \showarticletitle{Session-based {Recommendation} {Using} an {Ensemble} of {LSTM}-and {Matrix} {Factorization}-based {Models}}. In \bibinfo{booktitle}{\emph{{SIGIR} 2021 {Workshop} on {eCommerce}}}.
\newblock


\bibitem[Singer et~al\mbox{.}(2015)]%
        {singer_hyptrails_2015}
\bibfield{author}{\bibinfo{person}{Philipp Singer}, \bibinfo{person}{Denis Helic}, \bibinfo{person}{Andreas Hotho}, {and} \bibinfo{person}{Markus Strohmaier}.} \bibinfo{year}{2015}\natexlab{}.
\newblock \showarticletitle{{HypTrails}: {A} {Bayesian} {Approach} for {Comparing} {Hypotheses} {About} {Human} {Trails} on the {Web}}. In \bibinfo{booktitle}{\emph{Proceedings of the 24th {International} {Conference} on {World} {Wide} {Web}}} \emph{(\bibinfo{series}{{WWW} ’15})}. \bibinfo{publisher}{International World Wide Web Conferences Steering Committee}.
\newblock
\urldef\tempurl%
\url{https://doi.org/10.1145/2736277.2741080}
\showDOI{\tempurl}


\bibitem[Sun et~al\mbox{.}(2019)]%
        {sun_bert4rec_2019}
\bibfield{author}{\bibinfo{person}{Fei Sun}, \bibinfo{person}{Jun Liu}, \bibinfo{person}{Jian Wu}, \bibinfo{person}{Changhua Pei}, \bibinfo{person}{Xiao Lin}, \bibinfo{person}{Wenwu Ou}, {and} \bibinfo{person}{Peng Jiang}.} \bibinfo{year}{2019}\natexlab{}.
\newblock \showarticletitle{{BERT4Rec}: {Sequential} {Recommendation} with {Bidirectional} {Encoder} {Representations} from {Transformer}}. In \bibinfo{booktitle}{\emph{Proceedings of the 28th {ACM} {International} {Conference} on {Information} and {Knowledge} {Management} - {CIKM} 19}}. \bibinfo{publisher}{ACM Press}.
\newblock
\urldef\tempurl%
\url{https://doi.org/10.1145/3357384.3357895}
\showDOI{\tempurl}


\bibitem[Tagliabue et~al\mbox{.}(2021)]%
        {tagliabue_sigir_2021}
\bibfield{author}{\bibinfo{person}{Jacopo Tagliabue}, \bibinfo{person}{Ciro Greco}, \bibinfo{person}{Jean-Francis Roy}, \bibinfo{person}{Bingqing Yu}, \bibinfo{person}{Patrick~John Chia}, \bibinfo{person}{Federico Bianchi}, {and} \bibinfo{person}{Giovanni Cassani}.} \bibinfo{year}{2021}\natexlab{}.
\newblock \showarticletitle{{SIGIR} 2021 {E}-{Commerce} {Workshop} {Data} {Challenge}}. In \bibinfo{booktitle}{\emph{{SIGIR} 2021 {Workshop} on {eCommerce}}}.
\newblock
\newblock
\shownote{\_eprint: 2104.09423}.


\bibitem[Tan et~al\mbox{.}(2016)]%
        {tan_improved_2016}
\bibfield{author}{\bibinfo{person}{Yong~Kiam Tan}, \bibinfo{person}{Xinxing Xu}, {and} \bibinfo{person}{Yong Liu}.} \bibinfo{year}{2016}\natexlab{}.
\newblock \showarticletitle{Improved {Recurrent} {Neural} {Networks} for {Session}-based {Recommendations}}. In \bibinfo{booktitle}{\emph{Proceedings of the 1st workshop on deep learning for recommender systems}}, Vol.~\bibinfo{volume}{abs/1606.08117}. \bibinfo{pages}{17--22}.
\newblock
\urldef\tempurl%
\url{http://arxiv.org/abs/1606.08117}
\showURL{%
\tempurl}
\newblock
\shownote{arXiv: 1606.08117}.


\bibitem[Tang and Wang(2018)]%
        {tang_personalized_2018}
\bibfield{author}{\bibinfo{person}{Jiaxi Tang} {and} \bibinfo{person}{Ke Wang}.} \bibinfo{year}{2018}\natexlab{}.
\newblock \showarticletitle{Personalized {Top}-{N} {Sequential} {Recommendation} via {Convolutional} {Sequence} {Embedding}}. In \bibinfo{booktitle}{\emph{Proceedings of the {Eleventh} {ACM} {International} {Conference} on {Web} {Search} and {Data} {Mining}}} \emph{(\bibinfo{series}{{WSDM} '18})}. \bibinfo{publisher}{Association for Computing Machinery}, \bibinfo{address}{New York, NY, USA}, \bibinfo{pages}{565--573}.
\newblock
\showISBNx{978-1-4503-5581-0}
\urldef\tempurl%
\url{https://doi.org/10.1145/3159652.3159656}
\showDOI{\tempurl}
\newblock
\shownote{event-place: Marina Del Rey, CA, USA}.


\bibitem[Taylor(1953)]%
        {taylor_cloze_1953}
\bibfield{author}{\bibinfo{person}{Wilson~L. Taylor}.} \bibinfo{year}{1953}\natexlab{}.
\newblock \showarticletitle{“{Cloze} {Procedure}”: {A} {New} {Tool} for {Measuring} {Readability}}.
\newblock \bibinfo{journal}{\emph{Journalism \& Mass Communication Quarterly}} \bibinfo{volume}{30}, \bibinfo{number}{4} (\bibinfo{date}{Sept.} \bibinfo{year}{1953}), \bibinfo{pages}{415--433}.
\newblock
\urldef\tempurl%
\url{https://doi.org/10.1177/107769905303000401}
\showDOI{\tempurl}


\bibitem[Tuan and Phuong(2017)]%
        {tuan_3d_2017}
\bibfield{author}{\bibinfo{person}{Trinh~Xuan Tuan} {and} \bibinfo{person}{Tu~Minh Phuong}.} \bibinfo{year}{2017}\natexlab{}.
\newblock \showarticletitle{{3D} {Convolutional} {Networks} for {Session}-based {Recommendation} with {Content} {Features}}. In \bibinfo{booktitle}{\emph{Proceedings of the {Eleventh} {ACM} {Conference} on {Recommender} {Systems}}} \emph{(\bibinfo{series}{{RecSys} '17})}. \bibinfo{publisher}{ACM}, \bibinfo{address}{New York, NY, USA}, \bibinfo{pages}{138--146}.
\newblock
\urldef\tempurl%
\url{https://doi.org/10.1145/3109859.3109900}
\showDOI{\tempurl}
\newblock
\shownote{event-place: Como, Italy}.


\bibitem[van~der Maaten and Hinton(2008)]%
        {maaten_visualizing_2008}
\bibfield{author}{\bibinfo{person}{Laurens van~der Maaten} {and} \bibinfo{person}{Geoffrey Hinton}.} \bibinfo{year}{2008}\natexlab{}.
\newblock \showarticletitle{Visualizing {Data} using t-{SNE}}.
\newblock \bibinfo{journal}{\emph{Journal of Machine Learning Research}}  \bibinfo{volume}{9} (\bibinfo{year}{2008}), \bibinfo{pages}{2579--2605}.
\newblock
\urldef\tempurl%
\url{http://www.jmlr.org/papers/v9/vandermaaten08a.html}
\showURL{%
\tempurl}


\bibitem[Vaswani et~al\mbox{.}(2017)]%
        {vaswani_attention_2017}
\bibfield{author}{\bibinfo{person}{Ashish Vaswani}, \bibinfo{person}{Noam Shazeer}, \bibinfo{person}{Niki Parmar}, \bibinfo{person}{Jakob Uszkoreit}, \bibinfo{person}{Llion Jones}, \bibinfo{person}{Aidan~N Gomez}, \bibinfo{person}{Lukasz Kaiser}, {and} \bibinfo{person}{Illia Polosukhin}.} \bibinfo{year}{2017}\natexlab{}.
\newblock \showarticletitle{Attention is all you need}. In \bibinfo{booktitle}{\emph{Advances in neural information processing systems}}. \bibinfo{pages}{5998--6008}.
\newblock


\bibitem[Wattenberg et~al\mbox{.}(2016)]%
        {wattenberg_how_2016}
\bibfield{author}{\bibinfo{person}{Martin Wattenberg}, \bibinfo{person}{Fernanda Viégas}, {and} \bibinfo{person}{Ian Johnson}.} \bibinfo{year}{2016}\natexlab{}.
\newblock \showarticletitle{How to {Use} t-{SNE} {Effectively}}.
\newblock \bibinfo{journal}{\emph{Distill}} (\bibinfo{year}{2016}).
\newblock
\urldef\tempurl%
\url{https://doi.org/10.23915/distill.00002}
\showDOI{\tempurl}


\bibitem[Wu et~al\mbox{.}(2020)]%
        {wu_sse-pt_2020}
\bibfield{author}{\bibinfo{person}{Liwei Wu}, \bibinfo{person}{Shuqing Li}, \bibinfo{person}{Cho-Jui Hsieh}, {and} \bibinfo{person}{James Sharpnack}.} \bibinfo{year}{2020}\natexlab{}.
\newblock \showarticletitle{{SSE}-{PT}: {Sequential} {Recommendation} {Via} {Personalized} {Transformer}}. In \bibinfo{booktitle}{\emph{Proceedings of the 14th {ACM} {Conference} on {Recommender} {Systems}}} \emph{(\bibinfo{series}{{RecSys} '20})}. \bibinfo{publisher}{Association for Computing Machinery}, \bibinfo{address}{New York, NY, USA}, \bibinfo{pages}{328--337}.
\newblock
\showISBNx{978-1-4503-7583-2}
\urldef\tempurl%
\url{https://doi.org/10.1145/3383313.3412258}
\showDOI{\tempurl}
\newblock
\shownote{event-place: Virtual Event, Brazil}.


\bibitem[Xie et~al\mbox{.}(2022)]%
        {xie_decoupled_2022}
\bibfield{author}{\bibinfo{person}{Yueqi Xie}, \bibinfo{person}{Peilin Zhou}, {and} \bibinfo{person}{Sunghun Kim}.} \bibinfo{year}{2022}\natexlab{}.
\newblock \showarticletitle{Decoupled {Side} {Information} {Fusion} for {Sequential} {Recommendation}}. In \bibinfo{booktitle}{\emph{Proceedings of the 45th {International} {ACM} {SIGIR} {Conference} on {Research} and {Development} in {Information} {Retrieval}}} \emph{(\bibinfo{series}{{SIGIR} '22})}. \bibinfo{publisher}{Association for Computing Machinery}, \bibinfo{address}{New York, NY, USA}, \bibinfo{pages}{1611--1621}.
\newblock
\showISBNx{978-1-4503-8732-3}
\urldef\tempurl%
\url{https://doi.org/10.1145/3477495.3531963}
\showDOI{\tempurl}
\newblock
\shownote{event-place: Madrid, Spain}.


\bibitem[Xu et~al\mbox{.}(2019)]%
        {xu_recurrent_2019}
\bibfield{author}{\bibinfo{person}{Chengfeng Xu}, \bibinfo{person}{Pengpeng Zhao}, \bibinfo{person}{Yanchi Liu}, \bibinfo{person}{Jiajie Xu}, \bibinfo{person}{Victor~S.Sheng S.Sheng}, \bibinfo{person}{Zhiming Cui}, \bibinfo{person}{Xiaofang Zhou}, {and} \bibinfo{person}{Hui Xiong}.} \bibinfo{year}{2019}\natexlab{}.
\newblock \showarticletitle{Recurrent {Convolutional} {Neural} {Network} for {Sequential} {Recommendation}}. In \bibinfo{booktitle}{\emph{The {World} {Wide} {Web} {Conference} on - {WWW}'19}}. \bibinfo{publisher}{ACM Press}.
\newblock
\urldef\tempurl%
\url{https://doi.org/10.1145/3308558.3313408}
\showDOI{\tempurl}


\bibitem[Yang et~al\mbox{.}(2019)]%
        {yang_xlnet_2019}
\bibfield{author}{\bibinfo{person}{Zhilin Yang}, \bibinfo{person}{Zihang Dai}, \bibinfo{person}{Yiming Yang}, \bibinfo{person}{Jaime Carbonell}, \bibinfo{person}{Ruslan Salakhutdinov}, {and} \bibinfo{person}{Quoc~V. Le}.} \bibinfo{year}{2019}\natexlab{}.
\newblock \showarticletitle{{XLNet}: generalized autoregressive pretraining for language understanding}.
\newblock In \bibinfo{booktitle}{\emph{Proceedings of the 33rd {International} {Conference} on {Neural} {Information} {Processing} {Systems}}}. \bibinfo{publisher}{Curran Associates Inc.}, \bibinfo{address}{Red Hook, NY, USA}.
\newblock
\newblock
\shownote{Place: Red Hook, NY, USA Publisher: Curran Associates Inc.}.


\bibitem[Yuan et~al\mbox{.}(2019)]%
        {yuan_simple_2019}
\bibfield{author}{\bibinfo{person}{Fajie Yuan}, \bibinfo{person}{Alexandros Karatzoglou}, \bibinfo{person}{Ioannis Arapakis}, \bibinfo{person}{Joemon~M. Jose}, {and} \bibinfo{person}{Xiangnan He}.} \bibinfo{year}{2019}\natexlab{}.
\newblock \showarticletitle{A {Simple} {Convolutional} {Generative} {Network} for {Next} {Item} {Recommendation}}. In \bibinfo{booktitle}{\emph{Proceedings of the {Twelfth} {ACM} {International} {Conference} on {Web} {Search} and {Data} {Mining}}} \emph{(\bibinfo{series}{{WSDM} '19})}. \bibinfo{publisher}{Association for Computing Machinery}, \bibinfo{address}{New York, NY, USA}, \bibinfo{pages}{582--590}.
\newblock
\showISBNx{978-1-4503-5940-5}
\urldef\tempurl%
\url{https://doi.org/10.1145/3289600.3290975}
\showDOI{\tempurl}
\newblock
\shownote{\_eprint: 1808.05163}.


\bibitem[Zehe et~al\mbox{.}(2024)]%
        {zehe_adapting_2024}
\bibfield{author}{\bibinfo{person}{Albin Zehe}, \bibinfo{person}{Elisabeth Fischer}, \bibinfo{person}{Jonas Kaiser}, \bibinfo{person}{Toni Wagner}, {and} \bibinfo{person}{Andreas Hotho}.} \bibinfo{year}{2024}\natexlab{}.
\newblock \showarticletitle{Adapting {Sequential} {Recommender} {Models} to {Content} {Recommendation} in {Chat} {Data} using {Non}-{Item} {Page}-{Models}}. In \bibinfo{booktitle}{\emph{Proceedings of the {Sixth} {Knowledge}-aware and {Conversational} {Recommender} {Systems} {Workshop}}}. \bibinfo{publisher}{CEUR-WS.org}, \bibinfo{address}{Bari}, \bibinfo{pages}{66--84}.
\newblock
\urldef\tempurl%
\url{https://ceur-ws.org/Vol-3817/}
\showURL{%
\tempurl}


\bibitem[Zhang et~al\mbox{.}(2022)]%
        {zhang_neural_2022}
\bibfield{author}{\bibinfo{person}{Qi Zhang}, \bibinfo{person}{Longbing Cao}, \bibinfo{person}{Chongyang Shi}, {and} \bibinfo{person}{Zhendong Niu}.} \bibinfo{year}{2022}\natexlab{}.
\newblock \showarticletitle{Neural {Time}-{Aware} {Sequential} {Recommendation} by {Jointly} {Modeling} {Preference} {Dynamics} and {Explicit} {Feature} {Couplings}}.
\newblock \bibinfo{journal}{\emph{IEEE Transactions on Neural Networks and Learning Systems}} \bibinfo{volume}{33}, \bibinfo{number}{10} (\bibinfo{year}{2022}), \bibinfo{pages}{5125--5137}.
\newblock
\urldef\tempurl%
\url{https://doi.org/10.1109/TNNLS.2021.3069058}
\showDOI{\tempurl}


\bibitem[Zhao et~al\mbox{.}(2021)]%
        {zhao_recbole_2021}
\bibfield{author}{\bibinfo{person}{Wayne~Xin Zhao}, \bibinfo{person}{Shanlei Mu}, \bibinfo{person}{Yupeng Hou}, \bibinfo{person}{Zihan Lin}, \bibinfo{person}{Yushuo Chen}, \bibinfo{person}{Xingyu Pan}, \bibinfo{person}{Kaiyuan Li}, \bibinfo{person}{Yujie Lu}, \bibinfo{person}{Hui Wang}, \bibinfo{person}{Changxin Tian}, \bibinfo{person}{Yingqian Min}, \bibinfo{person}{Zhichao Feng}, \bibinfo{person}{Xinyan Fan}, \bibinfo{person}{Xu Chen}, \bibinfo{person}{Pengfei Wang}, \bibinfo{person}{Wendi Ji}, \bibinfo{person}{Yaliang Li}, \bibinfo{person}{Xiaoling Wang}, {and} \bibinfo{person}{Ji-Rong Wen}.} \bibinfo{year}{2021}\natexlab{}.
\newblock \showarticletitle{{RecBole}: {Towards} a {Unified}, {Comprehensive} and {Efficient} {Framework} for {Recommendation} {Algorithms}}. In \bibinfo{booktitle}{\emph{{CIKM}}}. \bibinfo{publisher}{ACM}, \bibinfo{pages}{4653--4664}.
\newblock


\end{thebibliography}
